# Finite Strain Robust Topology Optimization Considering Multiple Uncertainties

Nan Feng[1], Guodong Zhang[2] and Kapil Khandelwal[3]

[1]Graduate Student, Dept. of Civil & Env. Engg. & Earth Sci., University of Notre Dame, Notre Dame, IN 46556, United States.

[2]Associate Professor, School of Civil Engineering, Southeast University, 319 Civil Engineering BLDG, Jiulonghu campus, Nanjing, China, 211189.

[3]Associate Professor, Dept. of Civil & Env. Engg. & Earth Sci., 156 Fitzpatrick Hall, University of Notre Dame, Notre Dame, IN 46556, United States, Email: kapil.khandelwal@nd.edu
ORCID: 0000-0002-5748-6019, (Corresponding Author).

## Abstract

This paper presents a computational framework for the robust stiffness design of hyperelastic structures at finite deformations subject to various uncertain sources. In particular, the loading, material properties, and geometry uncertainties are incorporated within the topology optimization framework and are modeled by random vectors or random fields. A stochastic perturbation method is adopted to quantify uncertainties, and analytical adjoint sensitivities are derived for efficient gradient-based optimization. Moreover, the mesh distortion of low-density elements under finite deformations is handled by an adaptive linear energy interpolation scheme. The proposed robust topology optimization framework is applied to several examples, and the effects of different uncertain sources on the optimized topologies are systematically investigated. As demonstrated, robust designs are less sensitive to the variation of target uncertain sources than deterministic designs. Finally, it is shown that incorporating symmetry-breaking uncertainties in the topology optimization framework promotes stable designs compared to the deterministic counterpart, where – when no stability constraint is included – can lead to unstable designs.

**Keywords:** Robust topology optimization; Multiple uncertainties; Stochastic perturbation method; Finite deformation; Hyperelasticity.

## 1 Introduction

Following the pioneering work by Bendsøe and Kikuchi [1], topology optimization has undergone significant advancements in the past three decades; see, for instance, the review articles [2-4]. Moreover, the advancements in additive manufacturing (AM) technologies have enabled new pathways through which complex geometric features can be incorporated into the design and fabrication of novel materials and structures [5-7]. However, in the topology optimization of engineered structures and materials, there might be uncertainties in the loading conditions, material properties, and/or geometric features that should be considered in the design step. For instance, when using AM, there are inherent manufacturing uncertainties due to variations in material properties and geometric features that are related to the manufacturing processes [8-10]. If designed using deterministic topology optimization approaches, the performance of the optimized designs may be adversely affected by uncertainties in the manufacturing and/or operating environments. Therefore, to achieve better-performing designs under uncertainties, robust optimization methods or reliability-based optimization methods that canonically consider various types of uncertain sources are frequently employed [11-13]. This study is concerned with the topology optimization of nonlinear hyperelastic systems under multiple uncertainties.

For *linear elastic* stochastic design problems, various robust topology optimization (RTO) approaches based on different uncertainty quantification (UQ) methods [14, 15] have been proposed in the past decades [16-22]. For instance, Schevenels et al. [23] proposed an RTO framework where the manufacturing uncertainty is parameterized through the projection threshold filter value that is modeled by a *random variable* for uniform uncertainty or by a *random field* for the spatially varying manufacturing uncertainties and the UQ task was carried out via the Monte

Carlo simulation (MCS) where the objective function is a weighted sum of the mean and standard deviation of the structural compliance. Similarly, Lazarvo et al. [24] modeled geometric uncertainties by an anisotropic random field of the projection filter threshold and used the 2nd-order stochastic perturbation method for UQ. Tootkaboni et al. [25] formulated an RTO considering material uncertainties where the objective function was the total weight with constraints on the sum of the mean and standard deviation of structure deflection. The random material field was modeled by a lognormal random field and discretized via polynomial chaos expansion (PCE) together with the intrusive PCE for UQ. With a similar objective function as in Schevenels et al. [23], Silva et al. [26] considered stochastic topology optimization with material uncertainties (i.e., Young's modulus) modeled via an anisotropic random field together with the 2nd-order stochastic perturbation expansion for UQ. Keshavarzzadeh et al. [27] investigated robust and reliability-based topology optimization formulations for designing structures with minimum total material volume subjected to manufacturing uncertainties employing non-intrusive PCE for UQ, where the sum of expectation and standard deviation of compliance is used as a constraint for robust design, whereas the failure probability is used as a constraint for reliability-based design. Later, incorporating load uncertainties, Silva et al. [28] explored a stress-based RTO to minimize total material volume subject to a weighted sum of mean and standard deviation of the von Mises stress constraints wherein the 1st-order stochastic perturbation expansion is used as a surrogate for the von Mises stress field. Both material and geometric uncertainties were considered in Rostami et al. [29] and were modeled by two random fields for designing structures with optimal compliance, where the stochastic collocation method based on an adaptive sparse grid is used to construct a surrogate of a compliance function. Torres et al. [30] proposed an RTO approach via stochastic reduced order models (SROM) considering load uncertainties, where SROM is used to find a small number of optimal stochastic samples that minimize the error between empirical distribution function and the exact distribution of stochastic variables. The UQ was carried out non-intrusively by executing deterministic finite element analysis at these sample points.

On the other hand, for geometrically nonlinear topology optimization, the extension to the stochastic framework has received little attention due to the inherent challenges related to the mesh distortion of low-density elements under finite deformations and the design sensitivities computations. To the best of the authors' knowledge, relevant works can only be found in Jansen et al. [31] and Nishino and Kato [32]. Jansen et al. [31] considered an RTO formulation to minimize the end-compliance where the geometric uncertainties related to field random variables, i.e., nodal coordinates, were incorporated and modeled as a discrete random field using an expansion optimal linear estimation method. The second-order stochastic perturbation method was employed for UQ; however, the sensitivity analysis was conducted via a semi-analytical method that employs computationally expensive difference methods. More recently, Nishino and Kato [32] considered RTO under finite deformations with *loading uncertainties* and presented an analytical adjoint method for sensitivity analysis. The authors also emphasized the need to address mesh distortions that inevitably occur in finite strain topology optimization to obtain robust designs under large deformations.

The currently available finite strain robust topology optimization approaches only incorporate uncertainties related to a single uncertain source, e.g., loads [32] and geometry [31]. No design frameworks consistently incorporate multiple uncertain sources, i.e., material, loading, and geometric uncertainties, which might be present in manufacturing and/or operating environments. Such a framework is needed to study the influence of several types of uncertain sources on a robust design. To this end, a robust topology optimization framework based on a stochastic perturbation technique is proposed in this paper for designing robust topologies subjected to several types of uncertainties, which are modeled by either random vectors or random fields. In particular, the uncertainties related to loads, geometry, and material properties are consistently incorporated in a unified framework and can be simultaneously activated. An adaptive linear energy interpolation scheme is incorporated in the proposed stochastic optimization framework to address the low-density element distortion issues, a critical issue for finite deformation topology optimization. In addition, an analytical adjoint sensitivity analysis is presented for the considered robust topology optimization formulation with multiple uncertainties that obviate the need for computationally expensive difference methods.

The rest of this paper is organized as follows: Section 2 introduces the theoretical background for deterministic topology optimization of hyperelastic structures at finite deformations. Section 3 describes the uncertain sources considered in this study, and Section 4 presents the formulation for robust topology optimization. Section 5 and Section 6 describe the UQ and adjoint sensitivity methods for solving the robust design problem, respectively. Section 7 demonstrates the application of the proposed framework for obtaining robust designs under single and/or multiple uncertain sources for different examples. The conclusions of this work are given in Section 8.

## 2 Deterministic topology optimization

### *2.1 Topological parameterization*

Density-based topology optimization approach determines the optimal structural layout by optimizing material distribution in a design domain $\mathcal{D}_0$ which is discretized by $n_{ele}$ finite elements (FE). The material distribution is parameterized by an elementwise-constant density field represented by a vector $\{\rho_e\}_{e=1}^{n_{ele}}$, where $\rho_e = 1$ and $\rho_e = 0$ represents the presence and absence of material in the $e^{\text{th}}$ element, respectively. To facilitate gradient-based optimization, the element density variable $\rho_e$ is relaxed such that $\rho_e \in [0,1]$. Accordingly, the Solid Isotropic Material with Penalization (SIMP) method [33] is used for the interpolation of Young's modulus in terms of material density and is expressed as

$$E(\rho_e) = \left[\epsilon + (1-\epsilon)\rho_e^p\right]E_0 \tag{1}$$

where $\epsilon$ = $10^{-6}$ is the lower bound parameter and when multiplied with $E_0$ represents Young's modulus for the void material phase ($\rho_e = 0$), and $E_0$ represents Young's modulus for the solid material phase, where the $p \geq 1$ is an implicit penalty parameter to penalize intermediate densities.

To address mesh dependency and stability issues [34], the density filter is considered which replaces the design variable by a weighted average of its values [35-37], i.e.,

$$\widehat{\boldsymbol{\rho}} = \boldsymbol{W}\boldsymbol{x} \quad \text{with} \quad W_{pq} = \frac{w_{pq}v_q}{\sum_{q=1}^{n_{ele}} w_{pq}\, v_q} \quad \text{and} \quad w_{pq} = \max\left(r - \left\|\boldsymbol{X}_p - \boldsymbol{X}_q\right\|_2, 0\right) \tag{2}$$

where $\boldsymbol{x} \in \mathbb{R}^{n_{ele}}$ is a vector containing all design variables in the design domain with $x_e \in [0, 1]$ $e = 1, 2, \dots, n_{ele}$, and $\widehat{\boldsymbol{\rho}} \in \mathbb{R}^{n_{ele}}$ is the filtered density variable and $\boldsymbol{W} \in \mathbb{R}^{n_{ele} \times n_{ele}}$ is the filter matrix. In Eq. (2), $v_q$ represents the volume of the $q^{\text{th}}$ element, $\boldsymbol{X}_p$ and $\boldsymbol{X}_q$ are the centroidal coordinates of the $p^{\text{th}}$ and $q^{\text{th}}$ elements, respectively, and $r$ denotes the filter radius. To obtain the discrete topologies, a Heaviside projection function [38] is used to map filtered density variable $\hat{\rho}_e$ to a density variable $\rho_e$ as follows

$$\rho_e(\hat{\rho}_e) = \frac{\tanh(\beta\eta) + \tanh\big(\beta(\hat{\rho}_e - \eta)\big)}{\tanh(\beta\eta) + \tanh\big(\beta(1 - \eta)\big)} \tag{3}$$

where the parameters $\beta$ and $\eta$ (= 0.5) control the slope and cutoff location of the Heaviside projection function, respectively. Again, $\rho_e \in [0,1]$ and $\rho_e = 0$ indicates void, $\rho_e = 1$ signifies solid, while $0 < \rho_e < 1$ denotes intermediate density.

### 2.2 Nonlinear FEA addressing mesh distortions

For a continuum body in reference configuration $\mathcal{D}_0 \in \mathbb{R}^{\mathrm{sdim}}$, where sdim is the space dimension, with $\boldsymbol{X} \in \mathcal{D}_0$ denoting the position vector of an arbitrary material point in $\mathcal{D}_0$, it is assumed that the deformation of the body from a reference configuration to the current configuration $\mathcal{D}_0 \in \mathbb{R}^{\mathrm{sdim}}$ is described by a bijective smooth map $\boldsymbol{\varphi}: \boldsymbol{X} \to \boldsymbol{x}$ and the displacement field reads $\boldsymbol{u}(\boldsymbol{X}) = \boldsymbol{\varphi}(\boldsymbol{X}) - \boldsymbol{X}$. The strong form of the strain quasi-static boundary value problem in the total Lagrangian form reads: find displacement field ($\boldsymbol{u}(\boldsymbol{X})$) such that

$$\begin{cases} \boldsymbol{\nabla}_{\boldsymbol{X}}.\boldsymbol{P} = \boldsymbol{0} & \text{in } \mathcal{D}_0 \\ \boldsymbol{u} = \bar{\boldsymbol{u}} & \text{on } \partial\mathcal{D}_{0u} \\ \boldsymbol{P}.\widehat{\boldsymbol{N}} = \bar{\boldsymbol{T}} & \text{on } \partial\mathcal{D}_{0\sigma} \end{cases} \tag{4}$$

where $\boldsymbol{\nabla}_{\boldsymbol{X}}$ denotes the gradient w.r.t. the reference coordinates and the body forces are not considered. The boundary of the continuum body $\partial\mathcal{D}_0$ can be divided into the disjoint sets $\partial\mathcal{D}_u$ and $\partial\mathcal{D}_{0\sigma}$ such that $\partial\mathcal{D}_0 = \partial\mathcal{D}_{0u} \cup \partial\mathcal{D}_{0\sigma}$ and $\partial\mathcal{D}_{0u} \cap \partial\mathcal{D}_{0\sigma} = \emptyset$, $\boldsymbol{P}$ is the 1st Piola-Kirchhoff (1st-PK) stress tensor that depends on the deformation gradient $\boldsymbol{F} := \boldsymbol{\nabla}_{\boldsymbol{X}}\boldsymbol{\varphi} = \boldsymbol{I} + \boldsymbol{\nabla}_{\boldsymbol{X}}\boldsymbol{u}$ through a specified constitutive model, in which $\boldsymbol{I}$ is an identity tensor, $\bar{\boldsymbol{u}}$ and $\bar{\boldsymbol{T}}$ are the prescribed displacements and tractions on $\partial\mathcal{D}_{0u}$ and $\partial\mathcal{D}_{0\sigma}$, respectively, $\widehat{\boldsymbol{N}}$ is the outward unit normal on $\partial\mathcal{D}_{0\sigma}$.

The associated weak form of the problem can be written as: find $\boldsymbol{u} \in \mathcal{U}$ such that

$$\int_{\mathcal{D}_0} \boldsymbol{P}: \boldsymbol{\nabla}_{\boldsymbol{X}}\delta\boldsymbol{u}\, dV = \int_{\partial\mathcal{D}_{0\sigma}} \bar{\boldsymbol{T}}.\delta\boldsymbol{u}\, dS \qquad \forall\delta\boldsymbol{u} \in \mathcal{V} \tag{5}$$

where the appropriate functional spaces for solution and variation are

$$\begin{aligned} \mathcal{U} &= \left\{\boldsymbol{u}(\boldsymbol{X}): \mathcal{D}_0 \to \mathbb{R}^{\mathrm{sdim}} \middle| \boldsymbol{u}(\boldsymbol{X}) \in H^1(\mathcal{D}_0), \boldsymbol{u}(\boldsymbol{X}) = \bar{\boldsymbol{u}} \text{ on } \partial\mathcal{D}_{0u}\right\} \\ \mathcal{V} &= \left\{\boldsymbol{v}(\boldsymbol{X}): \mathcal{D}_0 \to \mathbb{R}^{\mathrm{sdim}} \middle| \boldsymbol{v}(\boldsymbol{X}) \in H^1(\mathcal{D}_0), \boldsymbol{v}(\boldsymbol{X}) = \boldsymbol{0} \text{ on } \partial\mathcal{D}_{0u}\right\} \end{aligned} \tag{6}$$

and $H^1(\mathcal{D}_0)$ is the Sobolev space given by

$$H^1(\mathcal{D}_0) = \left\{ \boldsymbol{v}: \mathcal{D}_0 \rightarrow \mathbb{R}^{\text{sdim}} \middle| \int_{\mathcal{D}_0} v_k^2 dV < \infty, \int_{\mathcal{D}_0} \left( \frac{\partial v_k}{\partial X_l} \right)^2 dV < \infty,\ k,l = 1,\ldots,\text{sdim} \right\} \tag{7}$$

Using the Galerkin method, the same finite element basis functions are used to discretize the weak form in Eq. (5) by constructing finite-dimensional approximations $\mathcal{U}^h$ and $\mathcal{V}^h$ of the spaces $\mathcal{U}$ and $\mathcal{V}$, respectively [39]. As a result, the system of the nonlinear equations in the finite element analysis (FEA) with a hyperelastic constitutive model can be expressed as

$$\begin{aligned} &\boldsymbol{R}_{NL}(\boldsymbol{u}) = \boldsymbol{F}_{int}(\boldsymbol{u}) - \boldsymbol{F}_{ext} = \boldsymbol{0} \qquad \text{with} \\ &\boldsymbol{F}_{int} = \mathop{\mathcal{A}}_{e=1}^{n_{ele}} \boldsymbol{F}_{int}^e \quad \text{and} \quad \boldsymbol{F}_{int}^e = \int_{\mathcal{D}_0^e} \boldsymbol{B}:\boldsymbol{P} dV \\ &\boldsymbol{F}_{ext} = \mathop{\mathcal{A}}_{e\in S_\sigma} \int_{\partial\mathcal{D}_{0\sigma}^e} \boldsymbol{N}^T \overline{\boldsymbol{T}}\, dV \end{aligned} \tag{8}$$

where the external force vector $\boldsymbol{F}_{ext}$ is assumed to be deformation-independent, $\boldsymbol{N}$ is the shape function matrix, and $\boldsymbol{B}$ is the displacement gradient tensor; $\mathcal{A}$ is the FE assembly operator; $S_\sigma$ is the set of elements on which external traction is applied; $\mathcal{D}_0^e$ is the $e^{\text{th}}$ element domain satisfying $\mathcal{D}_0 = \bigcup_{e=1}^{n_{ele}} \mathcal{D}_0^e$; $\partial\mathcal{D}_{0\sigma}^e$ denotes a part of the boundary $\partial\mathcal{D}_{0\sigma}$ shared with $e^{\text{th}}$ element and satisfying $\partial\mathcal{D}_{0\sigma} = \bigcup_{e\in S_\sigma} \partial\mathcal{D}_{0\sigma}^e$. Note that in Eq. $(8)_1$, it is assumed that essential boundary conditions on $\mathcal{D}_{0u}$ are enforced during assembly and $\boldsymbol{u} \in \mathbb{R}^{n_{\text{dof}}}$, where $n_{\text{dof}}$ is the total number of unknown displacement variables, i.e., the free degrees of freedom. Only 2-D plane strain problems are considered in this work, i.e., sdim $= 2$.

*2.2.1 Regularized neo-Hookean model*

A regularized neo-Hookean hyperelastic material is employed in nonlinear FEA for which the free energy is expressed as

$$\psi(\boldsymbol{C}) = \frac{1}{2}\kappa(J-1)^2 + \frac{1}{2}\mu(\bar{I}_1 - 3) \tag{9}$$

where $\bar{I}_1 = \mathrm{tr}(\bar{\boldsymbol{C}})$ with $\bar{\boldsymbol{C}} = J^{-2/3}\boldsymbol{C}$ and $\boldsymbol{C} = \boldsymbol{F}^T.\boldsymbol{F}$. Here, $\boldsymbol{C}$ is the right Cauchy-Green tensor and $J = \det \boldsymbol{F}$ is the volumetric Jacobian. The material properties are determined by the bulk and shear moduli $\kappa$ and $\mu$, which are related to Young's modulus $E$ and Poisson's ratio $\nu$ by

$$\kappa = \frac{E}{3(1-2\nu)} \quad \text{and} \quad \mu = \frac{E}{2(1+\nu)} \tag{10}$$

*2.2.2 FEA with linear energy interpolation*

For addressing the mesh distortion of low-density elements under large deformations, an element's internal force vector $\boldsymbol{F}_{int}^e$ that depends on the displacement field $\boldsymbol{u}^e$ is modified using a linear energy interpolation scheme, which was first proposed in Ref. [40], i.e.,

$$\boldsymbol{F}_{int}^e = \int_{\mathcal{D}_0^e} \gamma \boldsymbol{B}\colon \boldsymbol{P}\, dV + \int_{\mathcal{D}_0^e} (1-\gamma^2)\boldsymbol{B}_L\colon \mathbb{C}\colon \boldsymbol{\varepsilon}\, dV \tag{11}$$

in which $\boldsymbol{P}$ is the 1st-PK stress evaluated at an interpolated deformation gradient $\boldsymbol{F}(\gamma)$ with

$$\boldsymbol{F}(\gamma) = \boldsymbol{I} + \gamma(\rho)\boldsymbol{\nabla}_{\boldsymbol{X}}\boldsymbol{u}^e \quad \text{with} \quad \gamma(\rho) = \frac{\exp(\beta_0\rho)}{\exp(c\beta_0) + \exp(\beta_0\rho)} \tag{12}$$

where the parameters $\beta_0$ and $c$ control the slope and cutoff location of the projection function, $\gamma\colon \rho \to [0,1]$, in Eq. (12), where $\rho \equiv \rho_e$ is the density variable. This method was further developed in Ref. [41], where the $c$-value is adaptively changed to prevent analysis failures during the optimization process. Interested readers are referred to Ref. [40, 41] for more information on this linear energy interpolation scheme. In Eq. (11), $\boldsymbol{\varepsilon} = \boldsymbol{\nabla}_{\boldsymbol{X}}^s\boldsymbol{u}^e$ is the linear strain measure with $\boldsymbol{\nabla}_{\boldsymbol{X}}^s$ the symmetric gradient operator, and $\mathbb{C} = 3\kappa_L\mathbb{P}_{vol} + 2\mu_L\mathbb{P}_{dev}^s$ is the 4th-order (linear) isotropic elasticity tensor with $\mathbb{P}_{vol} = \frac{1}{3}\boldsymbol{I}\otimes\boldsymbol{I}$ and $\mathbb{P}_{dev}^s = \mathbb{I}_4^s - \frac{1}{3}\boldsymbol{I}\otimes\boldsymbol{I}$ the 4th-order volumetric and symmetric-deviatoric projection tensors, respectively [42]; $\boldsymbol{B}_L$ is the displacement gradient tensor for small deformations. For the linear isotropic material $\kappa_L = \frac{E_L}{3(1-2\nu_L)}$ and $\mu_L = \frac{E_L}{2(1+\nu_L)}$ with

$$E_L(\rho_e) = \left[\epsilon_L + (1-\epsilon_L)\rho_e^{p_l}\right]E_{L0} \tag{13}$$

where $\epsilon_L$ = $10^{-6}$ is the lower bound parameter and when multiplied with $E_{L0}$ represents Young's modulus for the void (linear) material phase ($\rho_e = 0$), and $E_{L0}$ represents Young's modulus for the solid (linear) material phase, where $p_l$ is an implicit penalty parameter to penalize intermediate densities.

With $\boldsymbol{F}_{int}^e$ being replaced by Eq. (11), the global equilibrium Eq. $(8)_1$ is solved using the Newton-Raphson (NR) method with the tangent stiffness matrix given by

$$\boldsymbol{K}_T = \mathop{\mathcal{A}}_{e=1}^{n_{ele}} \boldsymbol{K}_T^e \quad \text{with} \quad \boldsymbol{K}_T^e = \int_{\mathcal{D}_0^e} \gamma^2 \boldsymbol{B}:\mathbb{A}_4:\boldsymbol{B}\, dV + \int_{\mathcal{D}_0^e} (1-\gamma^2)\boldsymbol{B}_L:\mathbb{C}:\boldsymbol{B}_L\, dV \tag{14}$$

where $\mathbb{A}_4$ is the 4$^{\text{th}}$-order tangent moduli of the hyperelastic constitutive model, see Appendix B.2 for further details of the derivatives of the regularized neo-Hookean hyperelastic free energy function used in this study.

*Remark:* In this work, the adaptive scheme for the linear energy interpolation method proposed in [41] is used, i.e., $\beta_0$ is set to 120, whereas $c \in [0.1, 1.0]$ is varied adaptively, if needed. In the context of density-based topology optimization, this adaptive scheme is shown to be essential for the success of the optimization process in studies concerning finite deformation topology optimization [43-47]. In particular, the following adaptive update strategy for parameter $c$ is used in the nonlinear FEA. At first, a default $c$-value $c_0$ is used in the solution process. If the minimum step size is reached in the nonlinear solver and mesh distortion still occurs, the cutoff parameter is increased with a fixed increment $\Delta c$, which increases the contribution of linear energy and promotes convergence of nonlinear FEA. This process continues until the NR converges. The value of parameter $c$ is reset to $c_0$ in the next optimization iteration. Interested readers are referred to Ref. [41] for more information on this adaptive scheme. In this study, the values of $c_0$ and $\Delta c$ are set to be 0.1 and 0.02, respectively.

### 2.3 Optimization formulation

For geometrically nonlinear stiffness design, the end compliance minimization under the material volume constraint is considered, and the optimization formulation is expressed as

$$\begin{aligned} &\min_{\boldsymbol{x}} f_0(\boldsymbol{x}) = \boldsymbol{F}_{ext}^T \boldsymbol{u}(\boldsymbol{x}) \\ &\text{s.t. } f_1(\boldsymbol{x}) = \frac{1}{V}\left(\sum_{e=1}^{n_{ele}} \rho_e v_e\right) - V_f \leq 0 \\ &\quad\quad \boldsymbol{0} \leq \boldsymbol{x} \leq \boldsymbol{1} \end{aligned} \tag{15}$$

where $V$ and $V_f$ are the total volume of the design domain $D$ and the maximum allowable volume fraction, respectively, and $v_e$ is the volume of the $e^{\text{th}}$ element.

## 3 Uncertain sources

### 3.1 Load uncertainty

The load uncertainty is incorporated by considering the nodal force vector ($\boldsymbol{P}_0$) as stochastic variables. To this end, the nodal load is modeled by a Gaussian random vector, i.e., $\boldsymbol{P}_0 \sim \mathcal{N}(\boldsymbol{\mu}_P, \boldsymbol{\Sigma}_P) \in \mathbb{R}^{\text{sdim}}$, where $\boldsymbol{\mu}_P \in \mathbb{R}^{\text{sdim}}$ and $\boldsymbol{\Sigma}_P \in \mathbb{R}^{\text{sdim} \times \text{sdim}}$ are the prescribed mean vector and covariance matrix of the stochastic loading, respectively. This Gaussian random vector $\boldsymbol{P}_0$ is further parametrized by i.i.d. standard Gaussian random variables $\boldsymbol{\xi}_1 \sim \mathcal{N}(\boldsymbol{0}, \boldsymbol{I})$, i.e., $\boldsymbol{P}_0 = \boldsymbol{L}_c \boldsymbol{\xi}_1 + \boldsymbol{\mu}_P$, where $\boldsymbol{L}_c$ is the lower triangular matrix from the Cholesky decomposition of the covariance matrix, i.e., $[\boldsymbol{\Sigma}_P] = [\boldsymbol{L}_c][\boldsymbol{L}_c]^T$. The number of random variables induced by load uncertainty is denoted as $m_1$, i.e., $m_1 = \text{sdim}$.

### 3.2 Material uncertainty

In this study, material uncertainties are incorporated in topology optimization for obtaining robust design under material heterogeneities. In particular, Young's modulus for the solid material phase, i.e., $E_0$ in Eq. (1), is assumed to be a weakly stationary lognormal random field over the domain $\mathcal{D}_0$ with a probability space $\Omega_P$, i.e., $E_0(\boldsymbol{X}, \omega)$ with $\boldsymbol{X} \in \mathcal{D}_0$, $\omega \in \Omega_P$. The marginal distribution of the non-Gaussian random field $E_0(\boldsymbol{X}, \omega)$ is assumed to be lognormal with mean $\mu_0$ and standard deviation $\sigma_0$, i.e., $E_0(\boldsymbol{X}, \omega) \sim \text{Lognormal}(\mu_0, \sigma_0^2)$, $\forall\, \boldsymbol{X} \in \mathcal{D}_0$. This non-Gaussian random field is constructed by applying a nonlinear transformation to a Gaussian random field $Z(\boldsymbol{X}, \omega)$ which is characterized by a zero-mean function, and an anisotropic correlation function is given by

$$R_Z(\boldsymbol{X}_a, \boldsymbol{X}_b) = \exp\left(-\frac{|X_a - X_b|^2}{2l_{cx}^2} - \frac{|Y_a - Y_b|^2}{2l_{cy}^2} - \frac{|Z_a - Z_b|^2}{2l_{cz}^2}\right) \tag{16}$$

where $l_{cx}$, $l_{cy}$ and $l_{cz}$ are the correlation lengths in $x$, $y$, and $z$ directions, respectively; $\boldsymbol{X}_a = (X_a, Y_a, Z_a)$ and $\boldsymbol{X}_b = (X_b, Y_b, Z_b)$ are the position vectors of two points. Note that for the 2-D cases considered in this study $l_{cz} = \infty$. The transformation between random fields $E_0(\boldsymbol{X}, \omega)$ and $Z(\boldsymbol{X}, \omega)$ is given by

$$E_0(\boldsymbol{X}, \omega) = F_E^{-1}\left(\Phi\big(Z(\boldsymbol{X}, \omega)\big)\right) \tag{17}$$

where $F_E(y)$ is the cumulative distribution function (CDF) of the random variable $E_0(\boldsymbol{X}, \omega)$; $F_E^{-1}(\blacksquare)$ is the inverse of this CDF and $\Phi(z)$ is the CDF of the standard 1-D Gaussian distribution [48].

The spatial domain of a continuous random field is discretized by an elementwise constant finite element (FE) mesh, wherein the random variable evaluated at an element centroid is used for describing the randomness at any point within the element domain. Moreover, for computational expediency, the FE mesh is chosen to conform to the one used in the structural FE analysis. Thus, with $n_{ele}$ finite elements in the computational domain, the spatial random field $E(\boldsymbol{X}, \omega)$ is discretized to obtain a random vector

$$\boldsymbol{E}_0(\omega) = \begin{bmatrix} E_0(\boldsymbol{X}_1, \omega) \\ E_0(\boldsymbol{X}_2, \omega) \\ \vdots \\ E_0(\boldsymbol{X}_{n_{ele}}, \omega) \end{bmatrix} \in \mathbb{R}^{n_{ele}} \quad (18)$$

where $\boldsymbol{X}_q$ is the coordinate vector of the centroid of the $q^{\text{th}}$ element. Following the same discretization, the intermediate Gaussian random field $Z(\boldsymbol{X}, \omega)$ is also discretized and leads to a random vector

$$\boldsymbol{Z}(\omega) = \begin{bmatrix} Z(\boldsymbol{X}_1, \omega) \\ Z(\boldsymbol{X}_2, \omega) \\ \vdots \\ Z(\boldsymbol{X}_{n_{ele}}, \omega) \end{bmatrix} \in \mathbb{R}^{n_{ele}} \quad (19)$$

with the correlation matrix given by

$$\boldsymbol{R} = \begin{bmatrix} R_Z(\boldsymbol{X}_1, \boldsymbol{X}_1) & R_Z(\boldsymbol{X}_1, \boldsymbol{X}_2) & \cdots & R_Z(\boldsymbol{X}_1, \boldsymbol{X}_{n_{ele}}) \\ R_Z(\boldsymbol{X}_2, \boldsymbol{X}_1) & R_Z(\boldsymbol{X}_2, \boldsymbol{X}_2) & \cdots & R_Z(\boldsymbol{X}_2, \boldsymbol{X}_{n_{ele}}) \\ \vdots & \vdots & \cdots & \vdots \\ R_Z(\boldsymbol{X}_{n_{ele}}, \boldsymbol{X}_1) & R_Z(\boldsymbol{X}_{n_{ele}}, \boldsymbol{X}_2) & \cdots & R_Z(\boldsymbol{X}_{n_{ele}}, \boldsymbol{X}_{n_{ele}}) \end{bmatrix} \in \mathbb{R}^{n_{ele} \times n_{ele}} \quad (20)$$

As shown in the references [14, 27], the computational expense for constructing the discrete random field $\boldsymbol{Z}(\omega)$ modeling can be significantly reduced by using the truncated Karhunen–Loève (KL) expansion [49-52]. In KL expansion, an eigenanalysis of the correlation matrix $\boldsymbol{R}$ in Eq. (20) is first carried out to compute the eigenpairs $(\lambda_k, \boldsymbol{\gamma}_k)$, and the random vector $\boldsymbol{Z}$ in Eq. (19) is then approximated by

$$\boldsymbol{Z}(\boldsymbol{\xi}_2) \approx \sum_{k=1}^{m_2} \sqrt{\lambda_k} \boldsymbol{\gamma}_k \xi_{2k} \quad (21)$$

where $\boldsymbol{\xi}_2 = [\xi_{21} \quad \xi_{22} \quad \cdots \quad \xi_{2m_2}]^T$ is a standard Gaussian random vector, i.e., $\boldsymbol{\xi}_2 \sim \mathcal{N}(\boldsymbol{0}, \boldsymbol{I})$. In practice, $m_2$ is chosen such that a sufficient number of terms in the KL expansion (Eq. (21)) are included. In this study, the criterion for determining $m_2$ is that the sum of the first $m_2$ largest eigenvalues is greater than 90% of the sum of all eigenvalues, i.e., cumulative spectrum, which is equal to the trace of the correlation matrix. Finally, the discretized random field of interest $\boldsymbol{E}_0(\omega)$ in Eq. (18) is obtained by transforming $\boldsymbol{Z}$ (Eq. (21)) in an element-by-element fashion by using the inverse transformation in Eq. (17). It is remarked that the material properties in the linear energy part are constant, i.e., no uncertainty is considered, as they are included only to address low-density elements distortion.

### 3.3 Geometric uncertainty

Geometric uncertainty can be incorporated in numerous ways. Jansen et al. [16, 31] considered the nodal coordinates to be Gaussian random fields. Sato et al. [53] proposed to model the geometric uncertainty in the Eulerian description by taking the velocity fields of material points in a fictitious advection equation as Gaussian random fields. In this study, similar to the approach adopted in Refs. [23, 27], the geometric uncertainty that models the manufacturing imperfections is parameterized by a random-valued cutoff parameter ($\eta$) in the Heaviside projection in Eq. (3). The cutoff parameter is modeled as a random field $\eta(\boldsymbol{X}, \omega)$. The random vector $\boldsymbol{\eta}(\omega) = \left[\eta(\boldsymbol{X}_1, \omega) \quad \eta(\boldsymbol{X}_2, \omega) \quad \cdots \quad \eta\left(\boldsymbol{X}_{n_{ele}}, \omega\right)\right]^T$ represents a discrete random field obtained by using FE mesh to discretize a continuous random field $\eta(\boldsymbol{X}, \omega)$ with its marginal PDF following a uniform distribution $\mathcal{U}(\eta_{min}, \eta_{max})$. Using a similar approach for modeling non-Gaussian random field and its discretization and approximation in Section 3.2, the random variable $\eta(\boldsymbol{X}_e, \omega)$ is obtained by using the following transformation as

$$\eta(\boldsymbol{X}_e, \omega) = \eta_{min} + (\eta_{max} - \eta_{min})\Phi\left(\bar{Z}(\boldsymbol{X}_e, \omega)\right) \tag{22}$$

where the random vector $\bar{\boldsymbol{Z}}(\omega) = \left[\bar{Z}(\boldsymbol{X}_1, \omega) \quad \bar{Z}(\boldsymbol{X}_2, \omega) \quad \cdots \quad \bar{Z}(\boldsymbol{X}_{n_{ele}}, \omega)\right]^T$ follows multivariate Gaussian distribution with zero mean vector and covariance matrix $\boldsymbol{R}$ in Eq. (20) and is approximated by

$$\bar{\boldsymbol{Z}}(\boldsymbol{\xi}_3) \approx \sum_{k=1}^{m_3} \sqrt{\lambda_k} \boldsymbol{\gamma}_k \xi_{3k} \tag{23}$$

where $\boldsymbol{\xi}_3 = [\xi_{31} \quad \xi_{32} \quad \cdots \quad \xi_{3m_3}]^T$ follows a multivariate standard Gaussian distribution, i.e., $\boldsymbol{\xi}_3 \sim \mathcal{N}(\boldsymbol{0}, \boldsymbol{I})$. Using the same truncation criterion in Section 3.2, the first $m_3$ largest eigenvalues of the covariance matrix $\boldsymbol{R}$ takes up over 90% of the sum of all eigenvalues. As linear energy is added only to fix low-density mesh distortions, the geometric randomness is not considered in the linear energy evaluation, i.e., the density $\rho$ to interpolate the linear energy is projected using a constant cutoff parameter $\eta$ $(= 0.5)$.

## 4 Robust topology optimization

With the incorporation of multiple uncertain sources described in Section 3, a robust topology optimization formulation for stiffness design, in contrast to its deterministic counterpart in Eq. (15), is constructed as follows

$$\begin{aligned} &\min_{\boldsymbol{x}} \not{f}_0(\boldsymbol{x}) = \mathbb{E}[f(\boldsymbol{x}, \boldsymbol{\xi})] + \alpha \sqrt{Var[f(\boldsymbol{x}, \boldsymbol{\xi})]} \quad \text{with} \quad f(\boldsymbol{x}, \boldsymbol{\xi}) = [\boldsymbol{F}_{ext}(\boldsymbol{\xi}_1)]^T \boldsymbol{u}(\boldsymbol{x}, \boldsymbol{\xi}) \\ &\text{s.t. } \not{f}_1(\boldsymbol{x}) = \frac{1}{V} \sum_{e=1}^{n_{ele}} \bar{\rho}_e v_e - V_f \leq 0 \quad \text{with} \quad \bar{\rho}_e = \left. \frac{\tanh(\beta\eta) + \tanh\big(\beta(\hat{\rho}_e - \eta)\big)}{\tanh(\beta\eta) + \tanh\big(\beta(1 - \eta)\big)} \right|_{\eta = 0.5} \\ &\qquad \boldsymbol{0} \leq \boldsymbol{x} \leq \boldsymbol{1} \end{aligned} \tag{24}$$

where $\mathbb{E}[\blacksquare]$ and $Var[\blacksquare]$ denote the expectation and variance of a random variable, respectively, and the objective function $\not{f}_0$ considers a tradeoff between minimizing the mean and variance of a specified structural performance function $f(\boldsymbol{x}, \boldsymbol{\xi})$. Here, the stochastic structural response $\boldsymbol{u}(\boldsymbol{x}, \boldsymbol{\xi})$ is determined by the stochastic nonlinear FEA with the governing Eq. (8) replaced by

$$\boldsymbol{R}_{NL}(\boldsymbol{u}, \boldsymbol{x}, \boldsymbol{\xi}) = \boldsymbol{F}_{int}(\boldsymbol{u}, \boldsymbol{x}, \boldsymbol{\xi}_2, \boldsymbol{\xi}_3) - \boldsymbol{F}_{ext}(\boldsymbol{\xi}_1) = \mathbf{0} \quad (25)$$

in which $\boldsymbol{\xi} = [\boldsymbol{\xi}_1 \quad \boldsymbol{\xi}_2 \quad \boldsymbol{\xi}_3]^T$ represents the collection of the random variables of size $m$ with $m = m_1 + m_2 + m_3$ accounting for the load, material, and geometric uncertainties, respectively. In the KL expansion in Eq. (21) or (23), the truncation number $m_2$ or $m_3$ is determined based on the criterion that the ratio of the sum of the largest included eigenvalues to the sum of all the eigenvalues is no less than 90%. Specifically, with the exponential correlation function chosen for modeling the underlying standard Gaussian random field, this approach leads to a global error of less than 0.0857 across all the considered examples [14]. This level of error ensures an accurate representation of the considered random fields. The dependence of the internal force vector $\boldsymbol{F}_{int}$ on the random variables $\boldsymbol{\xi}_2$ and $\boldsymbol{\xi}_3$ comes from the stochastic Young's modulus for the solid phase $E(\boldsymbol{X}, \omega)$ and the stochastic density $\rho(\boldsymbol{X}_e, \omega)$ resulting from the stochastic cutoff parameter $\eta(\boldsymbol{X}_e, \omega)$. The dependence of the external force vector $\boldsymbol{F}_{ext}$ on the random variables $\boldsymbol{\xi}_1$ comes from load uncertainty. It should be noted that when geometric uncertainty is considered, the material volume becomes stochastic. However, only a deterministic volume constraint is considered in the optimization formulation. To this end, the material volume in the volume constraint is evaluated at $\eta = 0.5$, i.e., the mean value of the projection range [0,1]. In Eq. (24), a scalar weighting parameter $\alpha \geq 0$ is used to control the designs' robustness.

## 5 Uncertainty quantification (UQ)

For the quantification of the uncertainty of the structural response of interest, i.e., end-compliance, in terms of the random input parameters, i.e., applied loads, material property, and projection cutoff parameter, the $2^{nd}$-order stochastic perturbation expansion is used to approximate the end-compliance $f$ $(= \boldsymbol{F}_{ext}^T \boldsymbol{u})$ [14, 54, 55]. To increase the readability, the simplified notations listed in Table 1 are adopted in the following discussions and derivations.

Table 1. List of the notations used in the UQ and sensitivity analysis.

| | |
|---|---|
| $\boldsymbol{u}^{[0]} \stackrel{\text{def}}{=} \boldsymbol{u}(\boldsymbol{\xi}=\boldsymbol{0})$<br>$\boldsymbol{u}_k^{[1]} \stackrel{\text{def}}{=} \left.\frac{d\boldsymbol{u}}{d\xi_k}\right\rvert_{\boldsymbol{\xi}=\boldsymbol{0}} \qquad k=1,\dots,m$<br>$\boldsymbol{u}_{kl}^{[2]} \stackrel{\text{def}}{=} \left.\frac{d^2\boldsymbol{u}}{d\xi_k d\xi_l}\right\rvert_{\boldsymbol{\xi}=\boldsymbol{0}} \qquad k,l=1,\dots,m$ | $\boldsymbol{F}_{\text{int}}^{[0]} \stackrel{\text{def}}{=} \boldsymbol{F}_{\text{int}}\left(\boldsymbol{u}^{[0]},\boldsymbol{\xi}=\boldsymbol{0}\right)$<br>$\boldsymbol{F}_{\text{int}(k)}^{[1]} \stackrel{\text{def}}{=} \left.\frac{d\boldsymbol{F}_{\text{int}}}{d\xi_k}\right\rvert_{\boldsymbol{\xi}=\boldsymbol{0}} \qquad k=1,\dots,m$<br>$\boldsymbol{F}_{\text{int}(kl)}^{[2]} \stackrel{\text{def}}{=} \left.\frac{d^2\boldsymbol{F}_{\text{int}}}{d\xi_k d\xi_l}\right\rvert_{\boldsymbol{\xi}=\boldsymbol{0}} \qquad k,l=1,\dots,m$ |
| $\boldsymbol{F}_{\text{ext}}^{[0]} \stackrel{\text{def}}{=} \boldsymbol{F}_{\text{ext}}(\boldsymbol{\xi}=\boldsymbol{0})$<br>$\boldsymbol{F}_{\text{ext}(k)}^{[1]} \stackrel{\text{def}}{=} \left.\frac{d\boldsymbol{F}_{\text{ext}}}{d\xi_k}\right\rvert_{\boldsymbol{\xi}=\boldsymbol{0}} \qquad k=1,\dots,m$<br>$\boldsymbol{F}_{\text{ext}(kl)}^{[2]} \stackrel{\text{def}}{=} \left.\frac{d^2\boldsymbol{F}_{\text{ext}}}{d\xi_k d\xi_l}\right\rvert_{\boldsymbol{\xi}=\boldsymbol{0}} \qquad k,l=1,\dots,m$ | $f^{[0]} \stackrel{\text{def}}{=} {\boldsymbol{F}_{\text{ext}}^{[0]}}^T \boldsymbol{u}^{[0]}$<br>$f_k^{[1]} \stackrel{\text{def}}{=} \left.\frac{df}{d\xi_k}\right\rvert_{\boldsymbol{\xi}=\boldsymbol{0}} \qquad k=1,\dots,m$<br>$f_{kl}^{[2]} \stackrel{\text{def}}{=} \left.\frac{d^2 f}{d\xi_k d\xi_l}\right\rvert_{\boldsymbol{\xi}=\boldsymbol{0}} \qquad k,l=1,\dots,m$ |
| $\mathbb{A}_4 \stackrel{\text{def}}{=} \frac{\partial^2\bar{\psi}}{\partial\boldsymbol{F}\partial\boldsymbol{F}}$<br>$\mathbb{A}_6 \stackrel{\text{def}}{=} \frac{\partial^3\bar{\psi}}{\partial\boldsymbol{F}\partial\boldsymbol{F}\partial\boldsymbol{F}}$<br>$\mathbb{A}_8 \stackrel{\text{def}}{=} \frac{\partial^4\bar{\psi}}{\partial\boldsymbol{F}\partial\boldsymbol{F}\partial\boldsymbol{F}\partial\boldsymbol{F}}$ | $\boldsymbol{p}^e \stackrel{\text{def}}{=} \int_{\mathcal{D}_0^e} \boldsymbol{B}:\boldsymbol{P}dV$<br>$\boldsymbol{k}_L^e \stackrel{\text{def}}{=} \int_{\mathcal{D}_0^e} \boldsymbol{B}_L:\mathbb{C}_L:\boldsymbol{B}_L dV$ |
| $\boldsymbol{v}_k \stackrel{\text{def}}{=} \frac{d\gamma}{d\xi_k}\boldsymbol{u}^{e[0]} + \gamma\boldsymbol{u}_k^{e[1]}$<br>$\boldsymbol{v}_{kl} = \frac{\partial^2\gamma}{\partial\xi_k\partial\xi_l}\boldsymbol{u}^{e[0]} + \frac{\partial\gamma}{\partial\xi_k}\boldsymbol{u}_l^{e[1]} + \frac{\partial\gamma}{\partial\xi_l}\boldsymbol{u}_k^{e[1]} + \gamma\boldsymbol{u}_{kl}^{e[2]}$ | $\boldsymbol{K}_{T0}^{(1)} \stackrel{\text{def}}{=} \left.\frac{\partial\boldsymbol{F}_{\text{int}}}{\partial\boldsymbol{u}}\right\rvert_{\boldsymbol{\xi}=\boldsymbol{0}}$<br>$\boldsymbol{K}_{T0}^{(2)} \stackrel{\text{def}}{=} \left.\frac{\partial^2\boldsymbol{F}_{\text{int}}}{\partial\boldsymbol{u}\partial\boldsymbol{u}}\right\rvert_{\boldsymbol{\xi}=\boldsymbol{0}}$ |
| $(\boldsymbol{k}_1^e)_{\alpha\beta} \stackrel{\text{def}}{=} \int_{\mathcal{D}_0^e} (\mathbb{A}_4)_{ijkl} B_{kl\alpha} B_{ij\beta} dV$<br>$(\boldsymbol{k}_2^e)_{\alpha\beta\gamma} \stackrel{\text{def}}{=} \int_{\mathcal{D}_0^e} (\mathbb{A}_6)_{ijklmn} B_{mn\gamma} B_{kl\beta} B_{ij\alpha} dV$<br>$(\boldsymbol{k}_3^e)_{\alpha\beta\gamma\delta} \stackrel{\text{def}}{=} \int_{\mathcal{D}_0^e} (\mathbb{A}_8)_{ijklmnpq} B_{pq\delta} B_{mn\gamma} B_{kl\beta} B_{ij\alpha} dV$ | |

NOTE: (a) $\bar{\psi}$ is the free energy evaluated at the unit Young's modulus, i.e., $\psi = E\bar{\psi}$. Accordingly, $\boldsymbol{P}$, $\mathbb{A}_4$, $\mathbb{A}_6$ and $\mathbb{A}_8$ are all evaluated at $E=1$. (b) $\boldsymbol{B}$ is a 3rd-order tensor which when applied to the element displacement vector yields the displacement gradient, i.e., $\boldsymbol{\nabla}_{\boldsymbol{X}}\boldsymbol{u} = \boldsymbol{B}.\boldsymbol{u}^e$.

Using the second-order expansions, the end-compliance function can be expressed as

$$f(\boldsymbol{x},\boldsymbol{\xi}) \approx f^{[0]} + \sum_{k=1}^{m} f_k^{[1]}\xi_k + \frac{1}{2}\sum_{k=1}^{m}\sum_{l=1}^{m} f_{kl}^{[2]}\xi_k\xi_l \tag{26}$$

where the 1st order derivatives are calculated by ($k = 1, \dots, m$)

$$f_k^{[1]} = \left[\boldsymbol{F}_{\text{ext}(k)}^{[1]}\right]^T \boldsymbol{u}^{[0]} + \left[\boldsymbol{F}_{\text{ext}}^{[0]}\right]^T \boldsymbol{u}_k^{[1]} \tag{27}$$

and the 2nd order derivatives are given by ($k, l = 1, \dots, m$)

$$f_{kl}^{[2]} = \left[\boldsymbol{F}_{\text{ext}(kl)}^{[2]}\right]^T \boldsymbol{u}^{[0]} + \left[\boldsymbol{F}_{\text{ext}(k)}^{[1]}\right]^T \boldsymbol{u}_l^{[1]} + \left[\boldsymbol{F}_{\text{ext}(l)}^{[1]}\right]^T \boldsymbol{u}_k^{[1]} + \left[\boldsymbol{F}_{\text{ext}}^{[0]}\right]^T \boldsymbol{u}_{kl}^{[2]} \tag{28}$$

Therefore, the mean of the compliance $f$ can be computed as

$$\mathbb{E}[f] \approx f^{[0]} + \sum_{i=1}^{m} f_k^{[1]}\mathbb{E}[\xi_k] + \frac{1}{2}\sum_{k=1}^{m}\sum_{l=1}^{m} f_{kl}^{[2]}\mathbb{E}[\xi_k\xi_l] = f^{[0]} + \frac{1}{2}\sum_{k=1}^{m} f_{kk}^{[2]} \tag{29}$$

Again, using the second-order expansions, the variance of $f$ reads

$$\begin{aligned} Var[f(\boldsymbol{x},\boldsymbol{\xi})] &\approx \mathbb{E}\left[\left(f - f^{[0]} - \frac{1}{2}\sum_{k=1}^{m} f_{kk}^{[2]}\right)^2\right] \\ &\approx \mathbb{E}\left[\left(\sum_{k=1}^{m} f_k^{[1]}\xi_k + \frac{1}{2}\sum_{k=1}^{m}\sum_{l=1}^{m} f_{kl}^{[2]}\xi_k\xi_l - \frac{1}{2}\sum_{k=1}^{m} f_{kk}^{[2]}\right)^2\right] \\ &= \sum_{k=1}^{m}\left(f_k^{[1]}\right)^2 + \frac{1}{2}\sum_{k=1}^{m}\sum_{l=1}^{m}\left(f_{kl}^{[2]}\right)^2 \end{aligned} \tag{30}$$

From Eqns. (28) to (30), to obtain the mean and variance of the end compliance, the terms that need to be calculated are

$$\boldsymbol{F}_{\text{ext}}^{[0]},\ \boldsymbol{F}_{\text{ext}(k)}^{[1]},\ \boldsymbol{F}_{\text{ext}(kl)}^{[2]},\ \boldsymbol{u}^{[0]},\ \boldsymbol{u}_k^{[1]},\ \boldsymbol{u}_{kl}^{[2]}, \qquad k, l = 1, \dots, m \tag{31}$$

where the first three terms can be computed explicitly, while the last three terms $\boldsymbol{u}^{[0]}$, $\boldsymbol{u}_k^{[1]}$ and $\boldsymbol{u}_{kl}^{[2]}$ are determined by solving a sequence of a system of nonlinear/linear equations

$$\boldsymbol{R}^{[0]}\left(\boldsymbol{u}^{[0]}\right) = \boldsymbol{F}_{\text{int}}^{[0]} - \boldsymbol{F}_{\text{ext}}^{[0]} = \boldsymbol{0} \tag{32}$$

$$\boldsymbol{R}_{k}^{[1]}\left(\boldsymbol{u}^{[0]}, \boldsymbol{u}_{k}^{[1]}\right) = \boldsymbol{F}_{\text{int}(k)}^{[1]} - \boldsymbol{F}_{\text{ext}(k)}^{[1]} = \boldsymbol{0}\,, \quad k = 1, \dots, m \tag{33}$$

$$\boldsymbol{R}_{kl}^{[2]}\left(\boldsymbol{u}_{kl}^{[2]}\right) = \boldsymbol{F}_{\text{int}(kl)}^{[2]} - \boldsymbol{F}_{\text{ext}(kl)}^{[2]} = \boldsymbol{0}, \quad k, l = 1, \dots, m \tag{34}$$

where $\boldsymbol{u}^{[0]}$ is obtained by solving nonlinear Eq. (32) iteratively using the Newton-Raphson (NR) method, while $\boldsymbol{u}_{k}^{[1]}$ and $\boldsymbol{u}_{kl}^{[2]}$ are obtained using the system of linear equations in Eq. (33) and Eq. (34) sequentially, i.e.,

$$\begin{aligned} \boldsymbol{u}_{k}^{[1]} &= -{\boldsymbol{K}_{T0}^{(1)}}^{-1}\left(\boldsymbol{G}_{\text{int}(k)} - \boldsymbol{G}_{\text{ext}(k)}\right) \\ \boldsymbol{u}_{kl}^{[2]} &= -{\boldsymbol{K}_{T0}^{(1)}}^{-1}\left(\boldsymbol{G}_{\text{int}(kl)} - \boldsymbol{G}_{\text{ext}(kl)}\right) \end{aligned} \tag{35}$$

with

$$\begin{aligned} &\boldsymbol{G}_{\text{int}(k)} \stackrel{\text{def}}{=} \boldsymbol{F}_{\text{int}(k)}^{[1]} \text{ and } \boldsymbol{G}_{\text{ext}(k)} \stackrel{\text{def}}{=} \boldsymbol{F}_{\text{ext}(k)}^{[1]} \\ &\boldsymbol{G}_{\text{int}(kl)} \stackrel{\text{def}}{=} \boldsymbol{F}_{\text{int}(kl)}^{[2]} + \left.\frac{\partial \boldsymbol{K}_{T}^{(1)}}{\partial \xi_k}\right|_{\boldsymbol{\xi}=\boldsymbol{0}} \boldsymbol{u}_{l}^{[1]} + \left.\frac{\partial \boldsymbol{K}_{T}^{(1)}}{\partial \xi_l}\right|_{\boldsymbol{\xi}=\boldsymbol{0}} \boldsymbol{u}_{k}^{[1]} + \boldsymbol{K}_{T0}^{(2)} : \left(\boldsymbol{u}_{l}^{[1]} \otimes \boldsymbol{u}_{k}^{[1]}\right) \\ &\boldsymbol{G}_{\text{ext}(kl)} \stackrel{\text{def}}{=} \boldsymbol{F}_{\text{ext}(kl)}^{[2]} \end{aligned} \tag{36}$$

where the tangent stiffness matrix evaluated at $\boldsymbol{\xi} = \boldsymbol{0}$ and its derivative, i.e., $\boldsymbol{K}_{T0}^{(1)}$ and $\boldsymbol{K}_{T0}^{(2)}$, are computed at the converged displacement vector $\boldsymbol{u}^{[0]}$.

In Eq. (31), the calculation of the terms related to $\boldsymbol{F}_{ext}$ is straightforward. For example, $\partial \boldsymbol{F}_{ext}/\partial \xi_k = (d\boldsymbol{F}_{ext}/d\boldsymbol{P}_0)(d\boldsymbol{P}_0/d\xi_k)$ where $d\boldsymbol{F}_{ext}/d\boldsymbol{P}_0$ is a matrix with 0 and 1 entries in which nonzero entries are in the places corresponding to the degree-of-freedoms (DOFs) of applied loads; $\frac{d\boldsymbol{P}_0}{d\xi_k}$ is nonzero only if $\xi_k \in \boldsymbol{\xi}_1$ and in this case, its value is a column vector from $\boldsymbol{L}_c$; and finally, $\left.\frac{d^2\boldsymbol{F}_{ext}}{d\xi_k d\xi_l}\right|_{\boldsymbol{\xi}=\boldsymbol{0}} = \boldsymbol{0}$. Therefore, the terms that remain to be derived are $\boldsymbol{F}_{\mathrm{int}(k)}^{[1]}$ and $\boldsymbol{F}_{\mathrm{int}(kl)}^{[2]}$. The derivation details are provided below.

The 1st-order derivatives of $\boldsymbol{F}_{int}$ w.r.t. the random variables are calculated by

$$
\begin{aligned}
&\boldsymbol{F}_{\mathrm{int}(k)}^{[1]} = \mathop{\mathcal{A}}_{e=1}^{n_{ele}} \boldsymbol{F}_{\mathrm{int}(k)}^{e[1]} \qquad \text{with} \\
&\boldsymbol{F}_{\mathrm{int}(k)}^{e[1]} \stackrel{\text{def}}{\equiv} \frac{d\boldsymbol{F}_{int}^{e}}{d\xi_k} \\
&\qquad = \int_{\Omega_e} \frac{d(\gamma E)}{d\xi_k} \boldsymbol{B}:\boldsymbol{P} dV + \int_{\Omega_e} \gamma E \boldsymbol{B}:\frac{d\boldsymbol{P}}{d\xi_k} dV + \frac{d\left(E_L(1-\gamma^2)\right)}{d\xi_k} \boldsymbol{k}_L^e.\boldsymbol{u}^{e[0]} \\
&\qquad + E_L(1-\gamma^2)\boldsymbol{k}_L^e.\boldsymbol{u}_k^{e[1]}
\end{aligned}
\tag{37}
$$

where the superscript *e* represents the terms corresponding to the *e*$^{th}$ element and

$$
\begin{aligned}
&\frac{d\boldsymbol{P}}{d\xi_k} = \frac{\partial \boldsymbol{P}}{\partial \boldsymbol{F}}:\frac{d\boldsymbol{F}}{d\xi_k} = \mathbb{A}_4:\frac{d\boldsymbol{F}}{d\xi_k} \qquad \text{with} \\
&\frac{d\boldsymbol{F}}{d\xi_k} = \frac{d\gamma}{d\xi_k}\boldsymbol{B}.\boldsymbol{u}^e + \gamma \boldsymbol{B}.\frac{d\boldsymbol{u}^e}{d\xi_k} = \boldsymbol{B}.\boldsymbol{v}_k
\end{aligned}
\tag{38}
$$

where the scalar derivatives $\partial(\gamma E)/\partial\xi_k$ and $\partial\left(E_L(1-\gamma^2)\right)/\partial\xi_k$ are obtained by the chain rules of differentiation with the derivative of each item detailed in Appendix A.11. The 1st-PK stress $\boldsymbol{P}$ and material tangent $\mathbb{A}_4$ are given in Appendix B.

The 2nd-order derivatives of $\boldsymbol{F}_{int}$ w.r.t. the random variables are calculated by

$$
\begin{aligned}
&\boldsymbol{F}_{\mathrm{int}(kl)}^{[2]} = \mathop{\mathcal{A}}_{e=1}^{n_{ele}} \boldsymbol{F}_{\mathrm{int}(kl)}^{e[2]} \qquad \text{with} \\
&\boldsymbol{F}_{\mathrm{int}(kl)}^{e[2]} \stackrel{\text{def}}{\equiv} \frac{d^2\boldsymbol{F}_{int}^{e}}{d\xi_k d\xi_l} = \left(\frac{d^2\boldsymbol{F}_{int}^{e}}{d\xi_k d\xi_l}\right)_{NL} + \left(\frac{d^2\boldsymbol{F}_{int}^{e}}{d\xi_k d\xi_l}\right)_{L}
\end{aligned}
\tag{39}
$$

where the subscripts *NL* and *L* of the derivative $d^2\boldsymbol{F}_{int}^e/d\xi_k d\xi_l$ denotes the two parts coming from nonlinear and linear terms, respectively, and

$$\left(\frac{d^2\boldsymbol{F}_{int}^e}{d\xi_k d\xi_l}\right)_{NL} = \frac{d^2(\gamma E)}{d\xi_k d\xi_l}\boldsymbol{p}^e + \boldsymbol{k}_1^e.\left(\frac{d(\gamma E)}{d\xi_k}\boldsymbol{v}_l + \frac{d(\gamma E)}{d\xi_l}\boldsymbol{v}_k\right) + \gamma E \boldsymbol{k}_2^e:(\boldsymbol{v}_k \otimes \boldsymbol{v}_l) + \gamma E \boldsymbol{k}_1^e.\boldsymbol{v}_{kl}$$

$$\left(\frac{d^2\boldsymbol{F}_{int}^e}{d\xi_k d\xi_l}\right)_{L} = \frac{d^2\left(E_L(1-\gamma^2)\right)}{d\xi_k d\xi_l}\boldsymbol{k}_L^e.\boldsymbol{u}^{e[0]} + \frac{d\left(E_L(1-\gamma^2)\right)}{d\xi_k}\boldsymbol{k}_L^e.\boldsymbol{u}_l^{e[1]} \tag{40}$$

$$+\frac{d\left(E_L(1-\gamma^2)\right)}{d\xi_l}\boldsymbol{k}_L^e.\boldsymbol{u}_k^{e[1]} + E_L(1-\gamma^2)\boldsymbol{k}_L^e.\boldsymbol{u}_{kl}^{e[2]}$$

in which the scalar derivatives are given in Appendix A.11. The implementation of the uncertainty quantification method using 2nd-order perturbation is verified via the direct Monte Carlo method with a numerical example and the verification results can be checked in Appendix C.1.

# 6 Adjoint sensitivity analysis

This study employs the Method of Moving Asymptotes (MMA) [56] – a gradient-based optimization method – as an optimizer. To obtain the design sensitivities of the objective function with respect to the design variables that are needed by MMA, the adjoint function is constructed as

$$F = \oint_0 + \boldsymbol{\lambda}^{[0]^T}\boldsymbol{R}^{[0]} + \sum_{k=1}^{m}\boldsymbol{\lambda}_k^{[1]^T}\boldsymbol{R}_k^{[1]} + \sum_{k=1}^{m}\sum_{l=1}^{m}\boldsymbol{\lambda}_{kl}^{[2]^T}\boldsymbol{R}_{kl}^{[2]} \tag{41}$$

where $\boldsymbol{\lambda}^{[0]}$, $\boldsymbol{\lambda}_k^{[1]}$ and $\boldsymbol{\lambda}_{kl}^{[2]}$ $(k, l = 1, \ldots, m)$ are the adjoint variables. The adjoint variables $\boldsymbol{\lambda}^{[0]}$, $\boldsymbol{\lambda}_k^{[1]}$ and $\boldsymbol{\lambda}_{kl}^{[2]}$ are introduced to enforce the constraints in Eqns. (32), (33), and (34), which correspond to state variables $\boldsymbol{u}^{[0]}$, $\boldsymbol{u}_k^{[1]}$ and $\boldsymbol{u}_{kl}^{[2]}$, respectively. Interested readers are referred to [57] for the details about the choice and construction of such adjoint functions.

Total differentiation of the adjoint function gives

$$\frac{dF}{d\boldsymbol{x}} = \frac{d\oint_0}{d\boldsymbol{x}} + \boldsymbol{\lambda}^{[0]^T}\frac{d\boldsymbol{R}^{[0]}}{d\boldsymbol{x}} + \sum_{k=1}^{m}\boldsymbol{\lambda}_k^{[1]^T}\frac{d\boldsymbol{R}_k^{[1]}}{d\boldsymbol{x}} + \sum_{k=1}^{m}\sum_{l=1}^{m}\boldsymbol{\lambda}_{kl}^{[2]^T}\frac{d\boldsymbol{R}_{kl}^{[2]}}{d\boldsymbol{x}} \tag{42}$$

which can be expanded as

$$\frac{dF}{d\boldsymbol{x}} = \frac{\partial f_0}{\partial \boldsymbol{x}} + \boldsymbol{\lambda}^{[0]^T} \frac{\partial \boldsymbol{R}^{[0]}}{\partial \boldsymbol{x}} + \sum_{k=1}^{m} \boldsymbol{\lambda}_k^{[1]^T} \frac{\partial \boldsymbol{R}_k^{[1]}}{\partial \boldsymbol{x}} + \sum_{k=1}^{m}\sum_{l=1}^{m} \boldsymbol{\lambda}_{kl}^{[2]^T} \frac{\partial \boldsymbol{R}_{kl}^{[2]}}{\partial \boldsymbol{x}}$$
$$+ \left( \frac{\partial f_0}{\partial \boldsymbol{u}^{[0]}} + \boldsymbol{\lambda}^{[0]^T} \frac{\partial \boldsymbol{R}^{[0]}}{\partial \boldsymbol{u}^{[0]}} + \sum_{k=1}^{m} \boldsymbol{\lambda}_k^{[1]^T} \frac{\partial \boldsymbol{R}_k^{[1]}}{\partial \boldsymbol{u}^{[0]}} + \sum_{k=1}^{m}\sum_{l=1}^{m} \boldsymbol{\lambda}_{kl}^{[2]^T} \frac{\partial \boldsymbol{R}_{kl}^{[2]}}{\partial \boldsymbol{u}^{[0]}} \right) \frac{d\boldsymbol{u}^{[0]}}{d\boldsymbol{x}}$$
$$+ \sum_{q=1}^{m} \left( \frac{\partial f_0}{\partial \boldsymbol{u}_q^{[1]}} + \sum_{k=1}^{m} \boldsymbol{\lambda}_k^{[1]^T} \frac{\partial \boldsymbol{R}_k^{[1]}}{\partial \boldsymbol{u}_q^{[1]}} + \sum_{k=1}^{m}\sum_{l=1}^{m} \boldsymbol{\lambda}_{kl}^{[2]^T} \frac{\partial \boldsymbol{R}_{kl}^{[2]}}{\partial \boldsymbol{u}_q^{[1]}} \right) \frac{d\boldsymbol{u}_q^{[1]}}{d\boldsymbol{x}} \tag{43}$$
$$+ \sum_{k=1}^{m}\sum_{l=1}^{m} \left( \frac{\partial f_0}{\partial \boldsymbol{u}_{kl}^{[2]}} + \boldsymbol{\lambda}_{kl}^{[2]^T} \frac{\partial \boldsymbol{R}_{kl}^{[2]}}{\partial \boldsymbol{u}_{kl}^{[2]}} \right) \frac{d\boldsymbol{u}_{kl}^{[2]}}{d\boldsymbol{x}}$$

In the adjoint method [57, 58], the adjoint variables – $\boldsymbol{\lambda}^{[0]}$, $\boldsymbol{\lambda}_k^{[1]}$, and $\boldsymbol{\lambda}_{kl}^{[2]}$ – are chosen such the following adjoint equations are satisfied

$$\begin{cases} \dfrac{\partial f_0}{\partial \boldsymbol{u}^{[0]}} + \boldsymbol{\lambda}^{[0]^T} \dfrac{\partial \boldsymbol{R}^{[0]}}{\partial \boldsymbol{u}^{[0]}} + \sum_{k=1}^{m} \boldsymbol{\lambda}_k^{[1]^T} \dfrac{\partial \boldsymbol{R}_k^{[1]}}{\partial \boldsymbol{u}^{[0]}} + \sum_{k=1}^{m}\sum_{l=1}^{m} \boldsymbol{\lambda}_{kl}^{[2]^T} \dfrac{\partial \boldsymbol{R}_{kl}^{[2]}}{\partial \boldsymbol{u}^{[0]}} = \boldsymbol{0} \\ \dfrac{\partial f_0}{\partial \boldsymbol{u}_q^{[1]}} + \boldsymbol{\lambda}_q^{[1]^T} \dfrac{\partial \boldsymbol{R}_q^{[1]}}{\partial \boldsymbol{u}_q^{[1]}} + \sum_{k=1}^{m}\sum_{l=1}^{m} \boldsymbol{\lambda}_{kl}^{[2]^T} \dfrac{\partial \boldsymbol{R}_{kl}^{[2]}}{\partial \boldsymbol{u}_q^{[1]}} = \boldsymbol{0} & , \quad \begin{matrix} p = 1, \dots, m \\ q = 1, \dots, m \end{matrix} \\ \dfrac{\partial f_0}{\partial \boldsymbol{u}_{pq}^{[2]}} + \boldsymbol{\lambda}_{pq}^{[2]^T} \dfrac{\partial \boldsymbol{R}_{pq}^{[2]}}{\partial \boldsymbol{u}_{pq}^{[2]}} = \boldsymbol{0} \end{cases} \tag{44}$$

where $\partial \boldsymbol{R}_{kl}^{[2]} / \partial \boldsymbol{u}_q^{[1]} = \boldsymbol{0}$ if $q \neq k$ and $q \neq l$, $\partial \boldsymbol{R}^{[0]} / \partial \boldsymbol{u}^{[0]} = \partial \boldsymbol{R}_q^{[1]} / \partial \boldsymbol{u}_q^{[1]} = \boldsymbol{R}_{pq}^{[2]} / \partial \boldsymbol{u}_{pq}^{[2]} = \boldsymbol{K}_{T0}^{(1)}$. For solving the adjoint system in Eq. (44), $\boldsymbol{\lambda}_{pq}^{[2]}$ is computed first, followed by $\boldsymbol{\lambda}_q^{[1]}$, and $\boldsymbol{\lambda}^{[0]}$. Finally, the sensitivities are obtained as

$$\frac{dF}{d\boldsymbol{x}} = \frac{\partial f_0}{\partial \boldsymbol{x}} + \boldsymbol{\lambda}^{[0]^T} \frac{\partial \boldsymbol{R}^{[0]}}{\partial \boldsymbol{x}} + \sum_{k=1}^{m} \boldsymbol{\lambda}_k^{[1]^T} \frac{\partial \boldsymbol{R}_k^{[1]}}{\partial \boldsymbol{x}} + \sum_{k=1}^{m}\sum_{l=1}^{m} \boldsymbol{\lambda}_{kl}^{[2]^T} \frac{\partial \boldsymbol{R}_{kl}^{[2]}}{\partial \boldsymbol{x}} \tag{45}$$

Combining Eqns. (44) and (45), the derivatives that need to be derived to solve the adjoint system are

$$\frac{\partial \not{f}_0}{\partial \boldsymbol{x}}, \quad \frac{\partial \not{f}_0}{\partial \boldsymbol{u}^{[0]}}, \quad \frac{\partial \not{f}_0}{\partial \boldsymbol{u}_k^{[1]}}, \quad \frac{\partial \not{f}_0}{\partial \boldsymbol{u}_{kl}^{[2]}},$$

$$\frac{\partial \boldsymbol{R}^{[0]}}{\partial \boldsymbol{x}}, \quad \frac{\partial \boldsymbol{R}_k^{[1]}}{\partial \boldsymbol{x}}, \quad \frac{\partial \boldsymbol{R}_k^{[1]}}{\partial \boldsymbol{u}^{[0]}}, \tag{46}$$

$$\frac{\partial \boldsymbol{R}_{kl}^{[2]}}{\partial \boldsymbol{x}}, \quad \frac{\partial \boldsymbol{R}_{kl}^{[2]}}{\partial \boldsymbol{u}^{[0]}}, \quad \frac{\partial \boldsymbol{R}_{kl}^{[2]}}{\partial \boldsymbol{u}_q^{[1]}}$$

The analytical expressions for these derivatives are given in Appendices A.1 to A.10. The implementation of the design sensitivity analysis is verified via the central difference method and the verification results can be checked in Appendix C.2.

*Remark:* It is noted that there is *symmetry* for the second-order derivatives of displacements, i.e., $\boldsymbol{u}_{kl}^{[2]} = \boldsymbol{u}_{lk}^{[2]}$. Thus, only the constraint functions $\boldsymbol{R}_{kl}^{[2]}$ ($k = 1, \ldots, m$ and $l = k, \ldots, m$) are needed in the above adjoint formulation, which can significantly reduce the computational cost in the adjoint sensitivity analysis.

## 7 Numerical examples

In this section, the influence of distinct types of uncertain sources on the robust designs is studied via three numerical examples where the robust designs are compared with their deterministic counterparts. In all cases, a continuation scheme is used to avoid analysis failure during early optimization iterations and to relax the non-convexity of the optimization problem [34]. The specific continuation scheme, unless stated otherwise, is as follows: the penalization parameter $p$ (Eq. (1)) is increased from 1 to 4 @ 0.1 every 20 iterations while $p_l$ (Eq. (13)) is increased from 4 to 7 @ 0.1 every 20 iterations, and the filter parameter $\beta$ (Eq. (3)) is increased from 1 to 4 @ 0.1 every 20 iterations. After the maximum value of the parameters is reached, an additional 200 iterations are carried out, and the optimization is terminated. In the following examples, 2-D plane strain problems are considered, and in this setting, $\mathrm{sdim} = 2$ and $l_{cz} = \infty$. The MMA is used as an optimizer with default parameter settings unless stated otherwise. The optimization is initiated with a uniform density design such that $x_k = V_f$ , $k = 1, 2, \ldots, n_{ele}$ . Table 2 summarizes the parameters of the stochastic distributions considered for different robust designs in different numerical examples. All the numerical computations are carried out in a C++ based in-house finite element library *CPSSL-FEA* developed at the University of Notre Dame.

Table 2. Summary of distribution parameters considered for different robust designs (RD) in different numerical examples.

| Example | Design | Load uncertainty | Material uncertainty ($E$-field) | Geometric uncertainty ($\eta$-field) |
|---|---|---|---|---|
| Compression block | RD-1 | $\boldsymbol{\mu}_P = [0 \quad -0.08]^T$<br>$\boldsymbol{\Sigma}_P = 10^{-6}\boldsymbol{I}$ | - | - |
| | RD-2 | $\boldsymbol{\mu}_P = [0 \quad -0.08]^T$<br>$\boldsymbol{\Sigma}_P = 10^{-5}\boldsymbol{I}$ | - | - |
| | RD-3 | $\boldsymbol{\mu}_P = [0 \quad -0.08]^T$<br>$\boldsymbol{\Sigma}_P = 10^{-4}\boldsymbol{I}$ | - | - |
| | RD-4 | - | $E_0 \sim \text{Lognormal}(0.85, 0.0625)$<br>$l_{cx} = 20$, $l_{cy} = \infty$ | - |
| | RD-5 | - | $E_0 \sim \text{Lognormal}(0.85, 0.0625)$<br>$l_{cx} = 40$, $l_{cy} = \infty$ | - |
| | RD-6 | - | $E_0 \sim \text{Lognormal}(0.85, 0.0625)$<br>$l_{cx} = 60$, $l_{cy} = \infty$ | - |
| | RD-7 | - | - | $\eta \sim \mathcal{U}(0.3, 0.8)$<br>$l_{cx} = 20$, $l_{cy} = \infty$ |
| | RD-8 | - | - | $\eta \sim \mathcal{U}(0.3, 0.8)$<br>$l_{cx} = 60$, $l_{cy} = \infty$ |
| | RD-9 | - | - | $\eta \sim \mathcal{U}(0.3, 0.8)$<br>$l_{cx} = 100$, $l_{cy} = \infty$ |
| | RD-10 | $\boldsymbol{\mu}_P = [0 \quad -0.08]^T$<br>$\boldsymbol{\Sigma}_P = 10^{-5}\boldsymbol{I}$ | $E_0 \sim \text{Lognormal}(0.85, 0.0625)$<br>$l_{cx} = 20$, $l_{cy} = \infty$ | $\eta \sim \mathcal{U}(0.3, 0.8)$<br>$l_{cx} = 20$, $l_{cy} = \infty$ |
| Clamped beam | RD-1 | $\boldsymbol{\mu}_P = [0 \quad -90]^T$<br>$\boldsymbol{\Sigma}_P = 36\boldsymbol{I}$ | - | - |
| | RD-2 | $\boldsymbol{\mu}_P = [0 \quad -90]^T$<br>$\boldsymbol{\Sigma}_P = 64\boldsymbol{I}$ | - | - |
| | RD-3 | $\boldsymbol{\mu}_P = [0 \quad -90]^T$<br>$\boldsymbol{\Sigma}_P = 100\boldsymbol{I}$ | - | - |
| | RD-4 | - | $E_0 \sim \text{Lognormal}(10, 4)$<br>$l_{cx} = 30$, $l_{cy} = \infty$ | - |
| | RD-5 | - | $E_0 \sim \text{Lognormal}(10, 4)$<br>$l_{cx} = 45$, $l_{cy} = \infty$ | - |
| | RD-6 | - | $E_0 \sim \text{Lognormal}(10, 4)$<br>$l_{cx} = 90$, $l_{cy} = \infty$ | - |
| | RD-7 | $\boldsymbol{\mu}_P = [0 \quad -90]^T$<br>$\boldsymbol{\Sigma}_P = 36\boldsymbol{I}$ | $E_0 \sim \text{Lognormal}(10, 4)$<br>$l_{cx} = 90$, $l_{cy} = \infty$ | - |
| | RD-8 | $\boldsymbol{\mu}_P = [0 \quad -90]^T$<br>$\boldsymbol{\Sigma}_P = 36\boldsymbol{I}$ | $E_0 \sim \text{Lognormal}(10, 4)$<br>$l_{cx} = 90$, $l_{cy} = \infty$ | - |
| | RD-9 | $\boldsymbol{\mu}_P = [0 \quad -90]^T$<br>$\boldsymbol{\Sigma}_P = 36\boldsymbol{I}$ | $E_0 \sim \text{Lognormal}(10, 4)$<br>$l_{cx} = 90$, $l_{cy} = \infty$ | - |
| | RD-10 | $\boldsymbol{\mu}_P = [0 \quad -90]^T$<br>$\boldsymbol{\Sigma}_P = 36\boldsymbol{I}$ | $E_0 \sim \text{Lognormal}(10, 4)$<br>$l_{cx} = 90$, $l_{cy} = \infty$ | - |
| Pinned column | RD-1 | - | $E_0 \sim \text{Lognormal}(3, 0.5625)$<br>$l_{cx} = 60$, $l_{cy} = \infty$ | - |
| | RD-2 | - | $E_0 \sim \text{Lognormal}(3, 0.5625)$<br>$l_{cx} = 100$, $l_{cy} = \infty$ | - |
| | RD-3 | - | $E_0 \sim \text{Lognormal}(3, 0.5625)$ | - |
| | | | $l_{cx} = l_{cy} = 400$ | |
| | RD-4 | - | - | $\eta \sim \mathcal{U}(0.3, 0.8)$<br>$l_{cx} = 75$, $l_{cy} = \infty$ |
| | RD-5 | - | - | $\eta \sim \mathcal{U}(0.3, 0.8)$<br>$l_{cx} = 400$, $l_{cy} = \infty$ |
| | RD-6 | - | - | $\eta \sim \mathcal{U}(0.3, 0.8)$<br>$l_{cx} = l_{cy} = 400$ |
| | RD-7 | - | $E_0 \sim \text{Lognormal}(3, 0.5625)$<br>$l_{cx} = 60$, $l_{cy} = \infty$ | $\eta \sim \mathcal{U}(0.3, 0.8)$<br>$l_{cx} = 75$, $l_{cy} = \infty$ |

## *7.1 Compression block*

In the first example, a compression block discretized by 160×160 four-node quadrilateral finite elements (FE) is considered, as shown in Figure 1a. To avoid localized deformations, the elements in the top two FE layers are solid and are not designed. Poisson's ratio of the material is set to 0.4, and Young's modulus for deterministic optimization is set to $E_0$ = 0.85 MPa. Also, $E_{L0}$ = 0.85 MPa and $\nu_L$ = 0.4 for linear energy interpolation (Eq. (13)). The deterministic load is set to $\boldsymbol{P}_0 = [0 \quad -0.08]^T$ N. The load, material, and geometric uncertainties have been selectively included for different robust design optimizations; see Table 2 for the details. It is noted that the $\eta$-random field for the geometric uncertainty with correlation length ($l_{cx}$) 20 (robust design 7), 60 (robust design 8), and 100 (robust design 9) give 9, 4, and 2 random variables in the KL expansion, respectively, which cover 92.68%, 96.76%, and 90.71% of the cumulative spectrum. In addition, the $E$-random field for the material uncertainty with correlation length ($l_{cx}$) 20 (robust design 4), 40 (robust design 5), and 60 (robust design 6) lead to 9, 5, and 4 random variables in KL expansion, respectively, taking up 92.68%, 94.14%, and 96.76% of the cumulative spectrum. The density filter radius, in this case, is 9 mm, and the volume fraction is limited to 0.5. In addition, symmetry is enforced during optimization such that the left half of the domain mirrors the right half. In the robust topology optimization, for this problem, the penalty factor $\alpha$ in the objective function $f_0$ in Eq. (24) is set to 1. For comparison purposes, the deterministic optimization is first carried out, and the optimized design and convergence history are shown in Figure 1(b) and Figure 1(c), respectively.

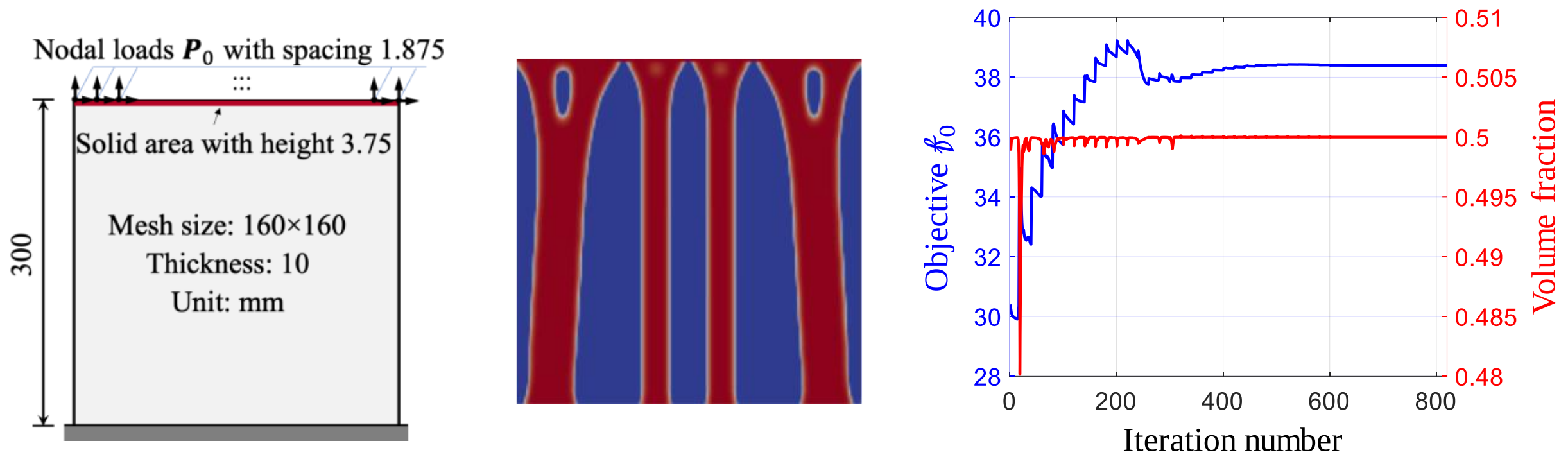

(a) Compression block (b) Deterministic design (c) Convergence history

Figure 1. Compression block and the optimization results of deterministic design.

### *7.1.1 Effect of load uncertainty*

Figure 2 and Figure 3 show the optimized designs and the objective and constraint evolution histories of the compression block under stochastic loads with different standard deviations $\sigma_P$. Figure 4 plots the end compliance of the designs under different loads, where the higher robustness of the designs optimized under loads with higher $\sigma_P$ can be observed. The robustness here is referred to as a change in the objective value, i.e., end compliance, under variations of loads. The higher robustness is achieved at the expense of an increase in objective value at the mean loads, i.e., the loads corresponding to the deterministic case, see Figure 4a. Next, the deformed shapes of the designs corresponding to the rightmost points on each curve in Figure 4a, i.e., load with a variation of 0.01 N in its horizontal component, are shown in Figure 5. It can be observed that the deterministic design undergoes much larger deformations compared to the other robust designs. Moreover, robust design 3 has the smallest deformation among the four designs, which confirms the results in Figure 4(a).

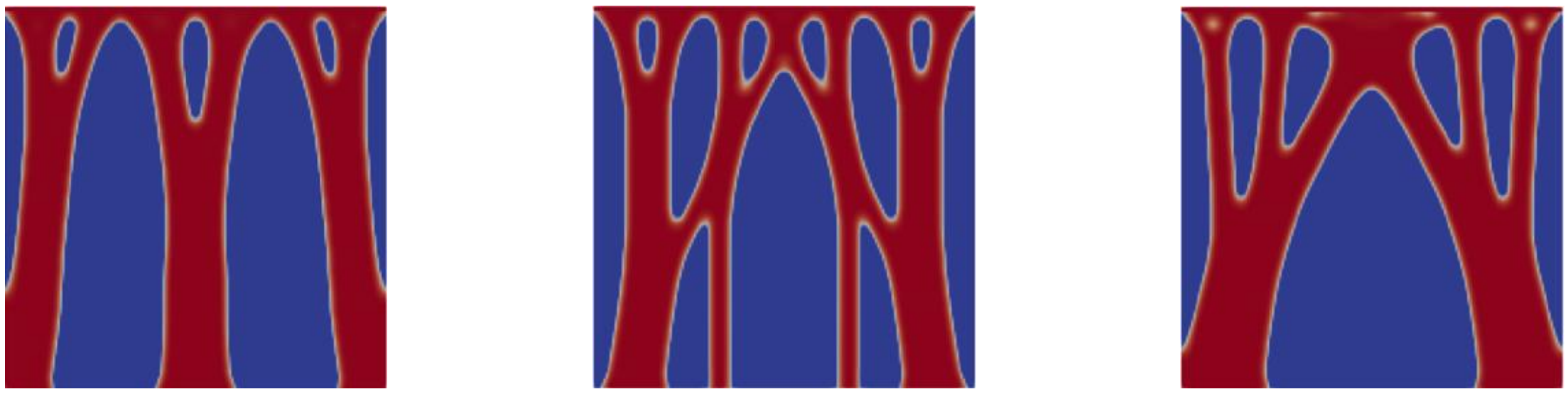

(a) Robust design 1 ($\sigma_P$ = $10^{-3}$ N)　(b) Robust design 2 ($\sigma_P$ = $3.162\times10^{-3}$ N)　(c) Robust design 3 ($\sigma_P$ = $10^{-2}$ N)

Figure 2. Robust designs with different standard deviations of the stochastic loads.

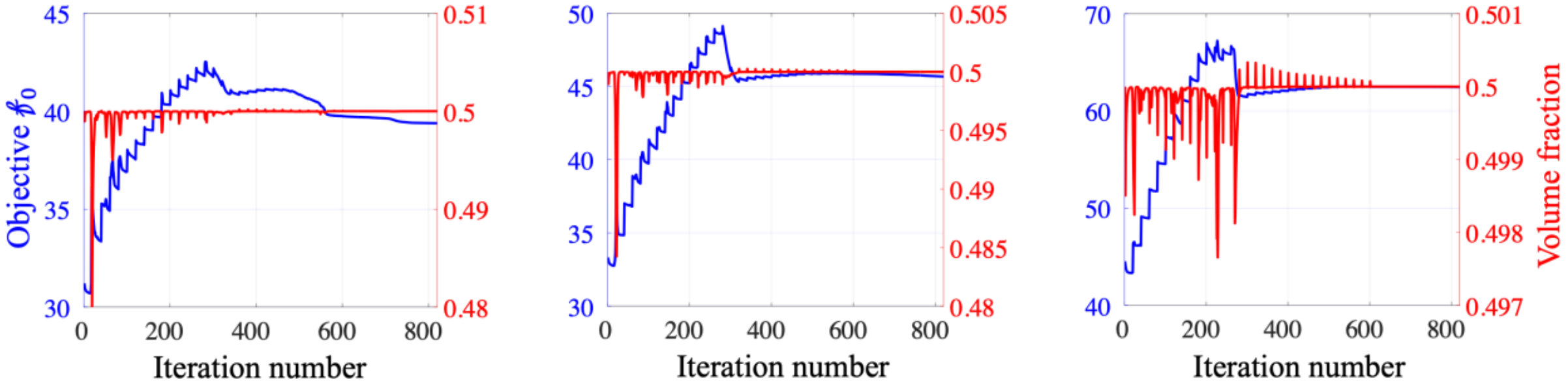

(a) Robust design 1 ($\sigma_P$ = $10^{-3}$ N)　(b) Robust design 2 ($\sigma_P$ = $3.162\times10^{-3}$ N)　(c) Robust design 3 ($\sigma_P$ = $10^{-2}$ N)

Figure 3. Convergence histories for the robust designs under stochastic loads.

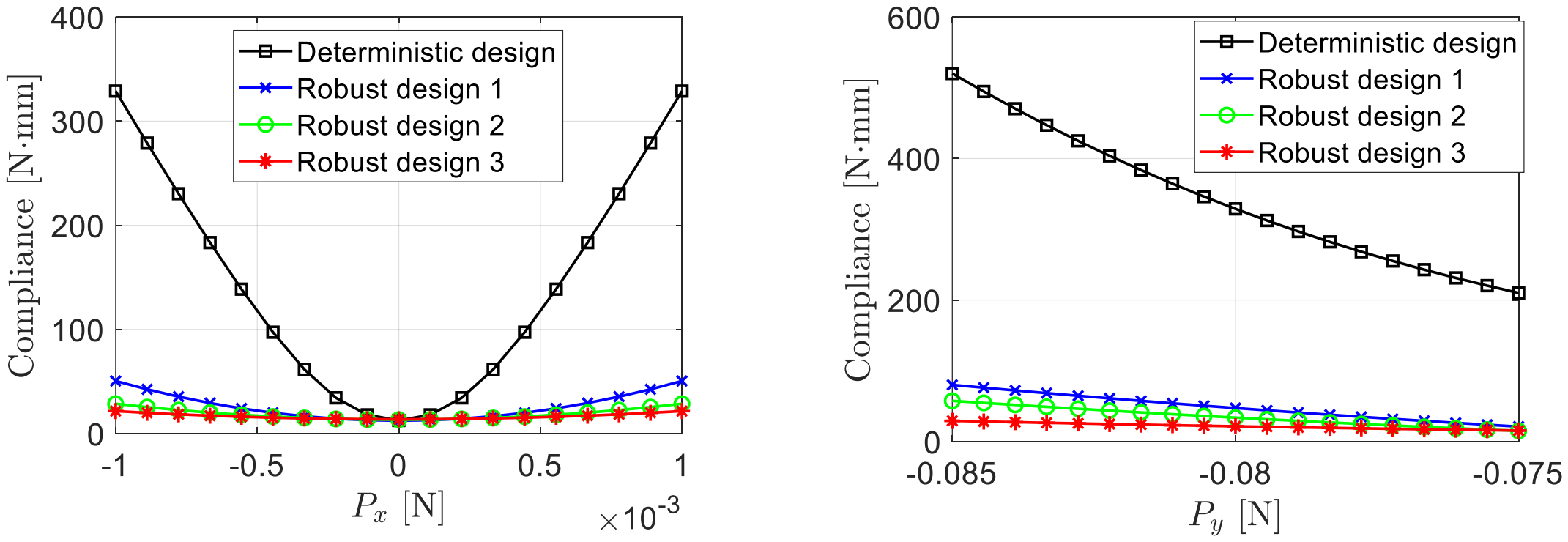

(a) $\boldsymbol{P}_0 = [P_x \quad -0.08]^T$ N　(b) $\boldsymbol{P}_0 = [0.001 \quad P_y]^T$ N

Figure 4. Compliance versus load curves of deterministic and robust designs under different stochastic loads.

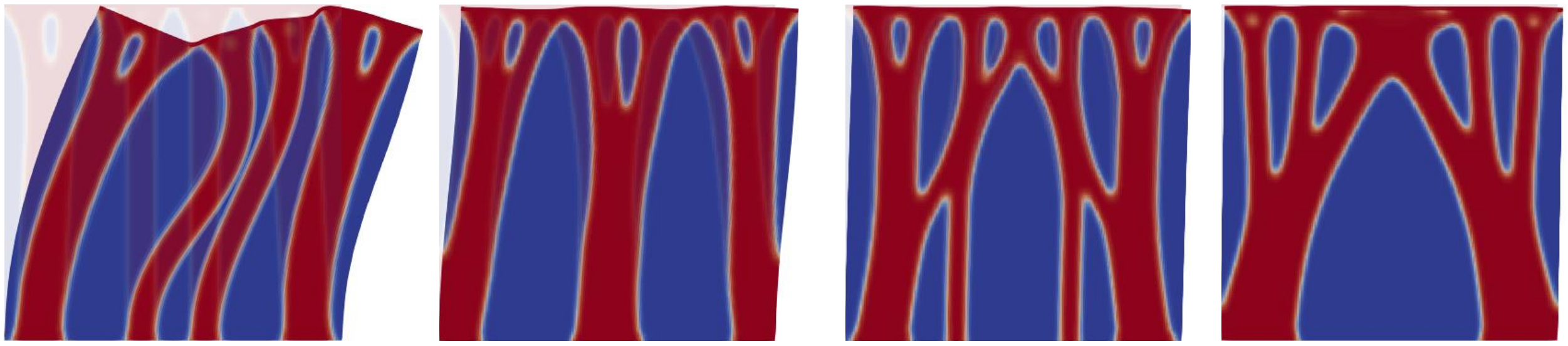

(a) Deterministic design (b) Robust design 1 (c) Robust design 2 (d) Robust design 3

Figure 5. Deformed shapes of deterministic and robust designs for the load $\boldsymbol{P}_0 = [0.001 \quad -0.08]^T$ N.

### *7.1.2 Effect of material uncertainty*

To study the effect of material uncertainty, Young's modulus is modeled by a lognormal random field with different values of the correlation length , $l_{cx}$ and their corresponding first two eigenmodes are obtained in Figure 6, i.e., $\boldsymbol{\gamma}_1$ and $\boldsymbol{\gamma}_2$ in Eq. (21) depicting the correlation structure of the random field. Figure 6 also indicates the symmetry and asymmetry of the underlying modes, i.e., $\boldsymbol{\gamma}_1$ is symmetric while $\boldsymbol{\gamma}_2$ is asymmetric. A larger magnitude in the color bar signifies larger variations, while the same sign indicates changes occurring in the same direction, i.e., simultaneous increase or decrease of the underlying field. The external load and projection cutoff value $\eta$ are deterministic in this case. When three different values of the correlation length $l_{cx}$ are chosen, different robust designs and their corresponding convergence histories are obtained in Figure 7 and Figure 8, respectively. To evaluate the robust performance of the deterministic and robust designs towards material uncertainty, a sequence of various Young's modulus fields is generated by preserving the terms related to the specified eigenmode (Figure 6) in Eq. (21) which is then transformed by Eq. (17). The curves of compliance versus eigenmode coefficient for 1$^{\text{st}}$ eigenmode (Figure 9) and 2$^{\text{nd}}$ eigenmode (Figure 10) are obtained by carrying out FEAs for different Young's modulus fields. It can be observed that when symmetric 1$^{\text{st}}$ eigenmode $\boldsymbol{\gamma}_1$ is considered for generating Young's modulus fields, robust designs are as sensitive to material variations as the deterministic design, while for the asymmetric 2nd eigenmode $\boldsymbol{\gamma}_2$ robust designs clearly show better robustness than the deterministic design. For the asymmetric material random field associated with the 2nd eigenmode with different settings of $\xi_{22}$, the deformed shapes of the deterministic and robust designs are shown in Figure 11. These results show that deterministic design can experience large deformations under perturbations of material fields, while the response of robust designs is considerably stiffer.

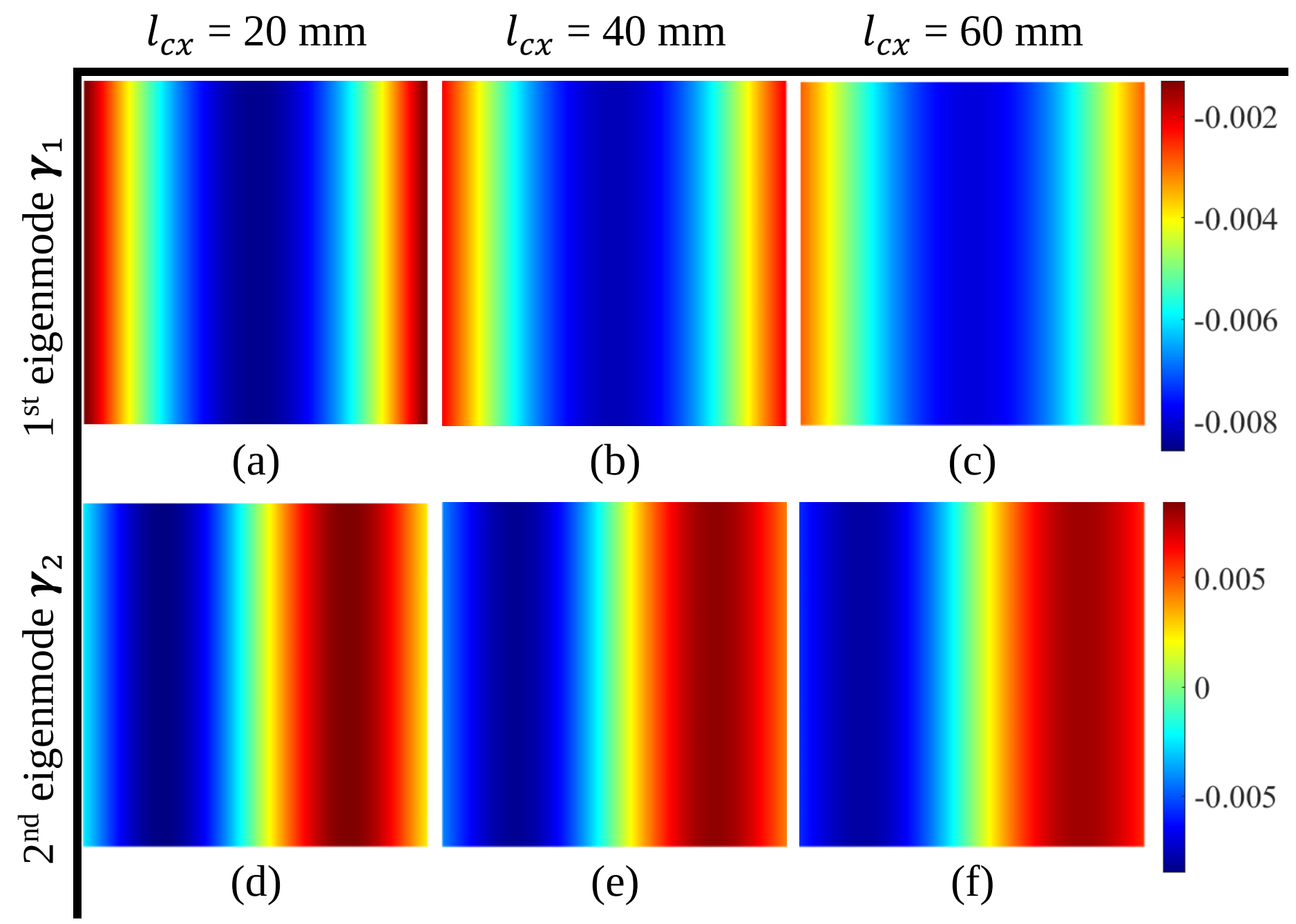

Figure 6. First two eigenmodes in the KL expansion of $E$-field with different values of the correlation length $l_{cx}$.

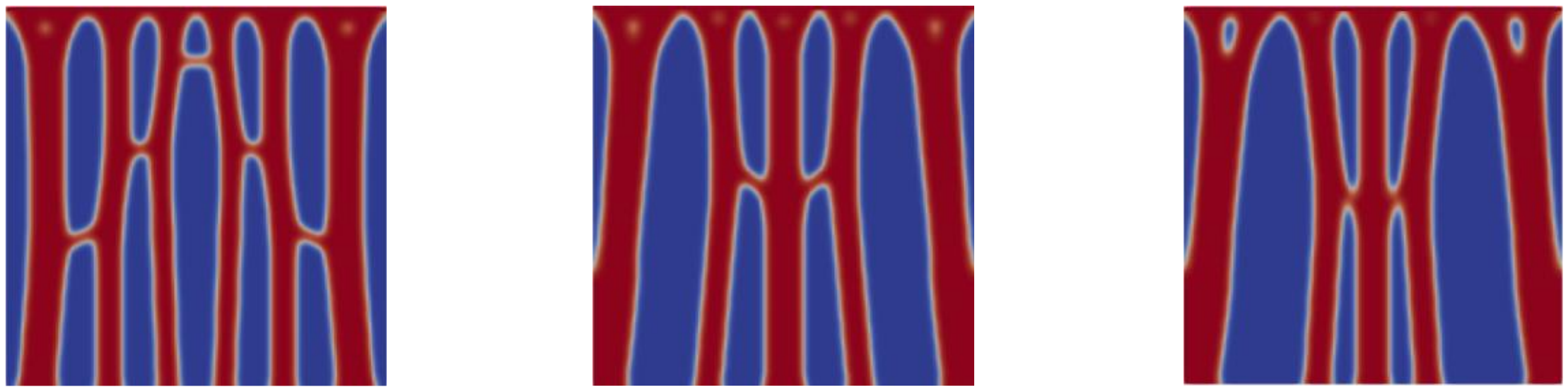

(a) Robust design 4 ($l_{cx}$ = 20 mm) (b) Robust design 5 ($l_{cx}$ = 40 mm) (c) Robust design 6 ($l_{cx}$ = 60 mm)

Figure 7. Robust designs with different values of the correlation length $l_{cx}$ in $E$-field.

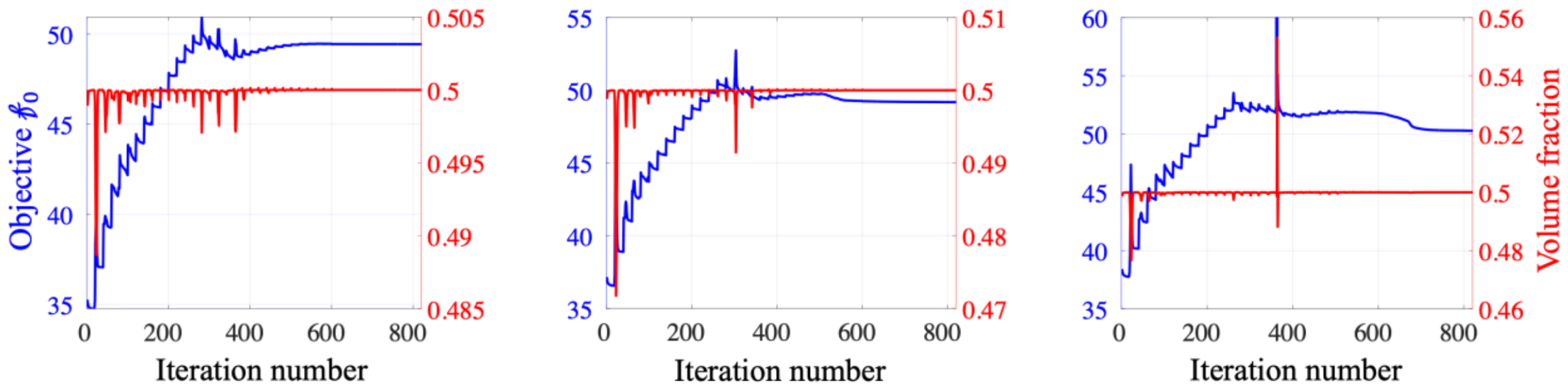

(a) Robust design 4 ($l_{cx}$ = 20 mm) (b) Robust design 5 ($l_{cx}$ = 40 mm) (c) Robust design 6 ($l_{cx}$ = 60 mm)

Figure 8. Convergence histories for robust designs with different values $l_{cx}$ in $E$-field.

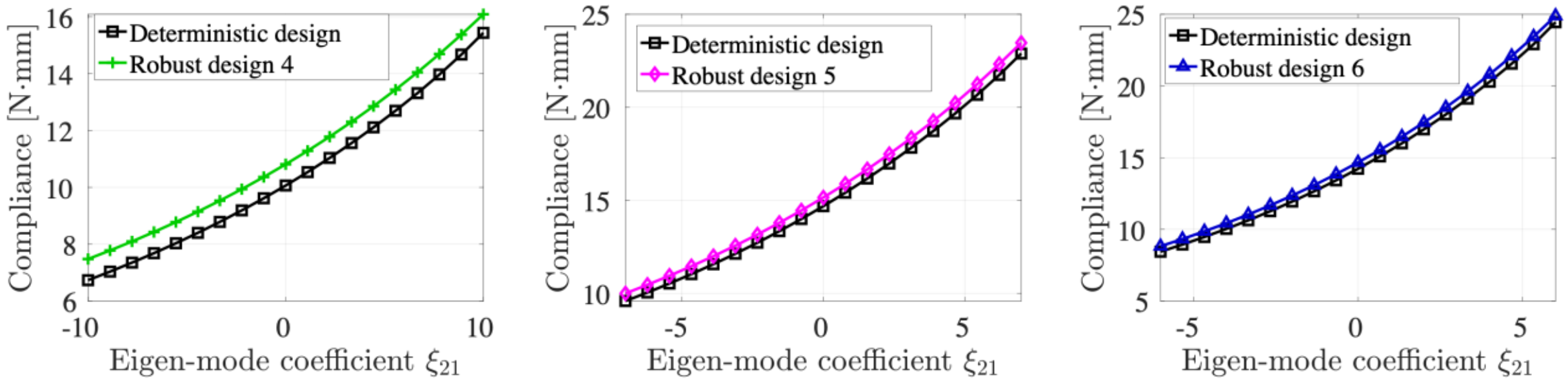

(a) $l_{cx}$ = 20 mm (b) $l_{cx}$ = 40 mm (c) $l_{cx}$ = 60 mm

Figure 9. Compliance versus eigenmode coefficient curves of deterministic and robust designs for 1st eigenmode $\boldsymbol{\gamma}_1$ in the KL expansion of $E$-field.

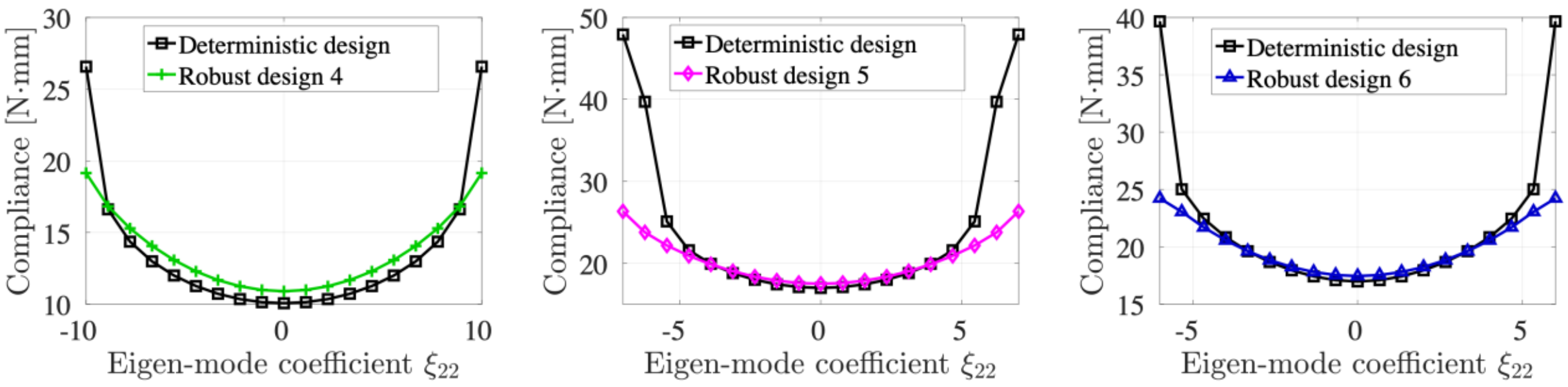

(a) $l_{cx}$ = 20 mm (b) $l_{cx}$ = 40 mm (c) $l_{cx}$ = 60 mm

Figure 10. Compliance versus eigenmode coefficient curves of deterministic and robust designs for 2nd eigenmode $\boldsymbol{\gamma}_2$ in the KL expansion of $E$-field.

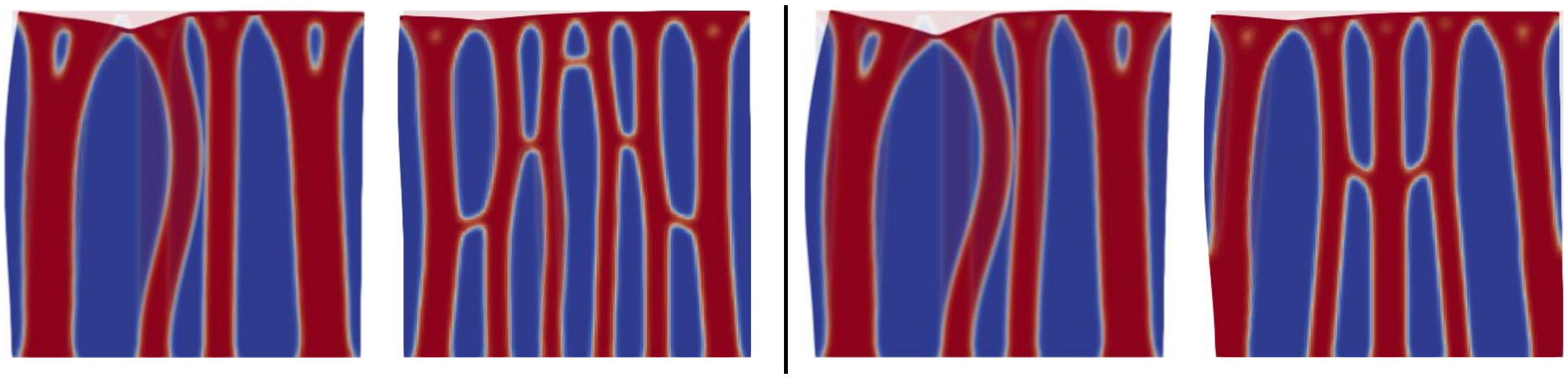

Deterministic design Robust design 4

(a) $l_{cx}$ = 20 mm, $\xi_{22}$ = 10

Deterministic design Robust design 5

(b) $l_{cx}$ = 40 mm, $\xi_{22}$ = 7

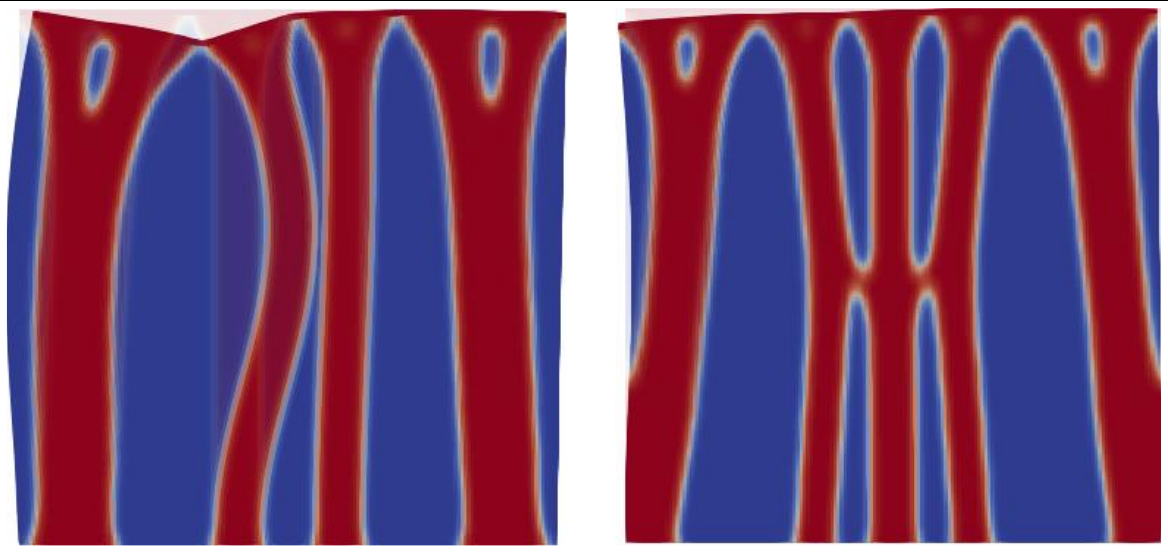

Deterministic design Robust design 6

(c) $l_{cx}$ = 60 mm, $\xi_{22}$ = 6

Figure 11. Deformed shapes of deterministic and robust designs for 2nd eigenmode $\boldsymbol{\gamma}_2$ in the KL expansion of $E$-field with different values of the eigenmode coefficient $\xi_{22}$.

### *7.1.3 Effect of geometric uncertainty*

To investigate the effect of geometric uncertainty, the projection cutoff value $\eta$ is modeled by a uniform random field with different values of the correlation length $l_{cx}$, and their corresponding first two eigenmodes are plotted in Figure 12. The loading and material properties are deterministic in this case. Different robust designs are obtained with three different correlation length values ($l_{cx}$) of the cutoff value $\eta$, as shown in Figure 13, and the corresponding convergence histories are shown in Figure 14. To study the robustness of the deterministic and robust designs under geometric uncertainty, a series of various projection cutoff value $\eta$ fields are generated by including only the terms related to the specified eigenmode (Figure 12) in Eq. (23) which is then transformed by Eq. (22). The curves of compliance versus eigenmode coefficient for 1st eigenmode (Figure 15) and 2nd eigenmode (Figure 16) are obtained by performing FEAs for different projection cutoff value $\eta$. As expected, all three robust designs are more robust than the deterministic design. This implies that incorporating geometric uncertainty in the design optimization phase leads to robust structures against the geometric spatial variations, and the optimized robust structures have different layouts when different geometric uncertainties are considered. As an illustration, Figure 17 plots the deformed shapes of the deterministic and robust designs with the asymmetric projection cutoff value $\eta$ field associated with the 2nd eigenmode with different settings of $\xi_{32}$ when different correlation length values ($l_{cx}$) are considered. Again, under geometric uncertainties, some members in the deterministic design undergo large deformations, while the performance of robust designs does not deteriorate under such perturbations.

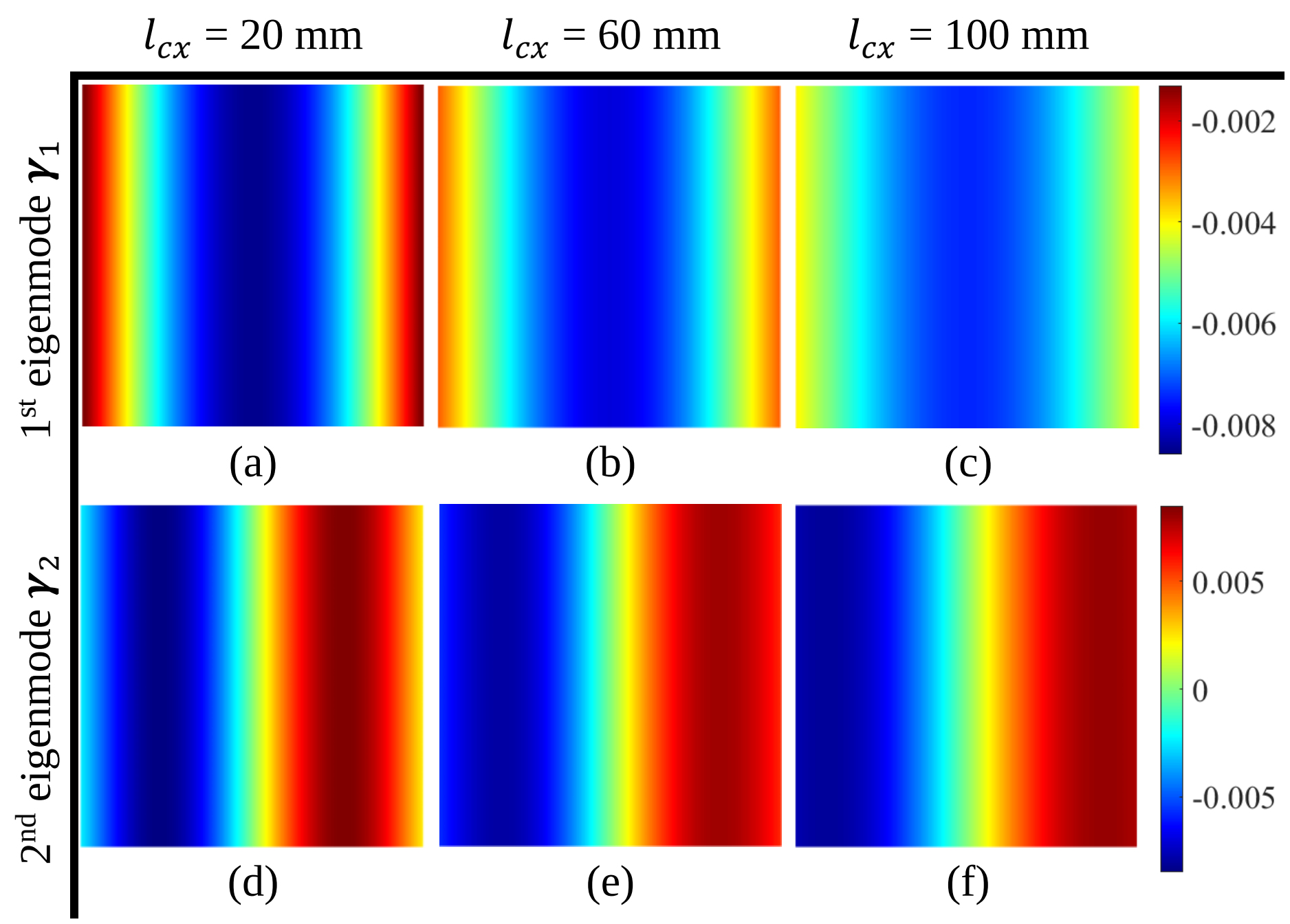

Figure 12. First two eigenmodes in the KL expansion of $\eta$-field.

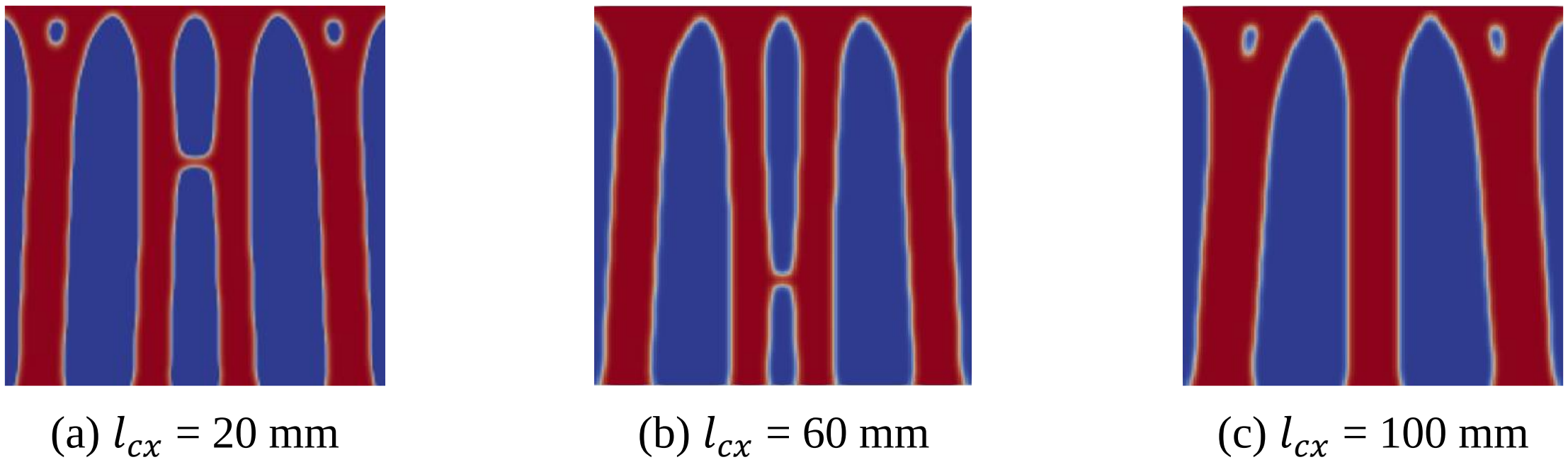

Figure 13. Robust designs with different values of the correlation length $l_{cx}$ in $\eta$-field.

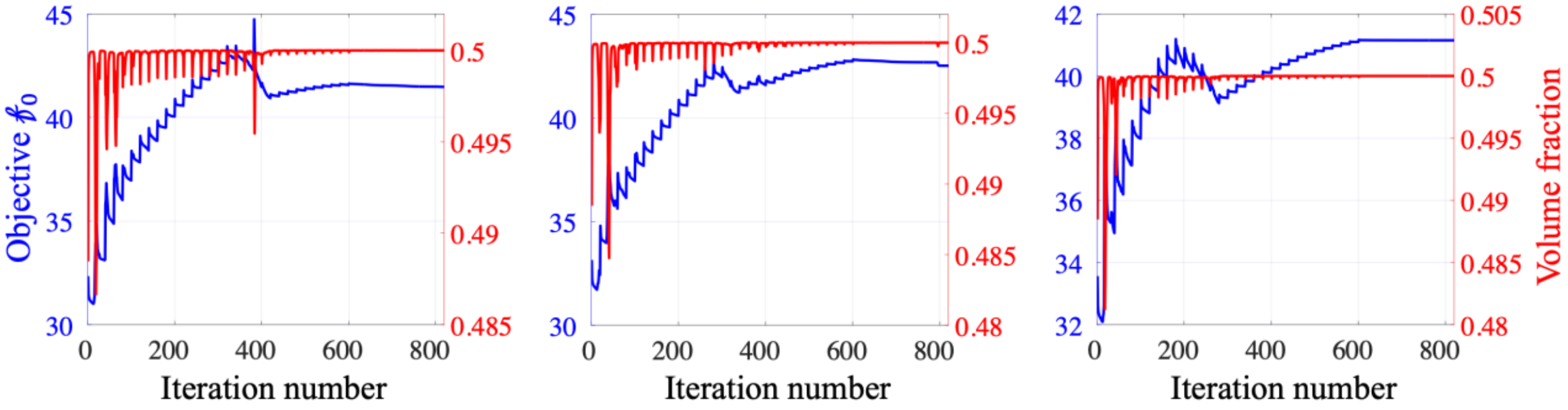

(a) Robust design 4 ($l_{cx}$ = 20 mm) (b) Robust design 5 ($l_{cx}$ = 60 mm) (c) Robust design 6 ($l_{cx}$ = 100 mm)

Figure 14. Convergence histories for the robust designs with different values of $l_{cx}$ in $\eta$-field.

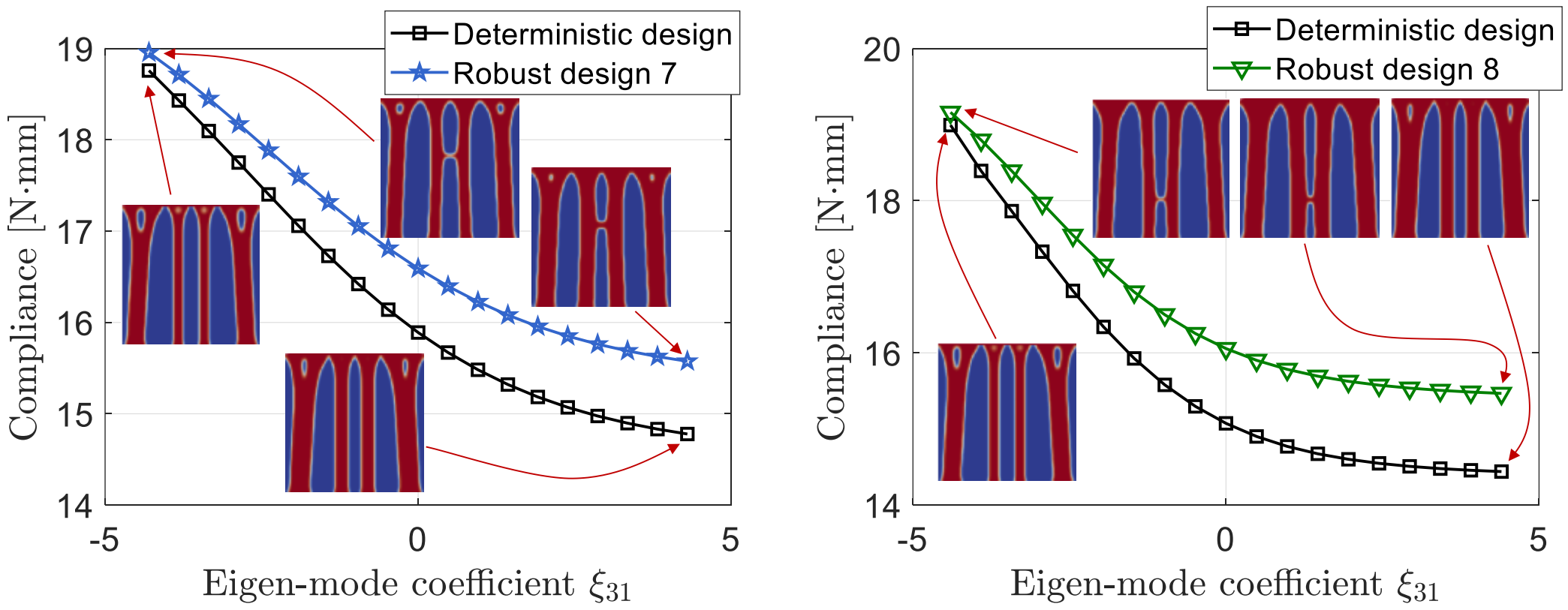

(a) $l_{cx}$ = 20 mm (b) $l_{cx}$ = 60 mm

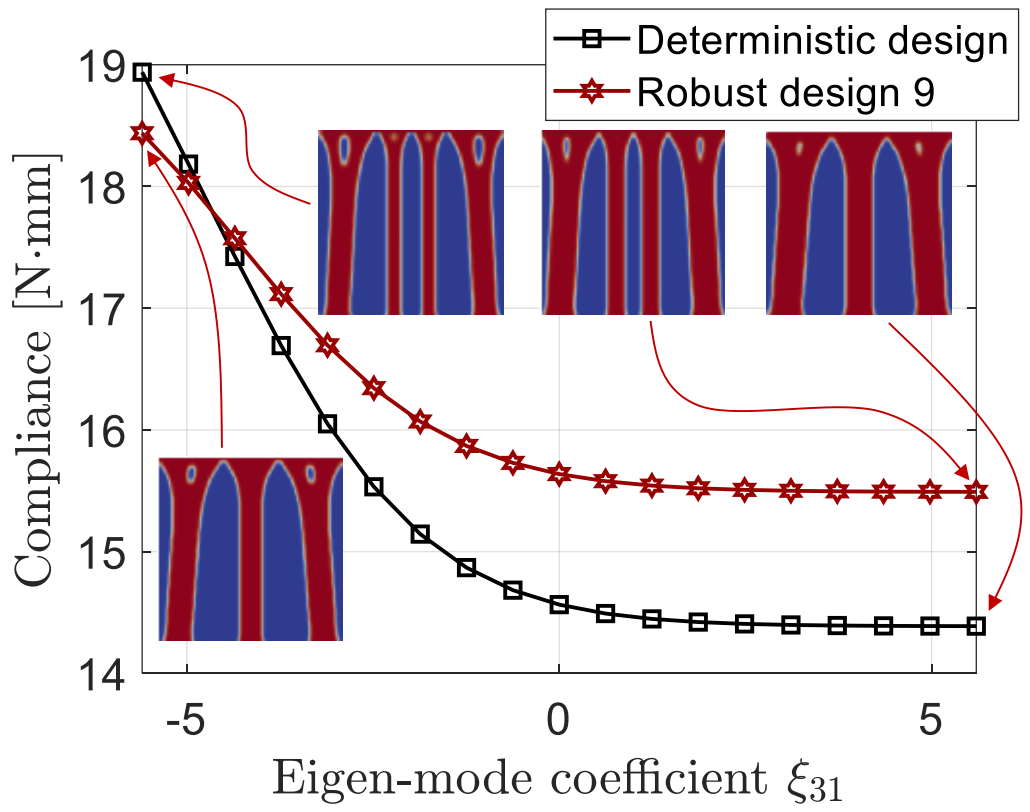

(c) $l_{cx}$ = 100 mm

Figure 15. Compliance versus eigenmode coefficient curves of deterministic and robust designs for 1st eigenmode $\boldsymbol{\gamma}_1$ in the KL expansion of $\eta$-field.

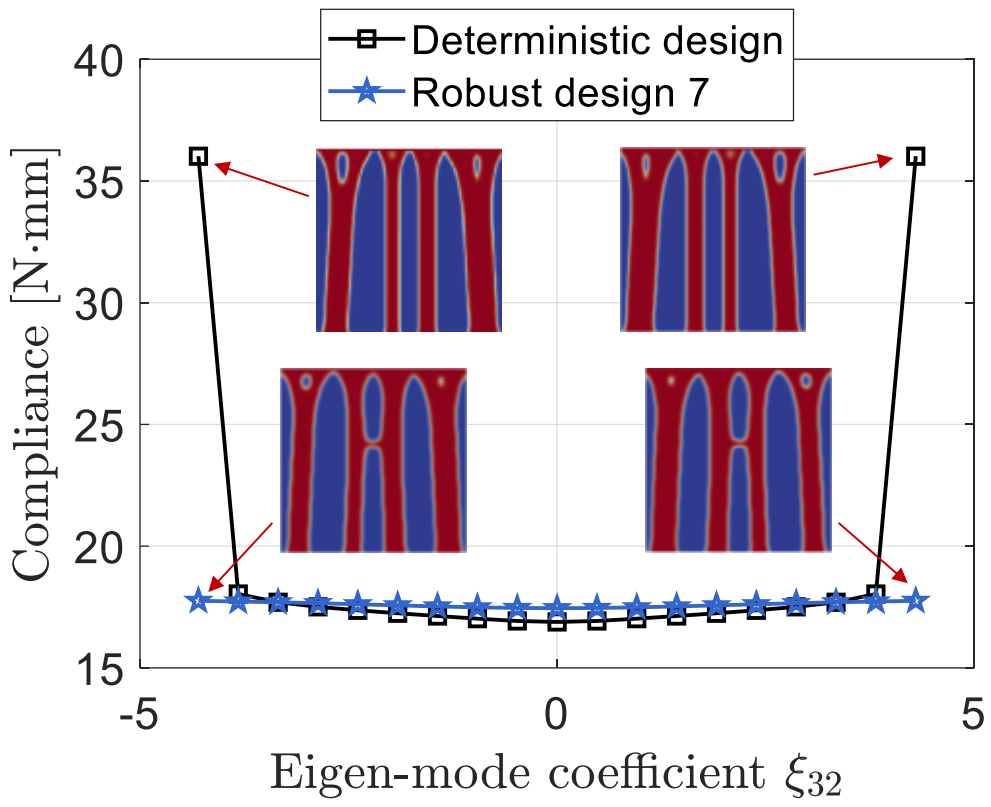

(a) $l_{cx}$ = 20 mm

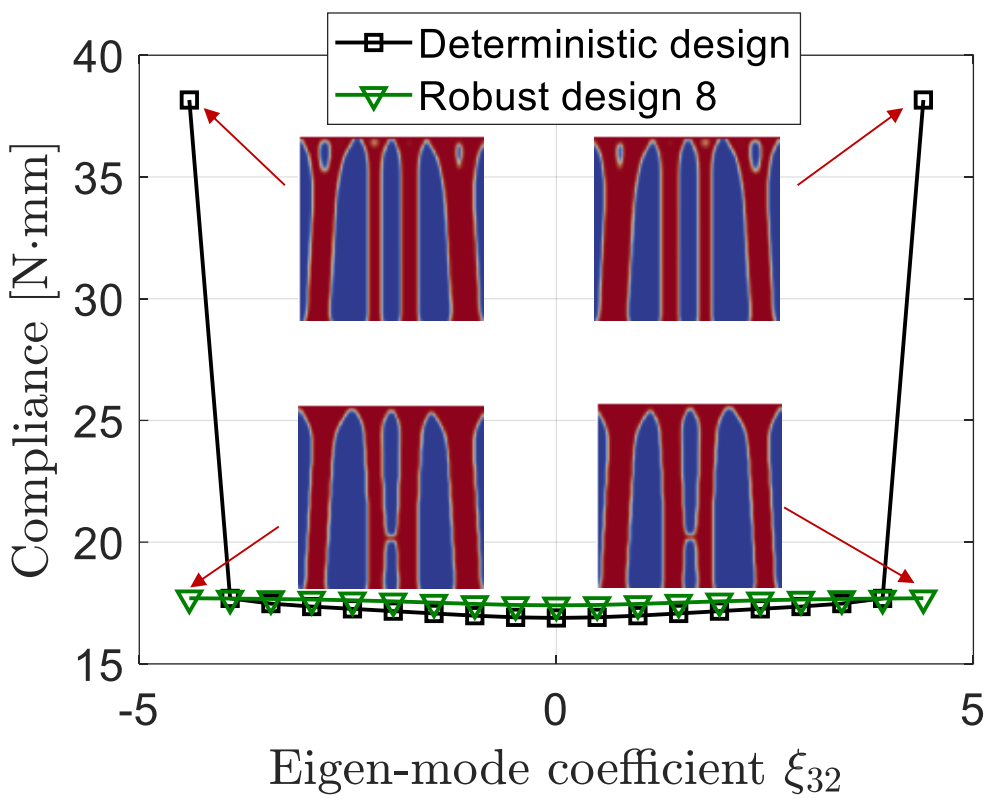

(b) $l_{cx}$ = 60 mm

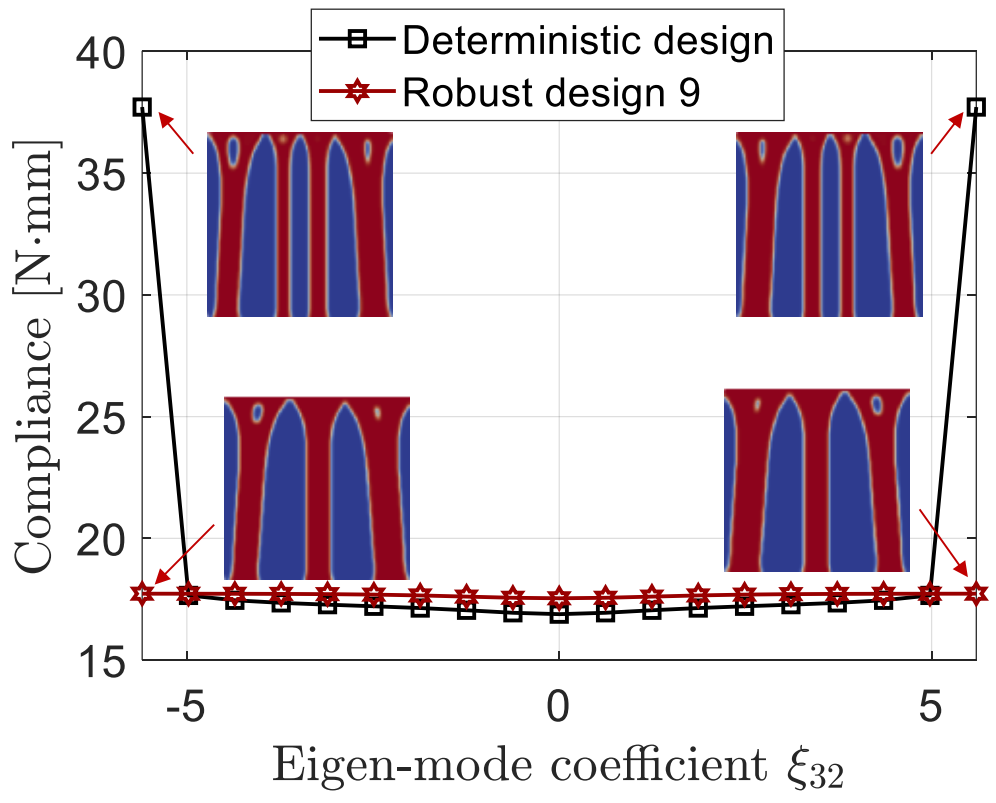

(c) $l_{cx}$ = 100 mm

Figure 16. Compliance versus eigenmode coefficient curves of deterministic and robust designs for 2nd eigenmode $\boldsymbol{\gamma}_2$ in the KL expansion of $\eta$-field.

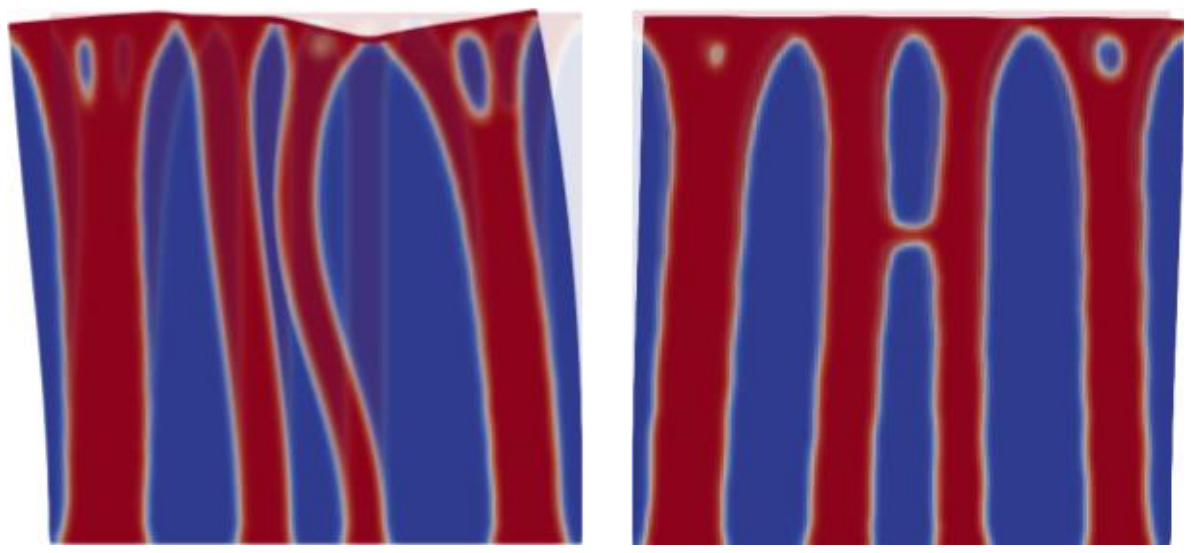

Deterministic design Robust design 7

(a) $l_{cx}$ = 20 mm, $\xi_{32}$ = 4.3

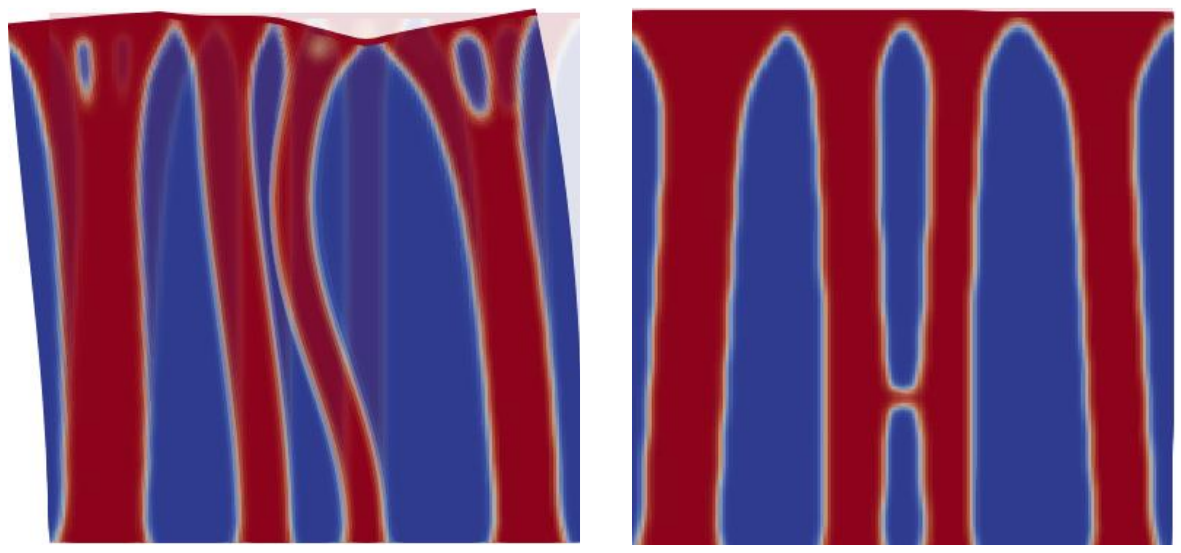

Deterministic design Robust design 8

(b) $l_{cx}$ = 60 mm, $\xi_{32}$ = 4.4

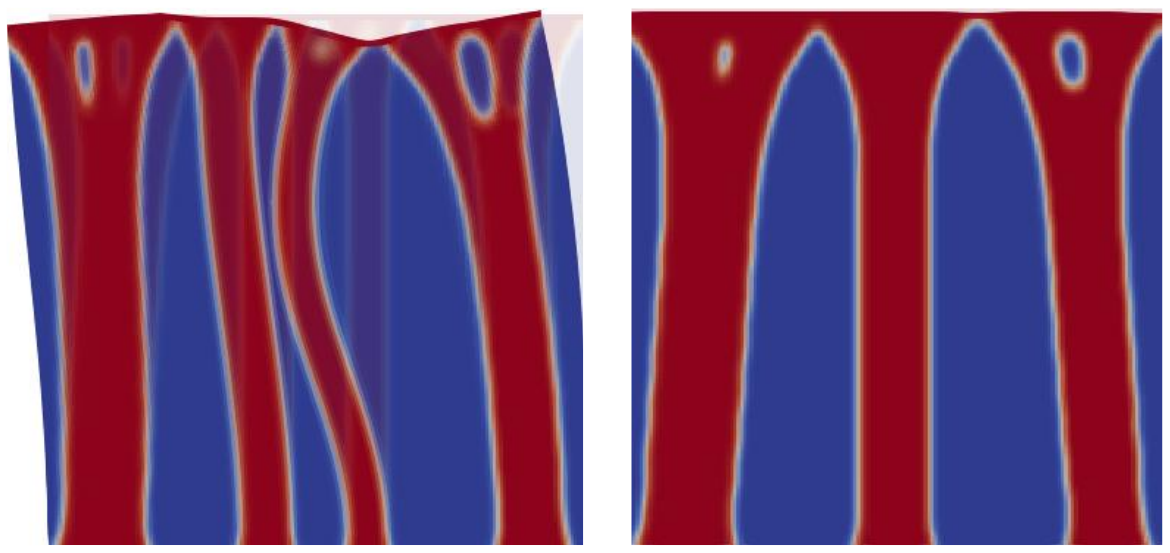

Deterministic design Robust design 9

(c) $l_{cx}$ = 100 mm, $\xi_{32}$ = 5.6

Figure 17. Deformed shapes of deterministic and robust designs for 2nd eigenmode $\boldsymbol{\gamma}_2$ in the KL expansion of the $\eta$-field with different values of eigenmode coefficient $\xi_{32}$.

### *7.1.4 Effect of combined loading, material, and geometric uncertainties*

The effect of combined loading, material, and geometric uncertainties is investigated in this case. Young's modulus is modeled by a lognormal random field with a correlation length of $l_{cx}$ = 20 mm. In addition, the stochastic loading with $\sigma_P$ = 3.162×10$^{-3}$ N and projection cutoff $\eta$ random field with a correlation length $l_{cx}$ = 20 mm are incorporated into optimization formulation, as well. After KL expansions, there are 20 random variables in this case. The optimized topology for this test case is shown in Figure 18.

To evaluate the performance of the robust design 2 and design 10 concerning material and geometric uncertainties, the curves of compliance versus eigenmode coefficient $\xi_{22}$ and $\xi_{32}$ related to the 2nd eigenmode $\boldsymbol{\gamma}_2$ are obtained by performing FEAs for the two robust designs with different Young's modulus ($E$) and projection cutoff ($\eta$) fields (Figure 19 and Figure 21), respectively. It can be observed that robust design 10 is less sensitive to the variations of Young's modulus and projection cutoff $\eta$ fields. Figure 20 and Figure 22 show the deformed shapes of the two designs for eigenmode coefficients of $\xi_{22}$ = $\xi_{32}$ = 10, and it can be seen that design 2 undergoes higher deformation than design 10, which confirms the results in Figure 19 and Figure 21. When the performance of the robust design 4 and design 10 towards loading and geometric uncertainties is evaluated, the curves of compliance versus loads and compliance versus eigenmode coefficient $\xi_{32}$ are plotted in Figure 23 and Figure 25, respectively. As expected, robust design 10 shows less sensitivity to the load and geometric variations than robust design 4. From Figure 24 and Figure 26, it can be observed that robust design 4 has larger deformation than robust design 10 when either load with a variation of 0.02 N in its horizontal component or projection cutoff $\eta$ field associated with eigenmode coefficient $\xi_{32} = 4$ is considered. For assessing the performance of robust design 7 and design 10 with respect to load and material uncertainties, the curves of compliance versus loads and compliance versus eigenmode coefficient $\xi_{22}$ are shown in Figure 27 and Figure 29, respectively. Compared to robust design 7, robust design 10 is less sensitive to load and material variations. For load with a 0.02 N variation in its horizontal component and Young's modulus field with eigenmode coefficient $\xi_{22} = 15$, the deformed shapes of the two designs are given in Figure 28 and Figure 30, respectively. It can be seen that robust design 7 has larger deformation than robust design 10 in both scenarios, which confirms the results in Figure 27 and Figure 29. These results demonstrate that optimization formulation incorporating multiple uncertain sources is necessary for obtaining designs that can perform robustly in an uncertain operating environment with multiple uncertainties.

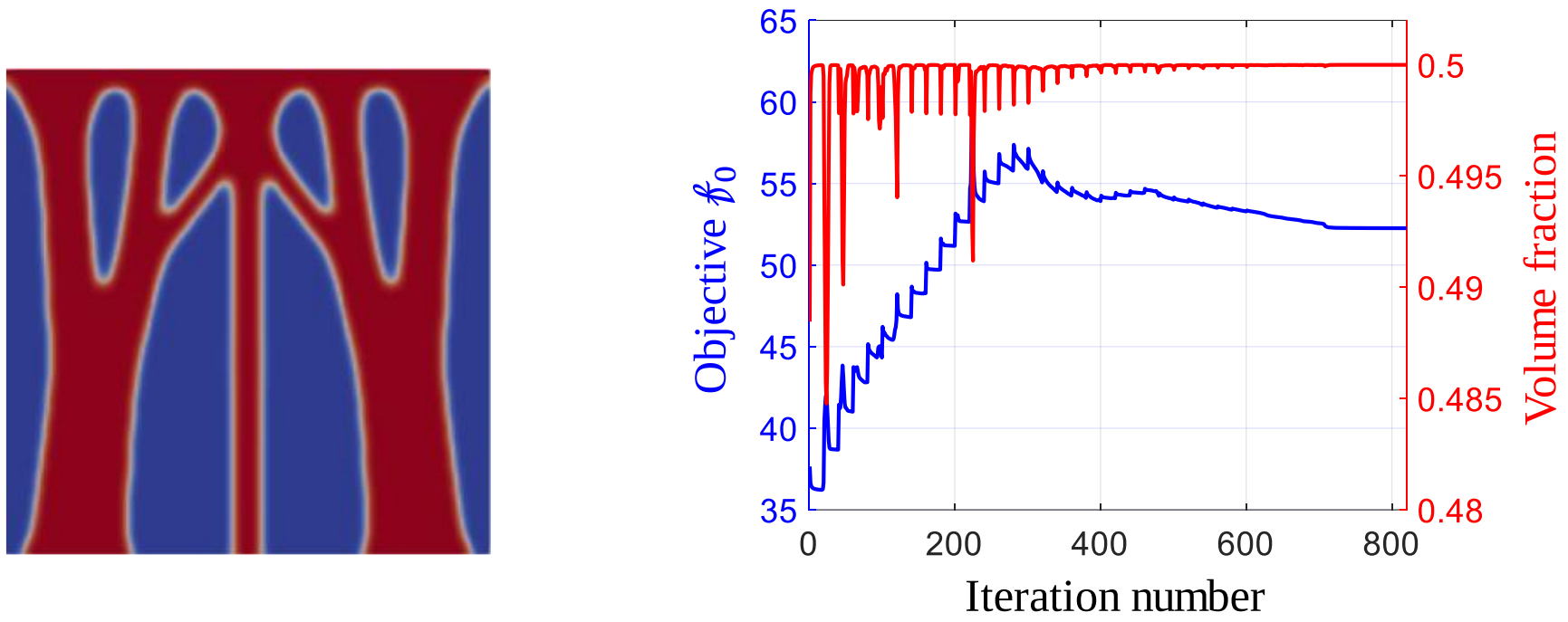

(a) Robust design 10 (b) Convergence history

Figure 18. Optimization results for the robust designs with loading, material, and geometric uncertainties.

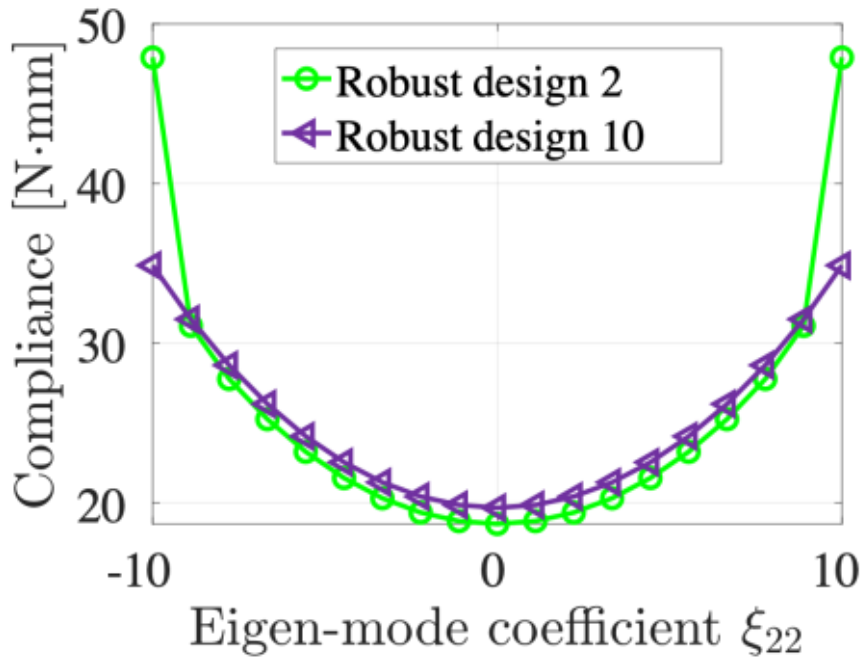

Figure 19. Compliance versus eigenmode coefficient curves of robust design 2 and robust design 10 for 2nd eigenmode $\boldsymbol{\gamma}_2$ in the KL expansion of $E$-field with $l_{cx}$ = 20 mm.

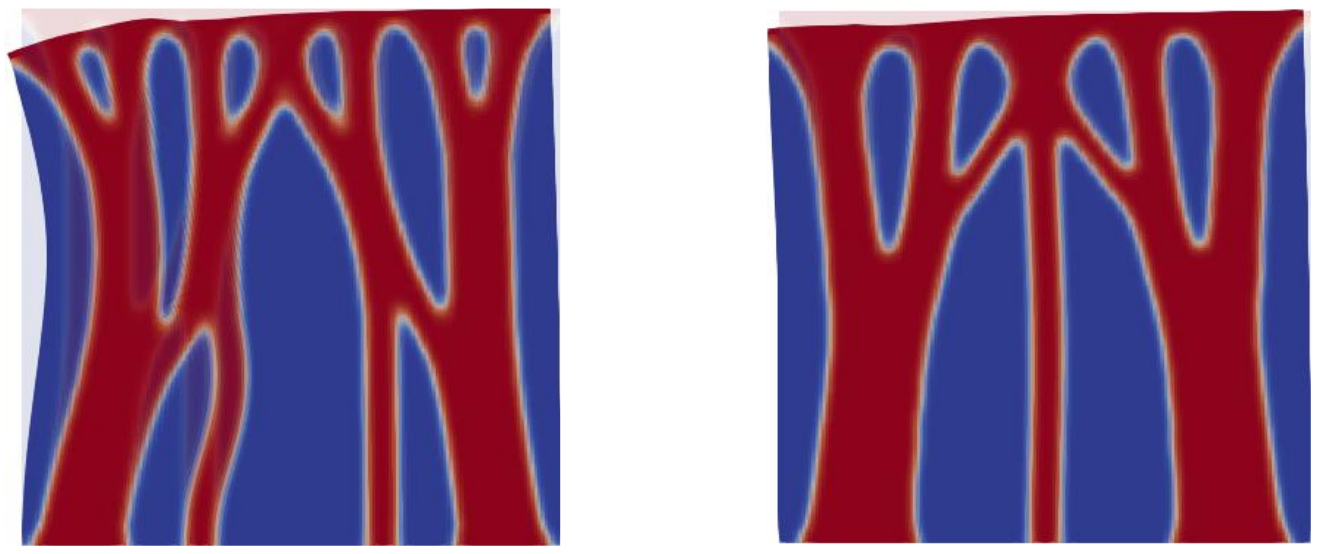

(a) Robust design 2 (b) Robust design 10

Figure 20. Deformed shapes of robust design 2 and robust design 10 for 2nd eigenmode $\boldsymbol{\gamma}_2$ in the KL expansion of $E$-field with $l_{cx}$ = 20 mm and $\xi_{22}$ = 10.

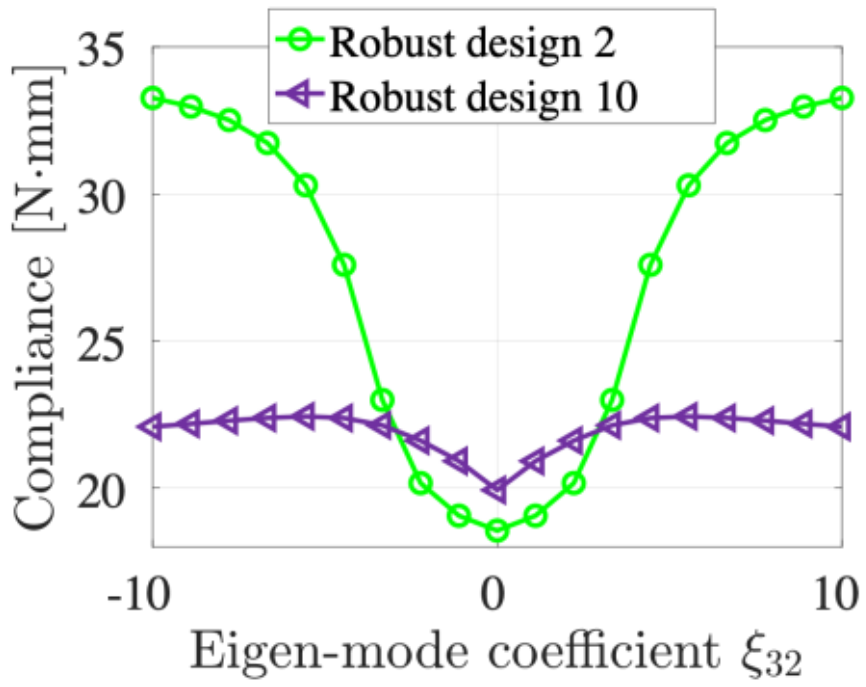

Figure 21. Compliance versus eigenmode coefficient curves of robust design 2 and robust design 10 for 2nd eigenmode $\boldsymbol{\gamma}_2$ in the KL expansion of $\eta$-field with $l_{cx}$ = 20 mm.

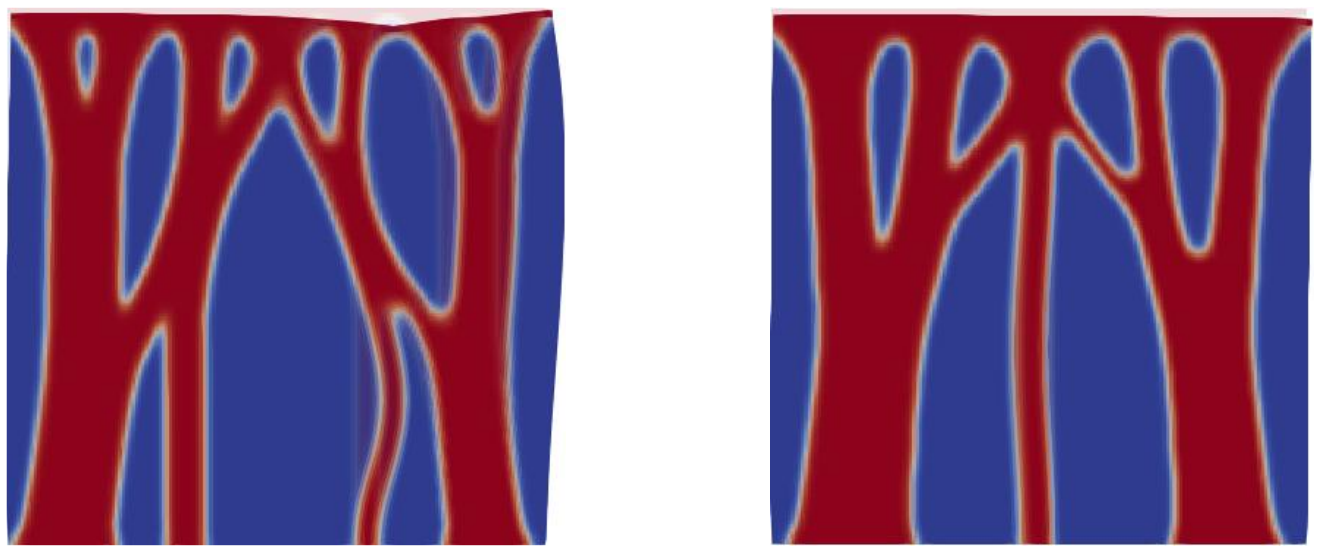

(a) Robust design 2 (b) Robust design 10

Figure 22. Deformed shapes of robust design 2 and robust design 10 for 2nd eigenmode $\boldsymbol{\gamma}_2$ in the KL expansion of $\eta$-field with $l_{cx}$ = 20 mm and $\xi_{32}$ = 10.

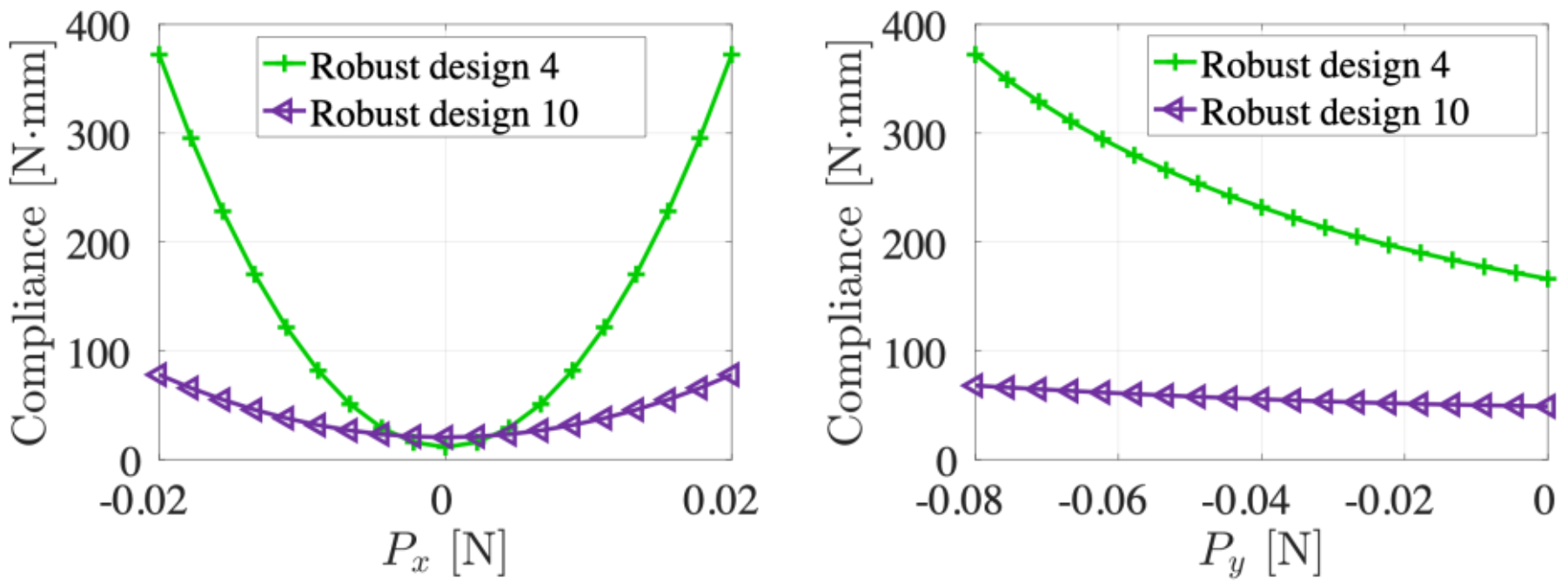

(a) $\boldsymbol{P}_0 = [P_x \quad -0.08]^T$ N (b) $\boldsymbol{P}_0 = [0.02 \quad P_y]^T$ N

Figure 23. Compliance versus load curves of robust design 4 and robust design 10 under different loads.

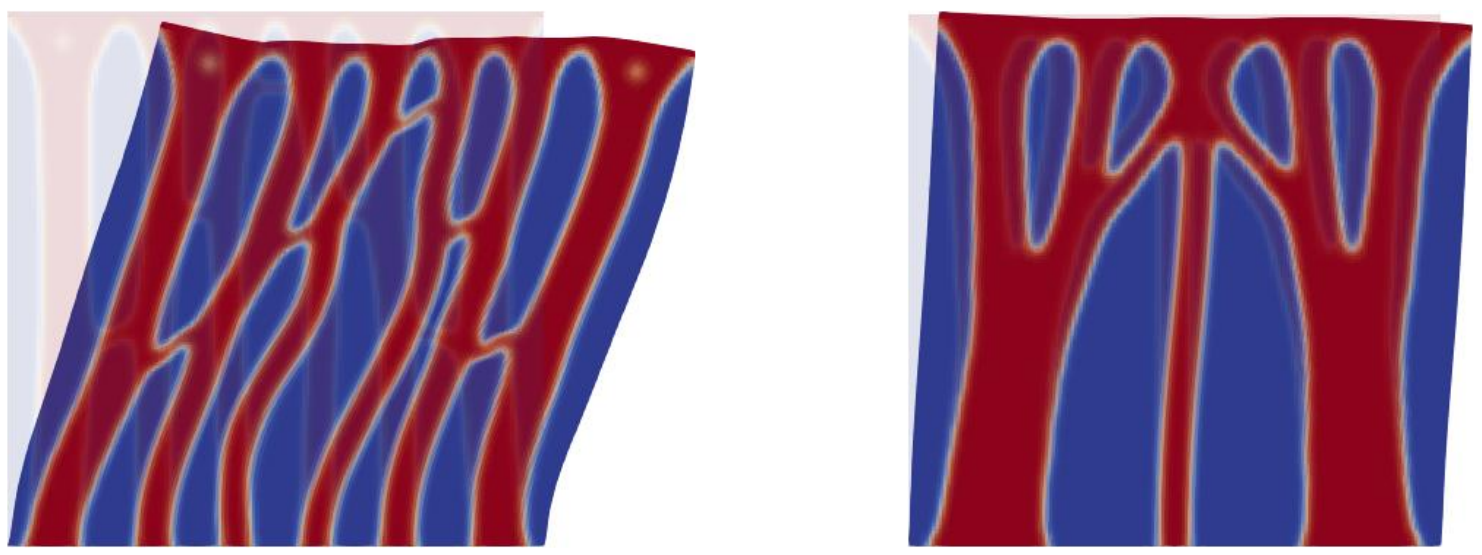

(a) Robust design 4 (b) Robust design 10

Figure 24. Deformed shapes of robust design 4 and robust design 10 for load $\boldsymbol{P}_0 = [0.02 \quad -0.08]^T$ N.

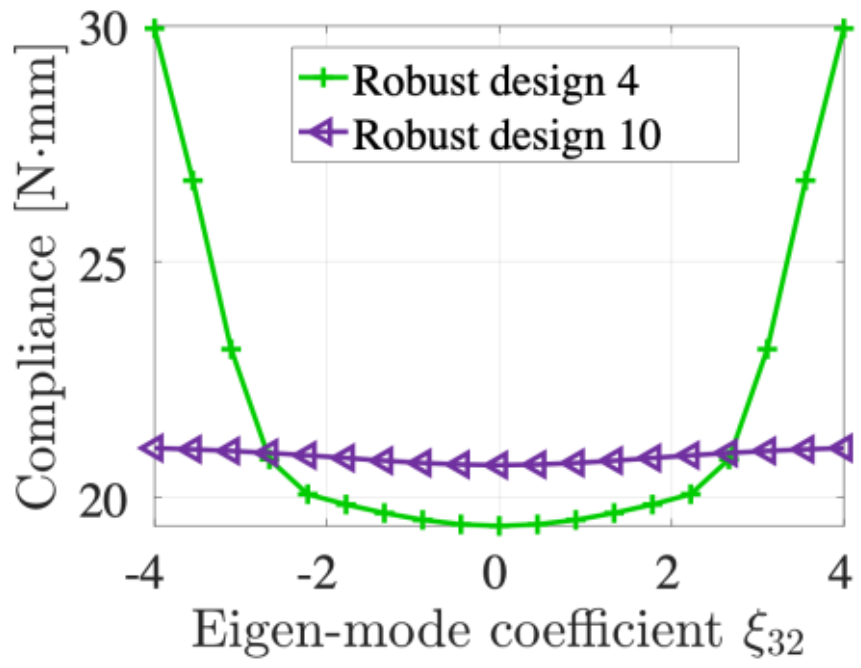

Figure 25. Compliance versus eigenmode coefficient curves of robust design 4 and robust design 10 for 2nd eigenmode $\boldsymbol{\gamma}_2$ in the KL expansion of $\eta$-field with $l_{cx}$ = 20 mm.

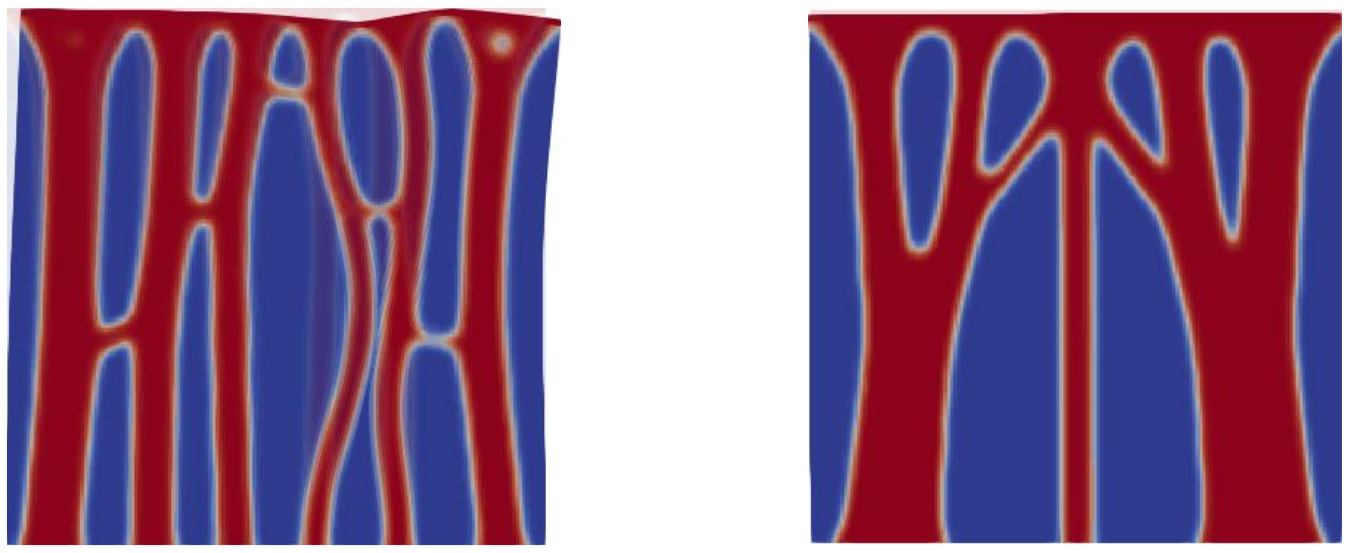

(a) Robust design 4 (b) Robust design 10

Figure 26. Deformed shapes of robust design 4 and robust design 10 for 2nd eigenmode $\boldsymbol{\gamma}_2$ in the KL expansion of $\eta$-field with $l_{cx}$ = 20 mm and $\xi_{32}$ = 4.

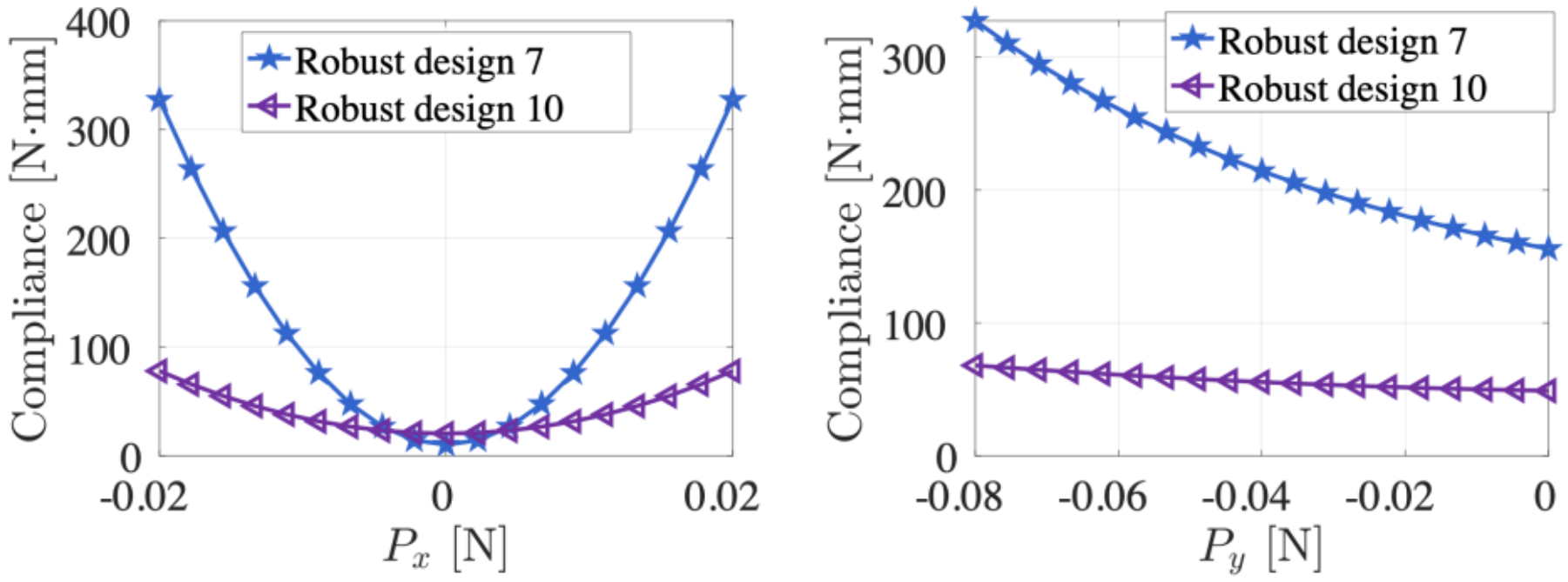

(a) $\boldsymbol{P}_0 = [P_x \quad -0.08]^T$ N (b) $\boldsymbol{P}_0 = [0.02 \quad P_y]^T$ N

Figure 27. Compliance versus load curves of robust design 7 and robust design 10 under different loads.

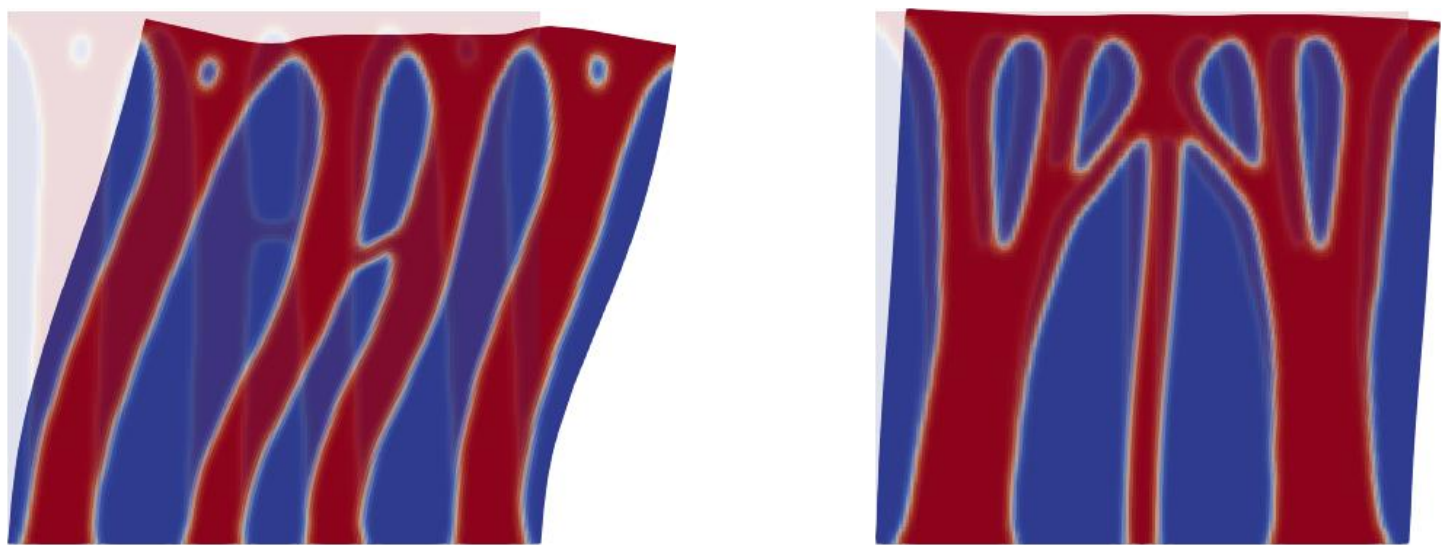

(a) Robust design 7 (b) Robust design 10

Figure 28. Deformed shapes of robust design 7 and robust design 10 for load vector $\boldsymbol{P}_0 = [0.02 \quad -0.08]^T$ N.

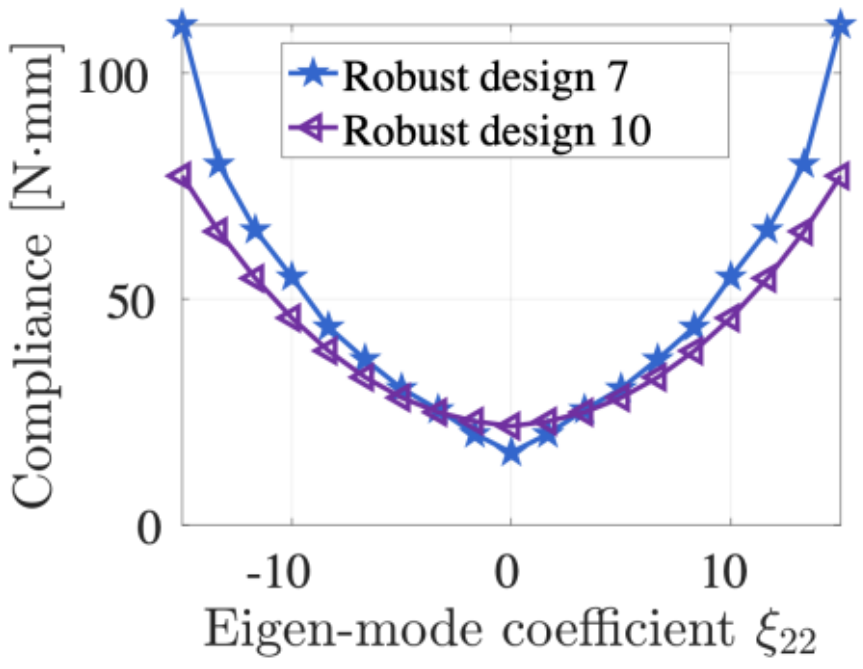

Figure 29. Compliance versus eigenmode coefficient curves of robust design 7 and robust design 10 for 2nd eigenmode $\boldsymbol{\gamma}_2$ in the KL expansion of $E$-field with $l_{cx}$ = 20 mm.

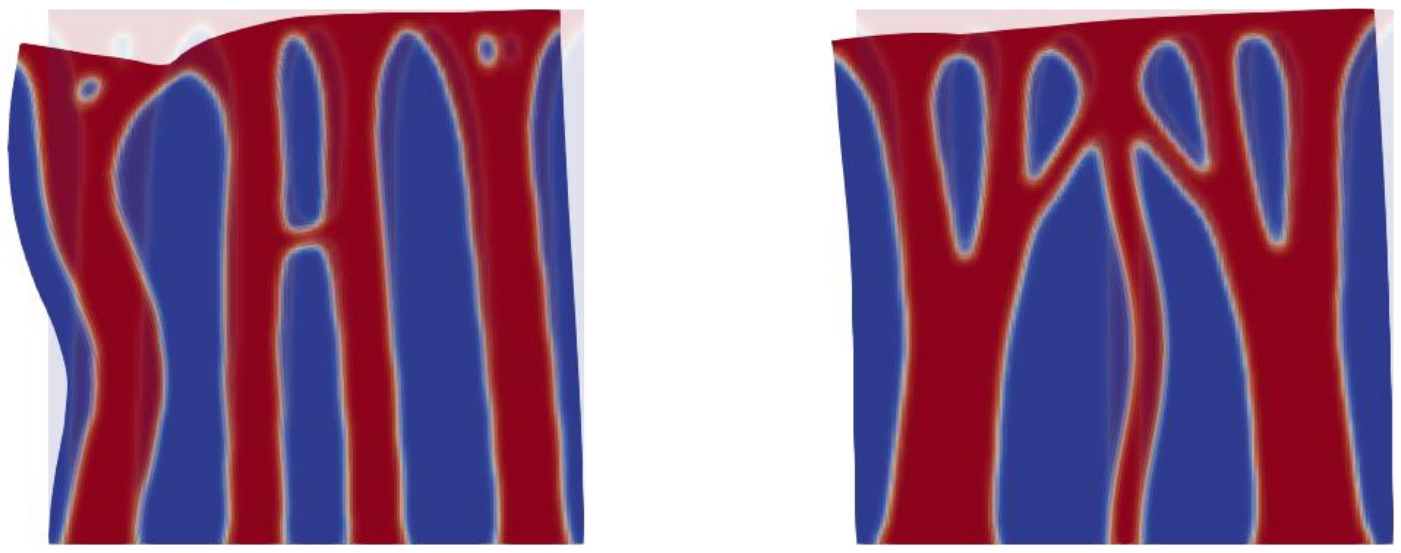

(a) Robust design 7 (b) Robust design 10

Figure 30. Deformed shapes of robust design 7 and robust design 10 for 2nd eigenmode $\boldsymbol{\gamma}_2$ in the KL expansion of $E$-field with $l_{cx}$ = 20 mm and $\xi_{22}$ = 15.

## 7.2 Clamped beam

In the second example, a clamped beam problem sketched in Figure 31(a) is studied, where 19200 four-node quadrilateral elements are used in the 240×80 FE mesh. To preclude localized deformations, the elements in the solid red area, where the loads are applied, are solid and are not designed. In addition, symmetry is enforced during optimization such that the left half of the domain mirrors the right half. Poisson's ratio is 0.4, the density filter radius is 7.5 mm, and the volume fraction is restricted to 0.2. In the deterministic scenario, Young's modulus is set to $E_0$ = 10 MPa, and the deterministic load is $\boldsymbol{P}_0 = [0 \quad -90]^T$ N. Also, $E_{L0}$ = 10 MPa and $\nu_L$ = 0.4 for linear energy interpolation (Eq. (13)). The stochastic scenarios with different load and/or material uncertainties are given in Table 2, wherein Young's modulus random field with correlation lengths $l_{cy} = \infty$ and $l_{cx}$ from the set {30, 45, 90} leads to 6, 4, and 3 random variables after KL expansion, respectively.

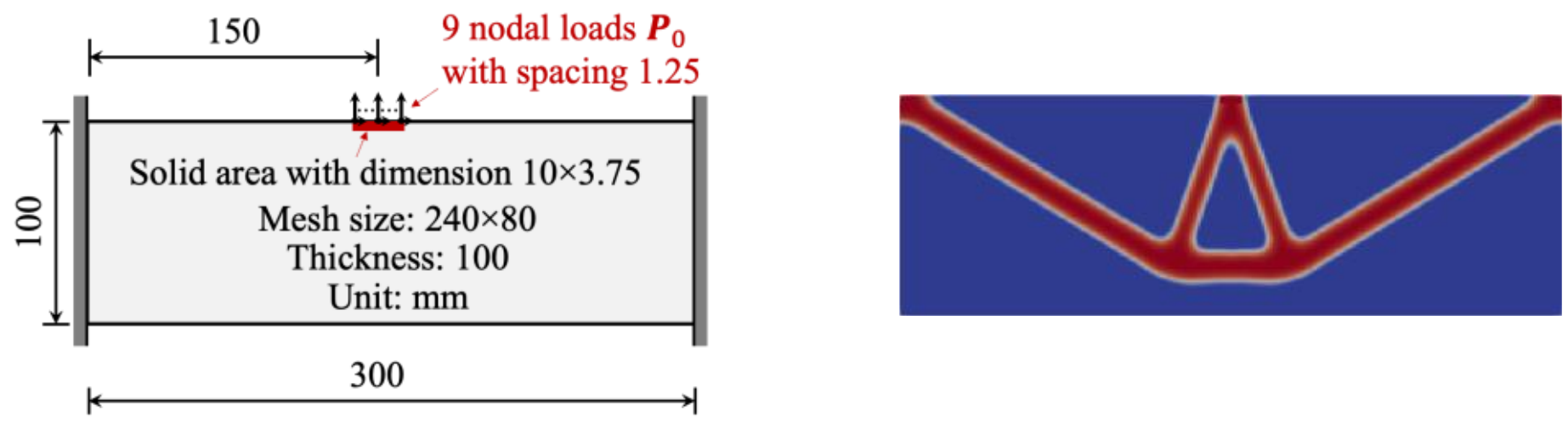

(a) Clamped beam (b) Deterministic design

Figure 31. Clamped beam and deterministic design.

### *7.2.1 Deterministic topology optimization*

This test case is used to demonstrate the efficacy of the incorporated adaptive $c$-update scheme in the energy interpolation scheme (Eq. (11)). The optimized design using the deterministic formulation (Eq. (15)) is shown in Figure 31(b). In this case, obtaining even the deterministic design is challenging and is only feasible with the considered adaptive scheme. To arrive at the optimized design in Figure 31(b), the topology needs to evolve from an unstable *linear-feature* design (Figure 32c) to the desired stable *nonlinear-feature* design (Figure 32k), see Figure 32. The *linear-feature* design is optimal at low $p$-values, where the material penalization is low and intermediate-density elements still have significant stiffness. However, as the $p$-value increases during continuation, material penalization increases, and the intermediate density elements lose stiffness, leading to the infeasibility of the *linear-feature* design. Thus, for a successful optimization process, the topology has to evolve from a *linear-feature* design (Figure 32c) to a *nonlinear-feature* design (Figure 32k). During this transition, significant mesh distortions occur, and to address this issue, in this study, the linear energy is increased by adaptively updating the cutoff parameter $c$ in Eq. (12) (Figure 33). The FE analysis will fail without such an adaptive scheme, and the optimization cannot proceed. Note that the $c$-updates are not needed after the nonlinear features emerge, i.e., Figure 32h to Figure 32k. This overall transition is an extreme example of how a stable topology can emerge even when starting with a low $p$-value in the continuation scheme. Such a transition is advantageous for the following robust design study since, when considering uncertainties, different topologies may emerge during this transition.

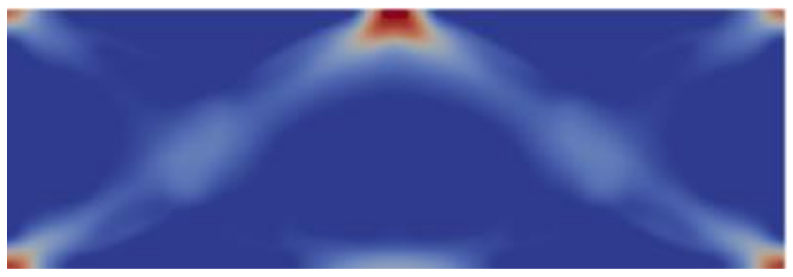

(a) Iteration number = 10, $p$ = 1.0

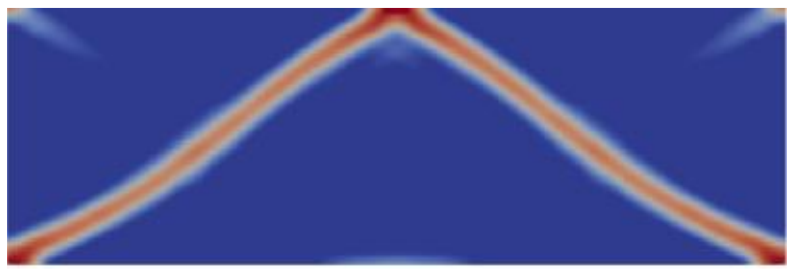

(b) Iteration number = 50, $p$ = 1.2

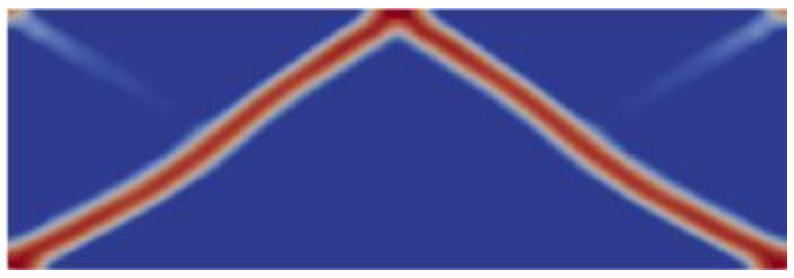

(c) Iteration number = 100, $p$ = 1.4

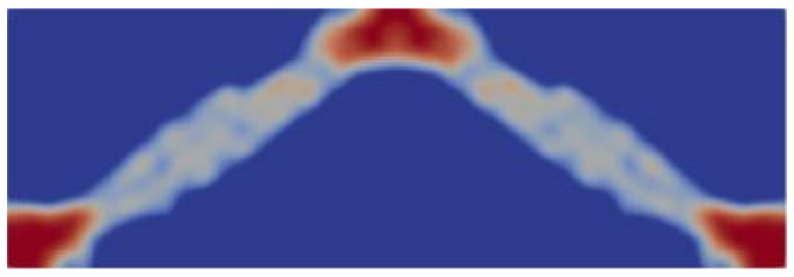

(d) Iteration number = 110, $p$ = 1.5

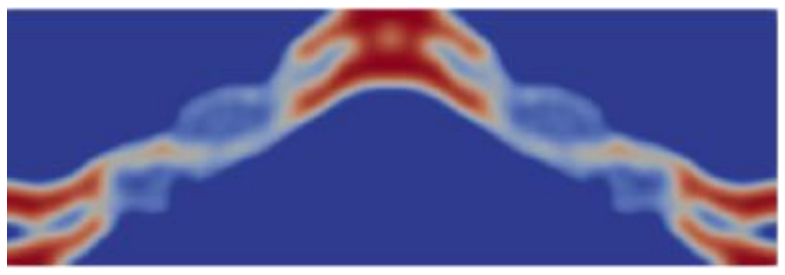

(e) Iteration number = 150, $p$ = 1.7

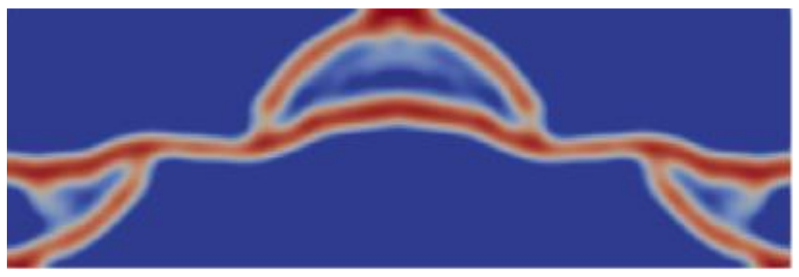

(f) Iteration number = 190, $p$ = 1.9

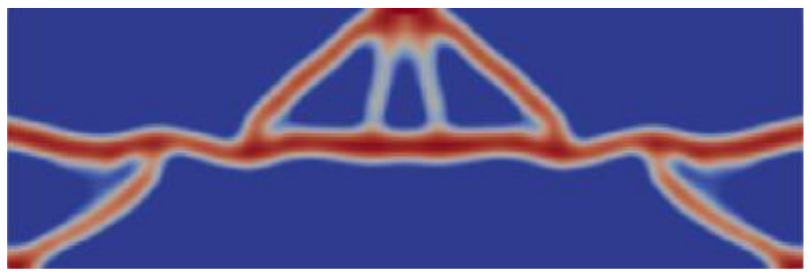

(g) Iteration number = 300, $p$ = 2.4

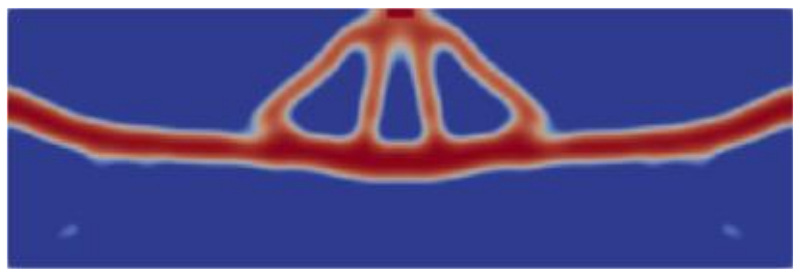

(h) Iteration number = 500, $p$ = 3.4

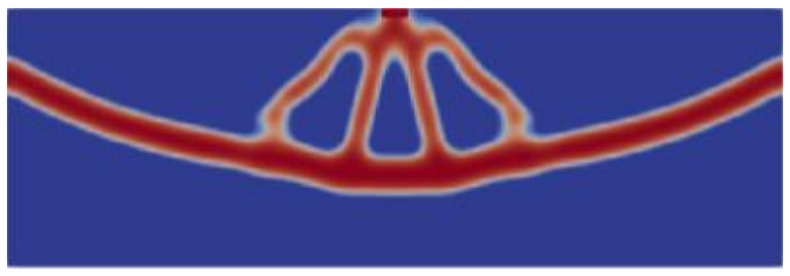

(i) Iteration number = 600, $p$ = 3.9

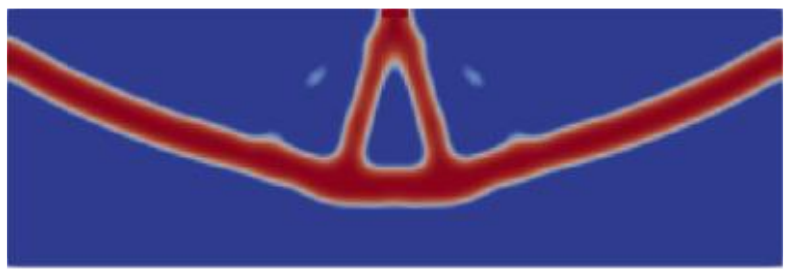

(j) Iteration number = 700, $p$ = 4.0

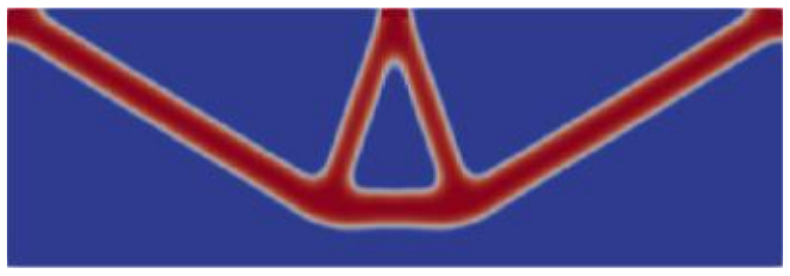

(k) Iteration number = 820, $p$ = 4.0

Figure 32. Evolution of topology for the clamped beam in deterministic optimization.

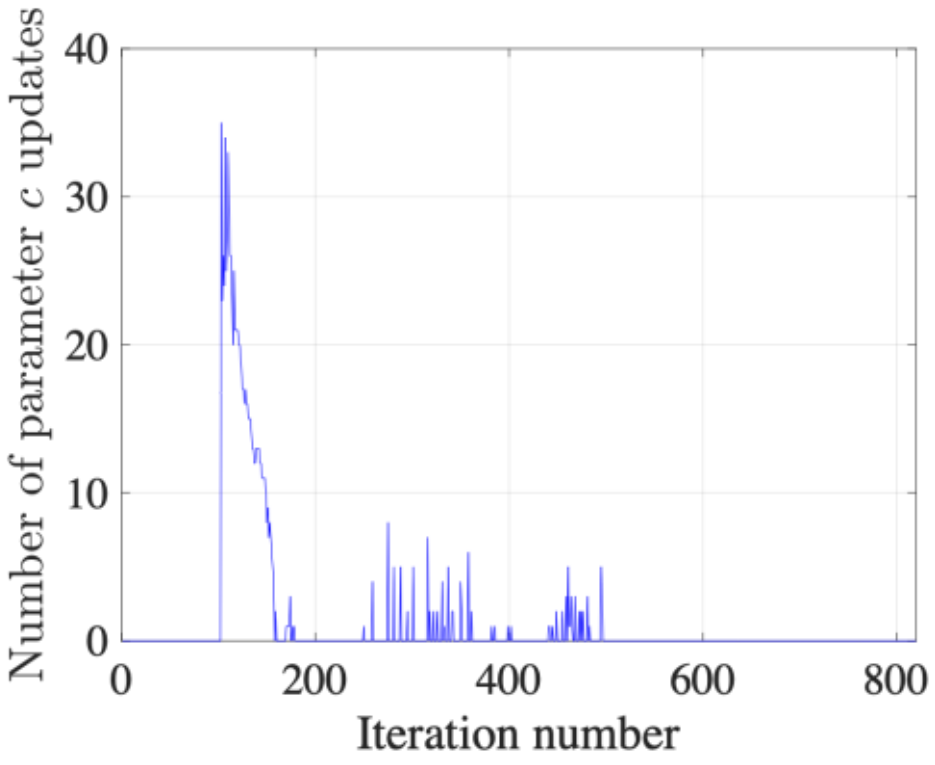

Figure 33. History of cutoff parameter $c$ updates for the clamped beam in deterministic optimization.

### *7.2.2 Effect of load uncertainty*

Incorporating only load uncertainty with different variances $\sigma_P$ ={6, 8, 10} N, with weighting parameter $\alpha$ = 1, the corresponding optimized designs are demonstrated in Figure 34. For this test case, an additional 600 iterations are carried out at the end of the continuation. To compare the performance of the optimized designs (in terms of end compliance) under the load variations, Figure 35(a) and Figure 35(b) plot the curves of compliance versus horizontal/vertical load, where it can be seen that the ranking of the sensitivity of structural performance w.r.t. the load variations is: robust design 3 < robust design 2 < robust design 1 < deterministic design. As an illustration, Figure 36 shows the deformed shapes of these designs under the load vector $\boldsymbol{P}_0 = [5 \quad -90]^T$ which can be seen as one realization of the prescribed load random vector. These deformed shapes again confirm the robustness trends of the optimized topologies as the topologies with higher robustness undergo smaller deformations.

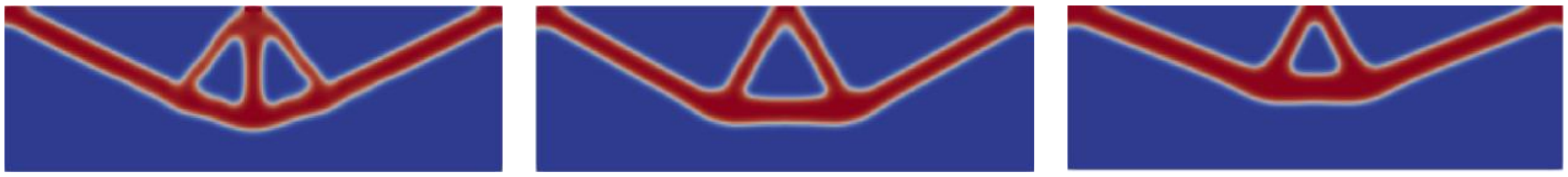

(a) Robust design 1 ($\sigma_P$ = 6 N) (b) Robust design 2 ($\sigma_P$ = 8 N) (c) Robust design 3 ($\sigma_P$ = 10 N)

Figure 34. Robust designs with different standard deviations $\sigma_P$ of the stochastic loads.

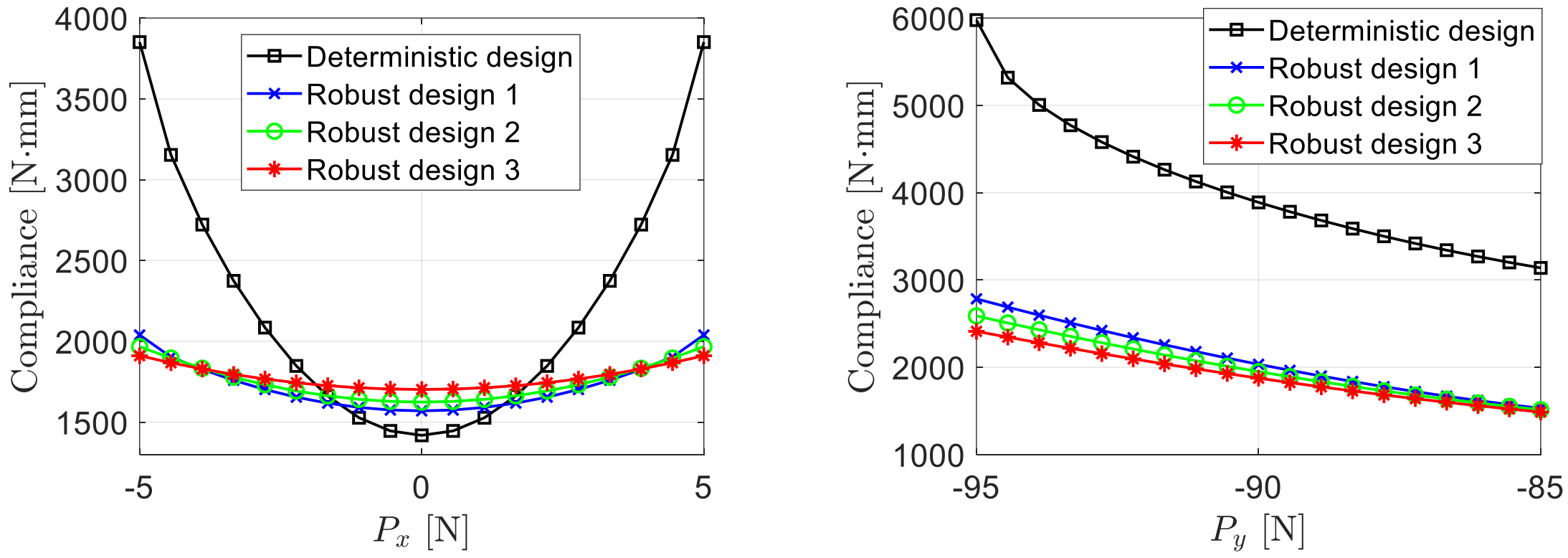

(a) $\boldsymbol{P}_0 = [P_x \quad -90]^T$ N (b) $\boldsymbol{P}_0 = [5 \quad P_y]^T$ N

Figure 35. Compliance versus load curves of deterministic and robust designs.

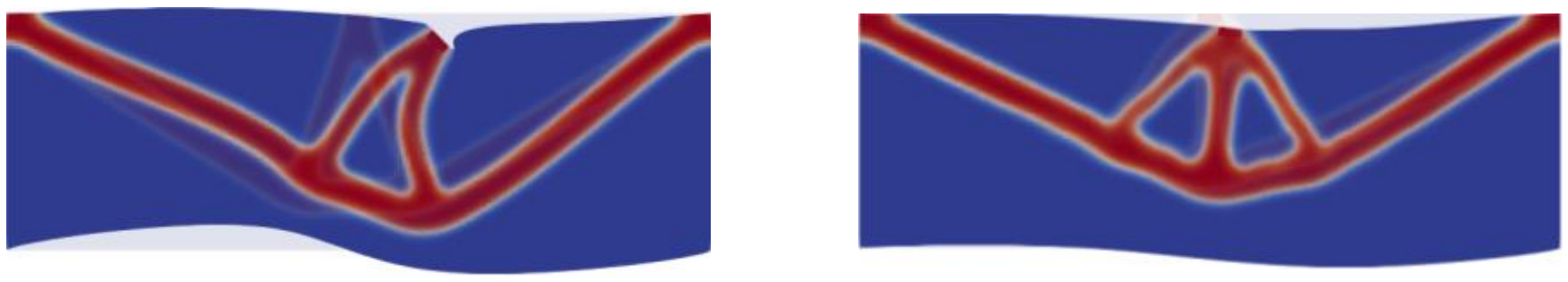

(a) Deterministic design (b) Robust design 1

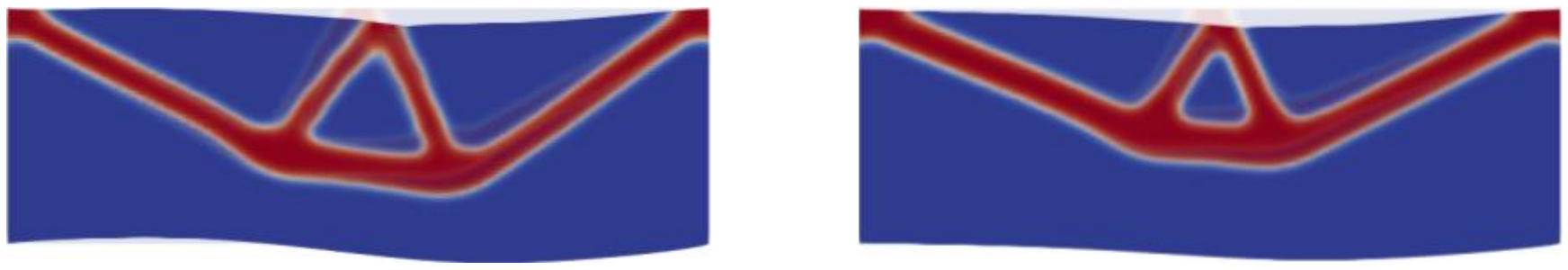

(c) Robust design 2 (d) Robust design 3

Figure 36. Deformed shapes of deterministic and robust designs for load $\boldsymbol{P}_0 = [5 \quad -90]^T$ N.

### 7.2.3 *Effect of material uncertainty*

In comparison, incorporating material uncertainty alone with different correlation lengths, i.e., $l_{cx}$ = {30, 45, 90} mm, with weighting parameter $\alpha$ = 1.0, the corresponding optimized designs are shown in Figure 37. The first two eigenmodes resulting from the KL expansion of Young's modulus random field with different correlation lengths are plotted in Figure 38, from which it is clear that the 1st eigenmode $\boldsymbol{\gamma}_1$ is symmetric while the 2nd eigenmode $\boldsymbol{\gamma}_2$ is antisymmetric.

To study the robustness of the deterministic and robust designs under material uncertainty, a series of Young's modulus fields are generated by keeping only the term related to the specified eigenmode (Figure 38) in Eq. (21) which is then transformed by Eq. (17). The curves of compliance versus the coefficient $\xi_{2i}$ for the 2nd eigenmode (Figure 39) are obtained by performing FEAs for different Young's modulus fields. As shown in the figure, with antisymmetric realizations (Figure 39), design 4, design 5, and design 6 optimized for the material uncertainty show higher robustness, while the deterministic design has the worst performance. As an illustration, Figure 40 plots the deformed shapes of different designs with the antisymmetric eigenmode $\boldsymbol{\gamma}_2$ with different values of correlation length $l_{cx}$ and coefficient $\xi_{22}$.

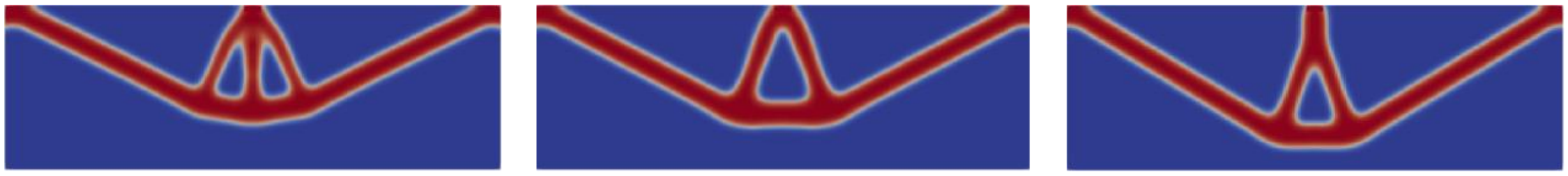

(a) Robust design 4 ($l_{cx}$ = 30 mm) (b) Robust design 5 ($l_{cx}$ = 45 mm) (c) Robust design 6 ($l_{cx}$ = 90 mm)

Figure 37. Robust designs with different values of $l_{cx}$ in $E$-field.

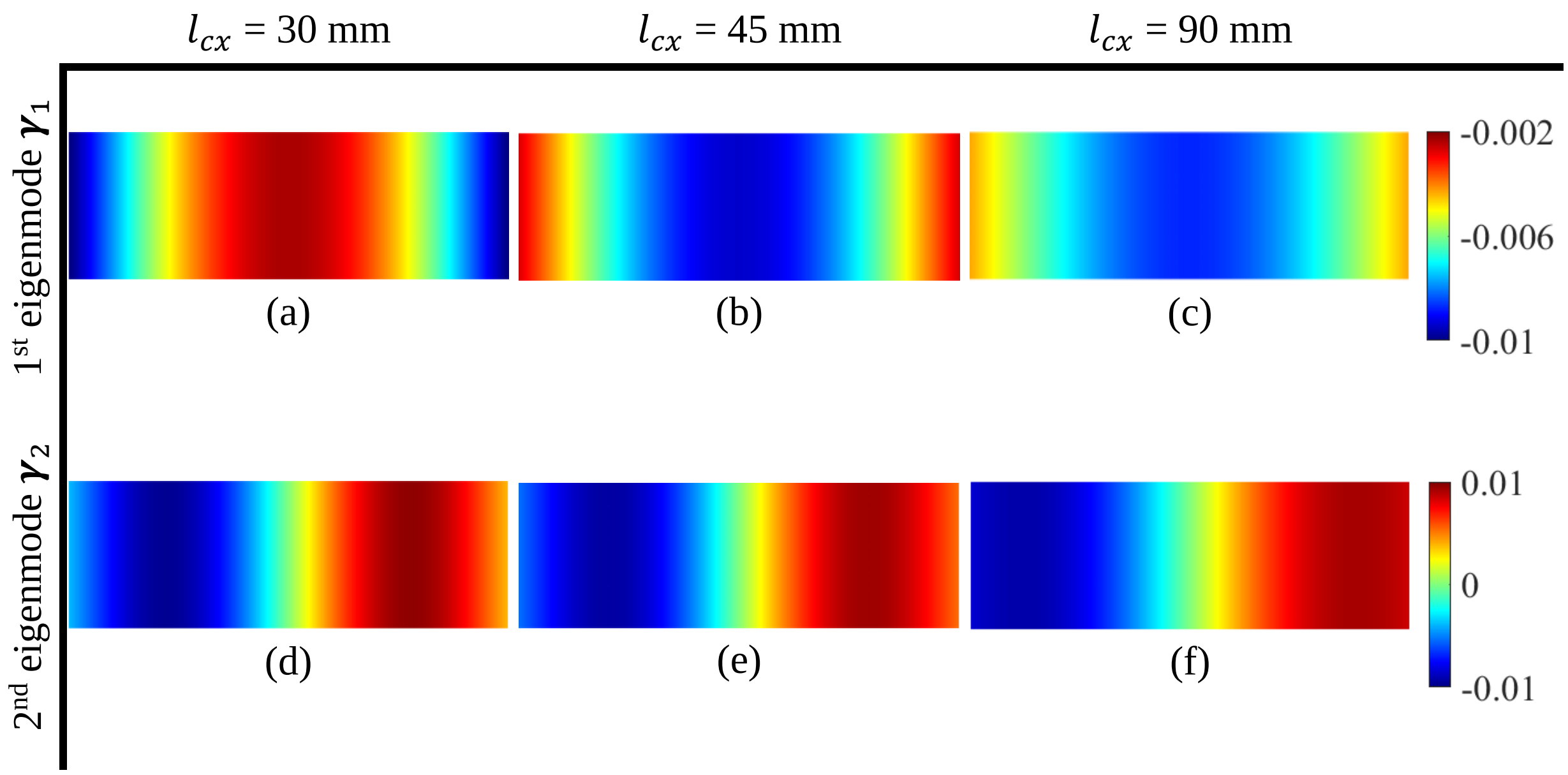

Figure 38. First two eigenmodes in the KL expansion of $E$-field.

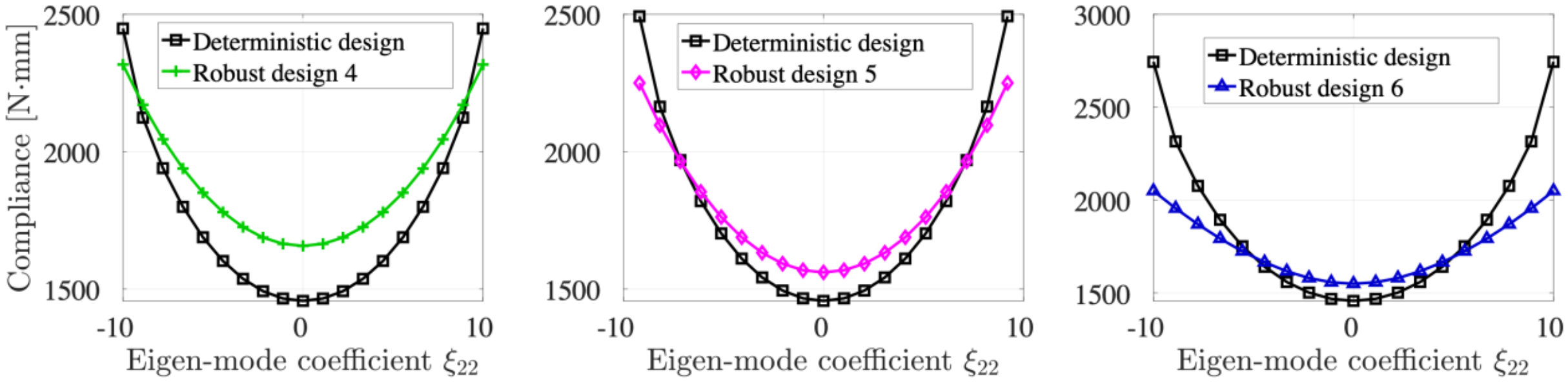

(a) $l_{cx}$ = 30 mm (b) $l_{cx}$ = 45 mm (c) $l_{cx}$ = 90 mm

Figure 39. Compliance versus eigenmode coefficient curves of deterministic and robust designs for 2nd eigenmode $\boldsymbol{\gamma}_2$ in the KL expansion of $E$-field.

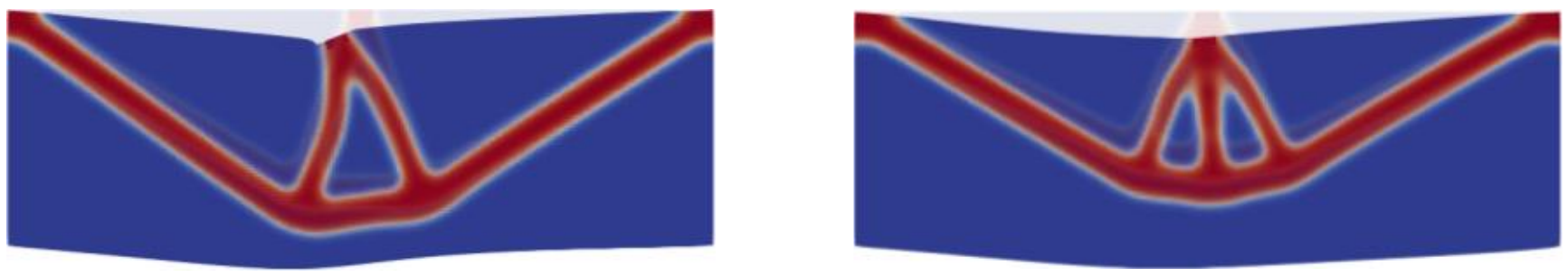

Deterministic design Robust design 4

(a) $l_{cx}$ = 30 mm, $\xi_{22}$ = 10

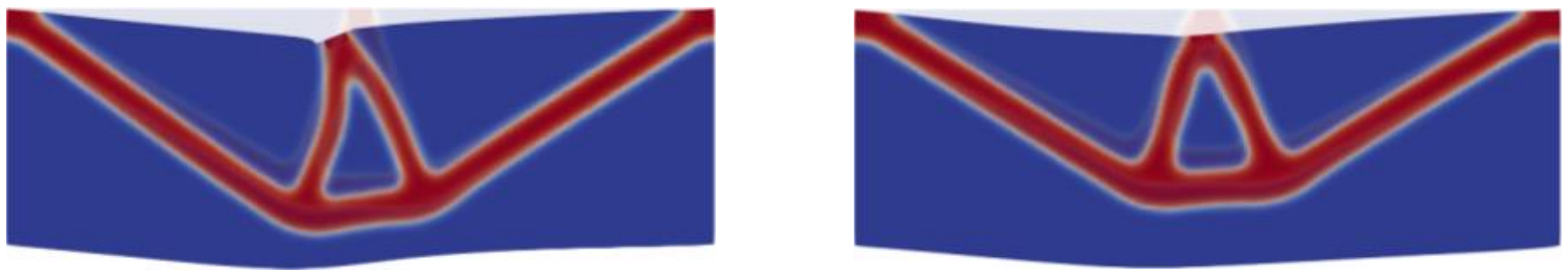

Deterministic design Robust design 5

(b) $l_{cx}$ = 45 mm, $\xi_{22}$ = 9.2

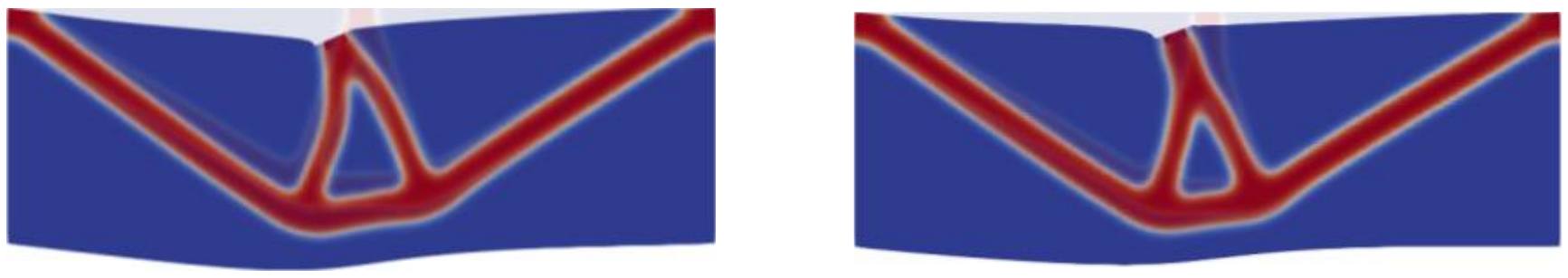

Deterministic design Robust design 6

(c) $l_{cx}$ = 90 mm, $\xi_{22}$ = 6

Figure 40. Deformed shapes of deterministic and robust designs for 2nd eigenmode $\boldsymbol{\gamma}_2$ in the KL expansion of $E$-field.

### 7.2.4 *Effect of combined load and material uncertainties*

Combining load uncertainty with $\sigma_P$ = 6 N and material uncertainty with $l_{cx}$ = 90 mm, the robust topology optimization is conducted for different values of the weighting parameter, i.e., $\alpha \in \{0, 0.5, 1, 2\}$. Compared to design 1, where only load uncertainty is incorporated, designs 7 to 10 are expected to exhibit less sensitivity w.r.t. the material variations. To demonstrate this, Figure 42 plots the curves of compliance versus coefficient $\xi_{22}$ in which only the antisymmetric eigenmodes in Figure 38(f) are kept in the KL expansion. These results confirm that the variation of the performance of design 1 due to the material uncertainties is indeed higher than that of all designs 7 to 9. Moreover, it also demonstrates that increasing the weight of the variation in the objective function ($\alpha$) helps to achieve higher robustness. As an illustration, Figure 43 shows the deformed shapes of the four robust designs for the antisymmetric Young's modulus field with coefficient $\xi_{22}$ = 10.

On the other hand, compared to design 6 where only material uncertainty is incorporated, designs 7 to 10 are expected to be less sensitive w.r.t. the load variations. This can be again confirmed in Figure 44 the higher flatness of the compliance curves versus horizontal/vertical load implies higher robustness against the load uncertainty. Finally, Figure 45 shows the deformations of designs 6 to 10 under the load vector $\boldsymbol{P}_0 = [1 \quad -90]^T$ where the deformation of design 6 is the largest among all the designs.

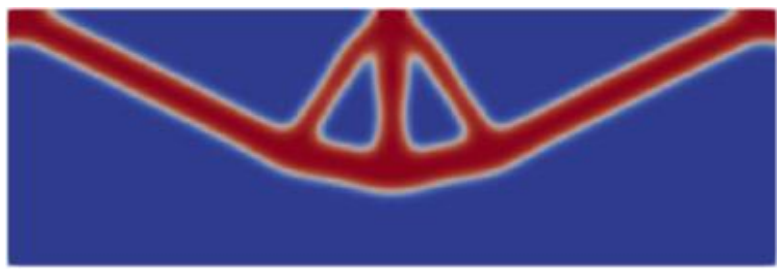

(a) Robust design 7 ($\alpha$ = 0)

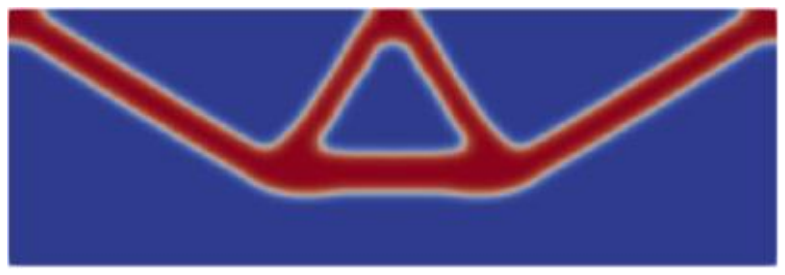

(b) Robust design 8 ($\alpha$ = 0.5)

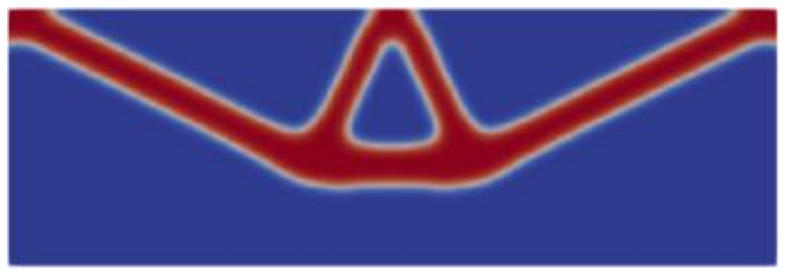

(c) Robust design 9 ($\alpha$ = 1)

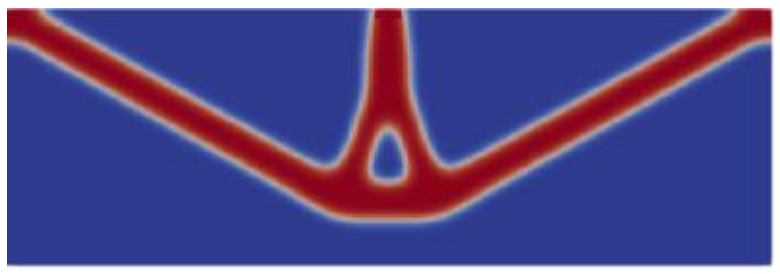

(d) Robust design 10 ($\alpha$ = 2)

Figure 41. Robust designs with different values of parameter $\alpha$ in the objective function $\not{f}_0$.

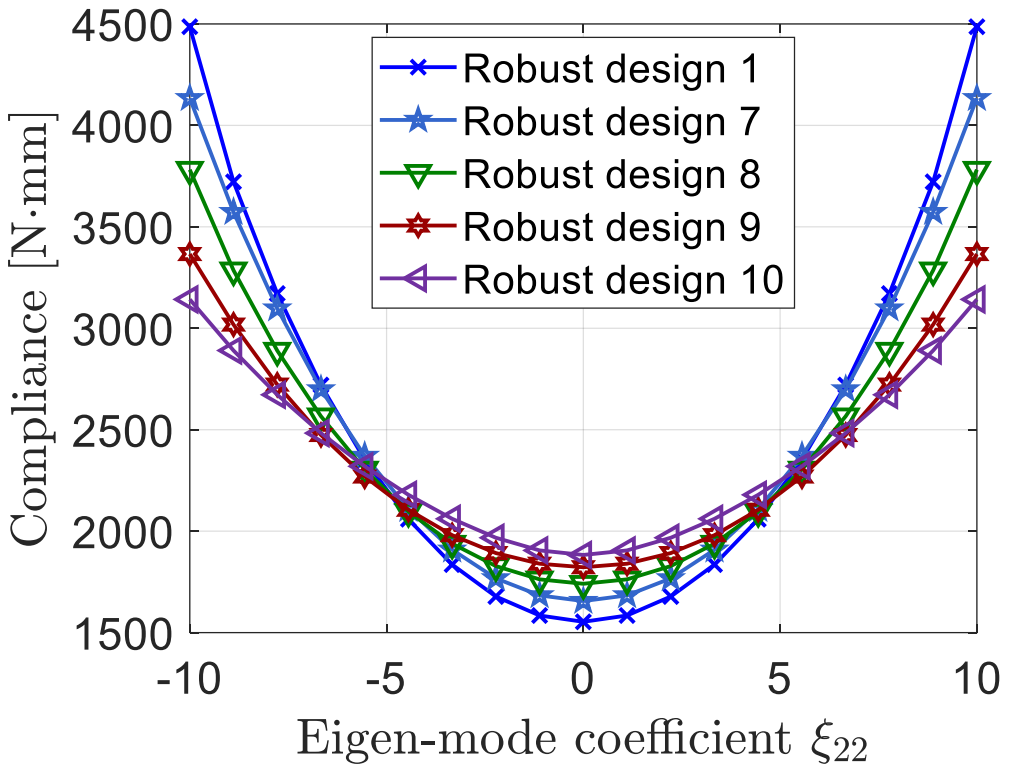

Figure 42. Compliance versus eigenmode coefficient curves of robust designs 1, 7, 8, 9, and 10 for 2nd eigenmode $\boldsymbol{\gamma}_2$ in the KL expansion of $E$-field with $l_{cx}$ = 90 mm.

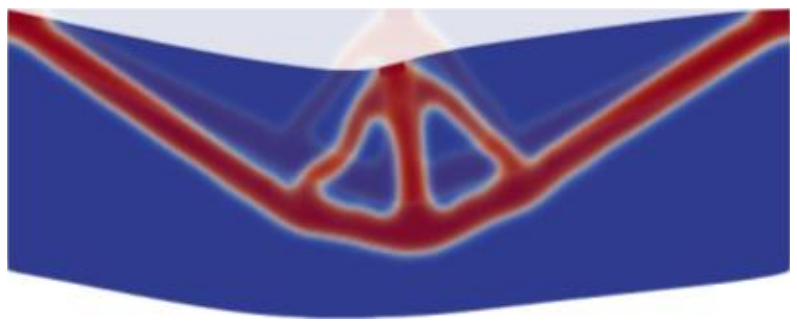

(a) Robust design 1

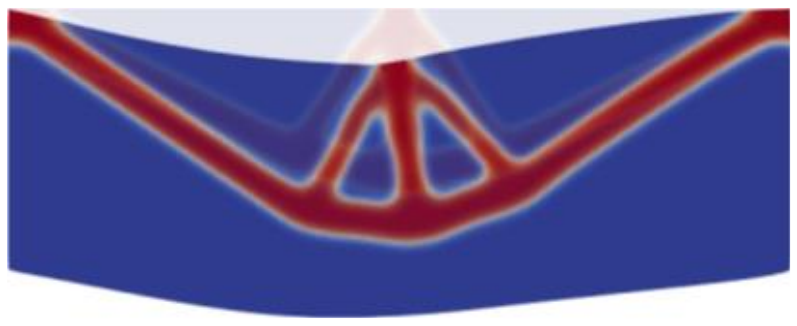

(b) Robust design 7

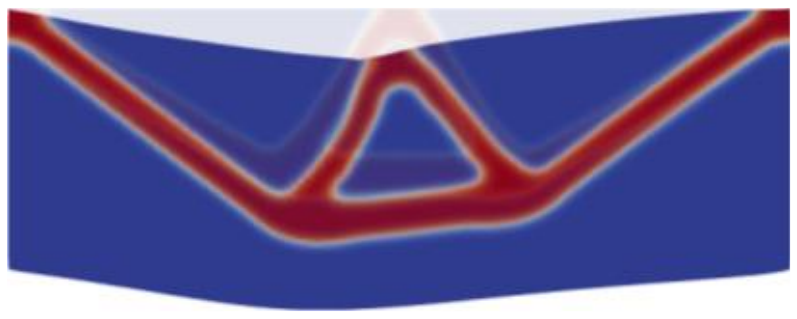

(c) Robust design 8

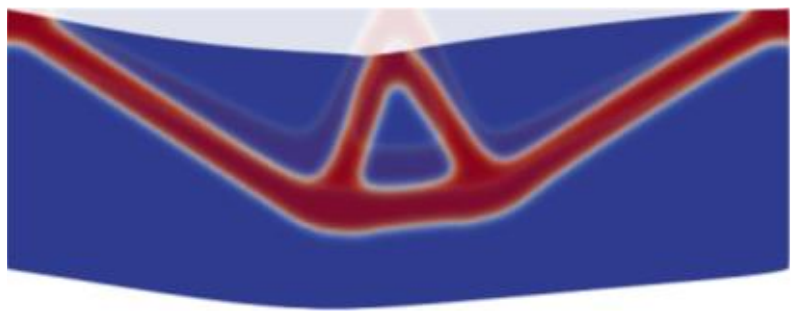

(d) Robust design 9

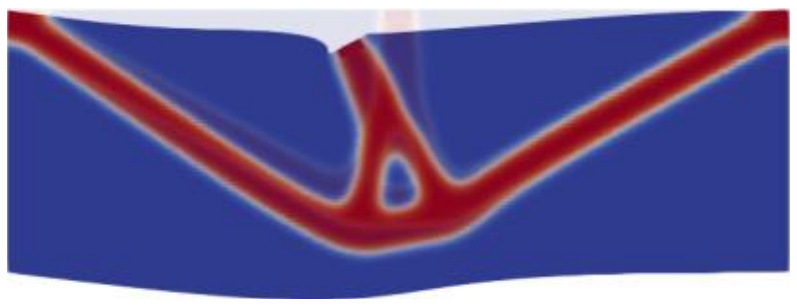

(e) Robust design 10

Figure 43. Deformed shapes of robust designs 1, 7, 8, 9, and 10 for 2nd eigenmode $\boldsymbol{\gamma}_2$ in the KL expansion of $E$-field with $l_{cx}$ = 90 mm and $\xi_{22}$ = 10.

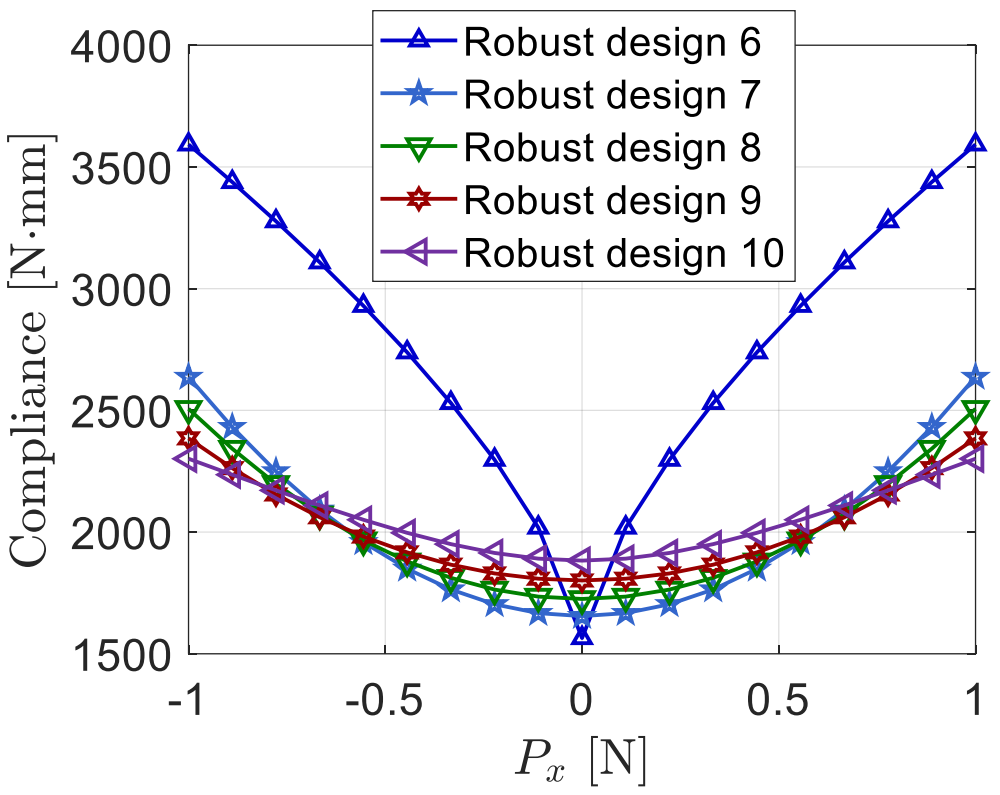

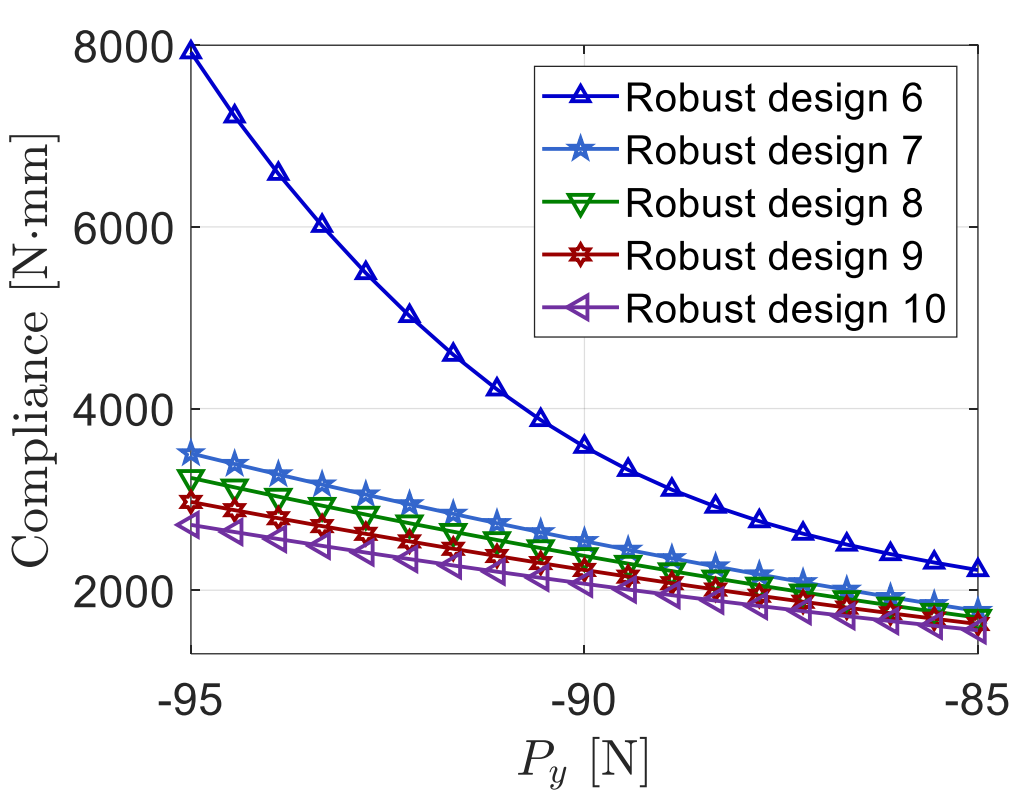

(a) $\boldsymbol{P}_0 = [P_x \quad -90]^T$ N (b) $\boldsymbol{P}_0 = [1 \quad P_y]^T$ N

Figure 44. Compliance of robust designs 6, 7, 8, 9, and 10 for two cases: (a) Compliance vs horizontal load with $\boldsymbol{P}_0 = [P_x \quad -90]^T$N; (b) Compliance vs vertical load with $\boldsymbol{P}_0 = [1 \quad P_y]^T$N.

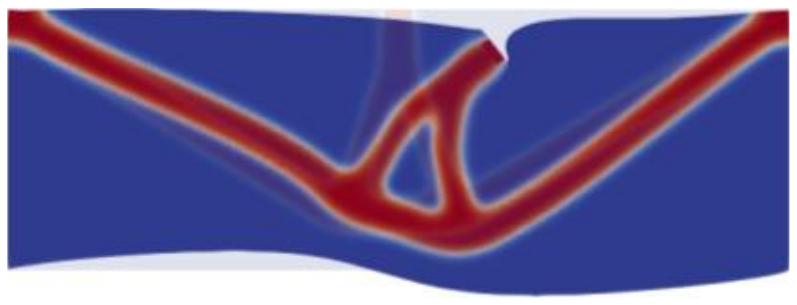

(a) Robust design 6

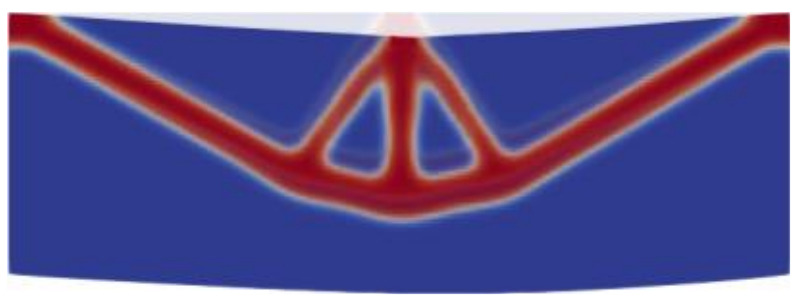

(b) Robust design 7

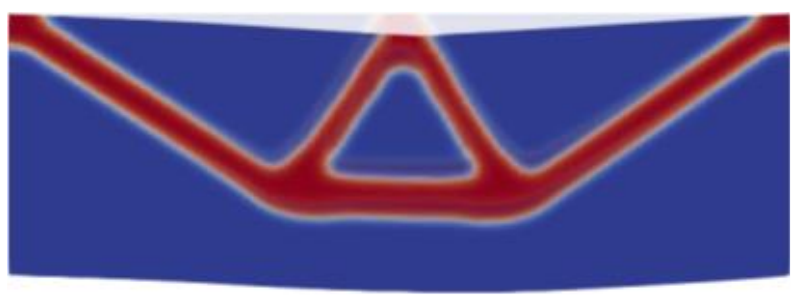

(c) Robust design 8

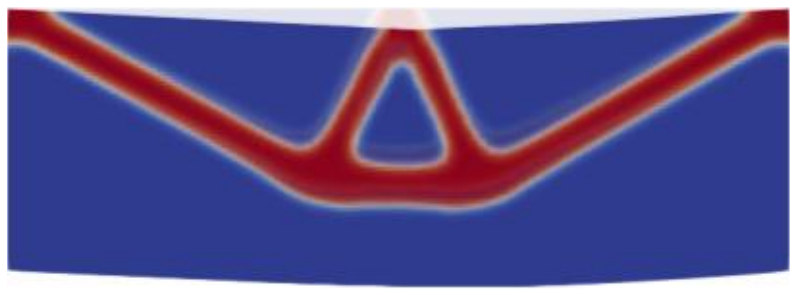

(d) Robust design 9

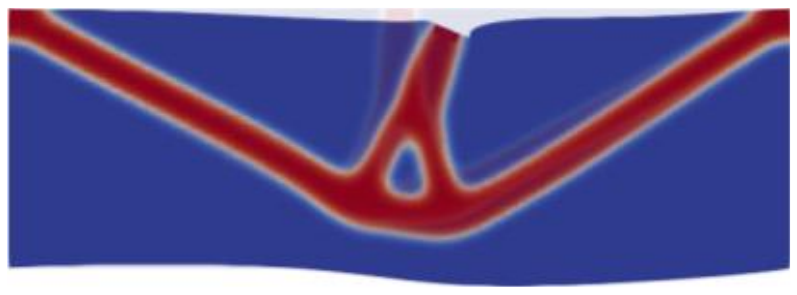

(e) Robust design 10

Figure 45. Deformed shapes of robust designs 6, 7, 8, 9, and 10 with load $\boldsymbol{P}_0 = [1 \quad -90]^T$ N.

### *7.3 Pinned column*

The last example considers a pinned column design problem shown in Figure 46(a), which is discretized by a 100×300 FE mesh. To avoid localized deformations, the elements in the top and bottom red areas are fixed to a solid phase, and a quarter symmetry is enforced during optimization. Poisson's ratio and Young's modulus for the deterministic scenario are 0.4 and 3 MPa, respectively. Also, $E_{L0}$ = 3 MPa and $\nu_L$ = 0.4 for linear energy interpolation (Eq. (13)). The density filter radius and maximum volume fraction are set to 36 mm and 0.25, respectively. The continuation scheme employed in this case is as follows: the penalization parameter $p$ (Eq. (1)) is increased from 1.8 to 4 @ 0.1 every 20 iterations, while $p_l$ (Eq. (13)) is increased from 4.8 to 7 @ 0.1 every 20 iterations, and the filter parameter $\beta$ (Eq. (3)) is increased from 1 to 4 @ 0.1 every 20 iterations. An additional 200 iterations are carried out after the maximum value of the parameters is reached. Moreover, the MMA move limit parameter is set to 0.1 to further stabilize the optimization process. In the stochastic scenario, the material and geometric uncertainties are investigated; see Table 2 for the details. In particular, Young's modulus random field with correlation lengths chosen from the set {($l_{cx}$ = 60 mm, $l_{cy}$ = ∞), ($l_{cx}$ = 100 mm, $l_{cy}$ = ∞), ($l_{cx}$ = 400 mm, $l_{cy}$ = 400 mm)} leads to 11, 7, and 15 random variables (robust designs 1-3) after the KL expansion, respectively. For geometric uncertainty, the uniform $\eta$ random field with correlation lengths chosen from the set {($l_{cx}$ = 75 mm, $l_{cy}$ = ∞), ($l_{cx}$ = 400 mm, $l_{cy}$ = ∞), ($l_{cx}$ = 400 mm, $l_{cy}$ = 400 mm)} leads to 9, 2, and 15 random variables (robust designs 4-6) after the KL expansion, respectively. The weighting parameter $\alpha$ in robust optimization formulation is set to 1.0. Figure 46(b) shows the optimized topology using the deterministic design formulation, a vertical column member, as expected.

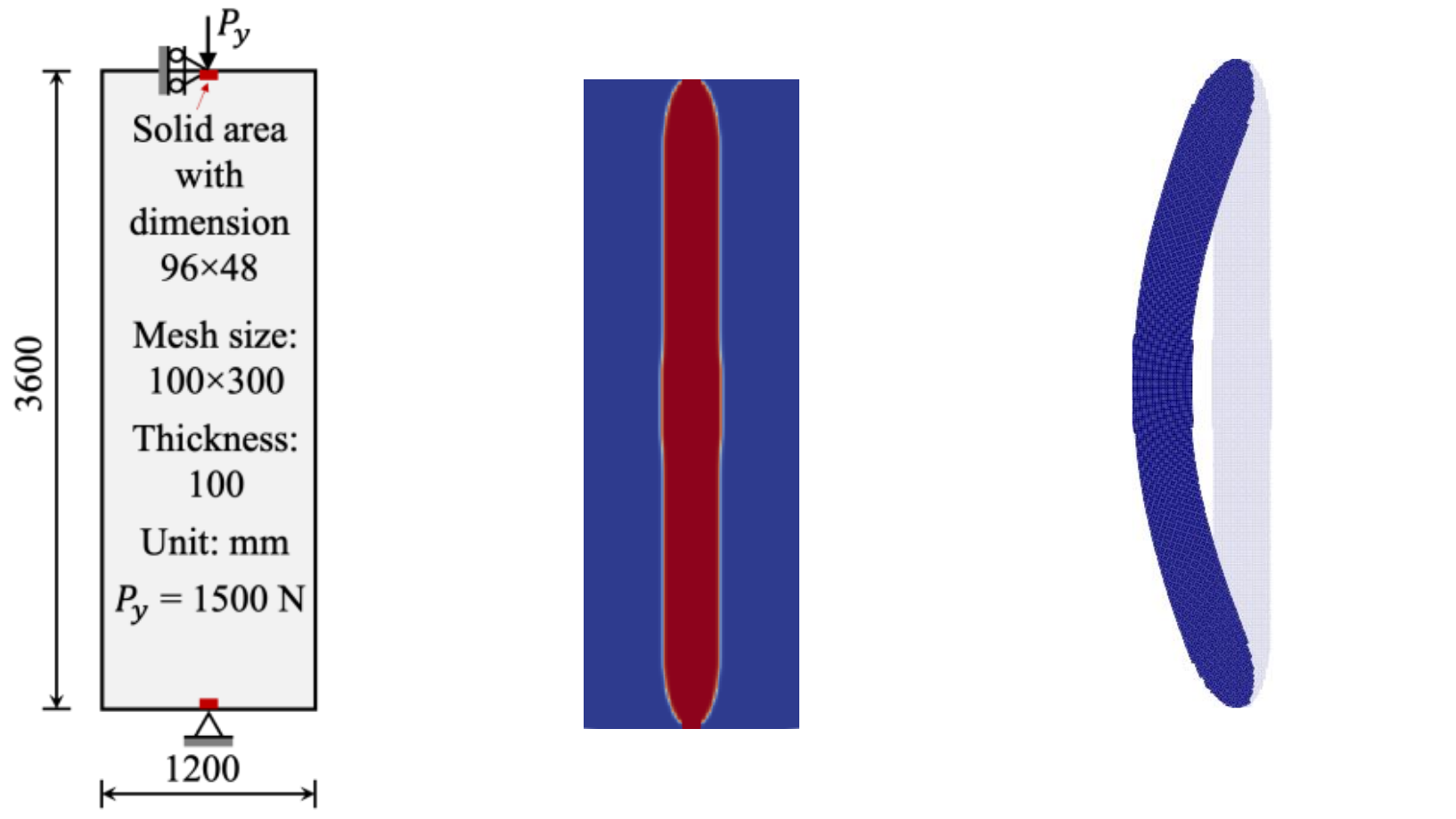

(a) Pinned column (b) Deterministic design (c) Buckling mode

Figure 46. Problem setup of a pinned column and deterministic design.

### *7.3.1 Effect of material uncertainty*

Incorporating only material uncertainty with varying correlation lengths $l_{cx}$ and $l_{cy}$ the optimized robust designs are shown in Figure 47, which are different from the deterministic design in Figure 46(b). The KL expansion of Young's modulus random fields with varying correlation lengths results in the first two eigenmodes in Figure 48, one symmetric ($\boldsymbol{\gamma}_1$) and the other one antisymmetric ($\boldsymbol{\gamma}_2$). To investigate the robustness of the optimized designs w.r.t. Young's modulus random field corresponding to asymmetric eigenmode, Figure 49 plots the curves of compliance versus eigenmode coefficient $\xi_{2i}$ with subscript $i = 2$ denoting the 2$^{nd}$ eigenmode. Like the previous examples, random field realizations are generated by including only the specified eigenmode in Eq. (21) with a specified coefficient $\xi_{22}$. Figure 49 demonstrates that with antisymmetric fields, the stiffness of the deterministic design decreases significantly after the $\xi_{22}$ reaches around $\pm$0.22 for robust design 1 and 2, and $\pm$0.44 for robust design 3. The drop in the stiffness of the deterministic design is illustrated by the deformations at $\xi_{22}$ = 1 plotted in Figure 50 — the deterministic design buckles in this case. To investigate this case further, a stability analysis is carried out on all four designs by removing void elements and by examining the positive-definiteness of the tangent stiffness matrix at each loading step [59]. The stability analyses show that the deterministic design encounters 1$^{st}$ bifurcation point with its buckling mode plotted in Figure 46(c) around 730 N, below the design load of 1500 N, while all the other robust designs remain stable up to the design load. As the stability constraints are not explicitly included during the optimization process, the deterministic design is unstable, which is not surprising. However, optimized topologies are stable when the uncertainties are appropriately incorporated. This result demonstrates that considering uncertainty in optimization formulation may help design stable topologies.

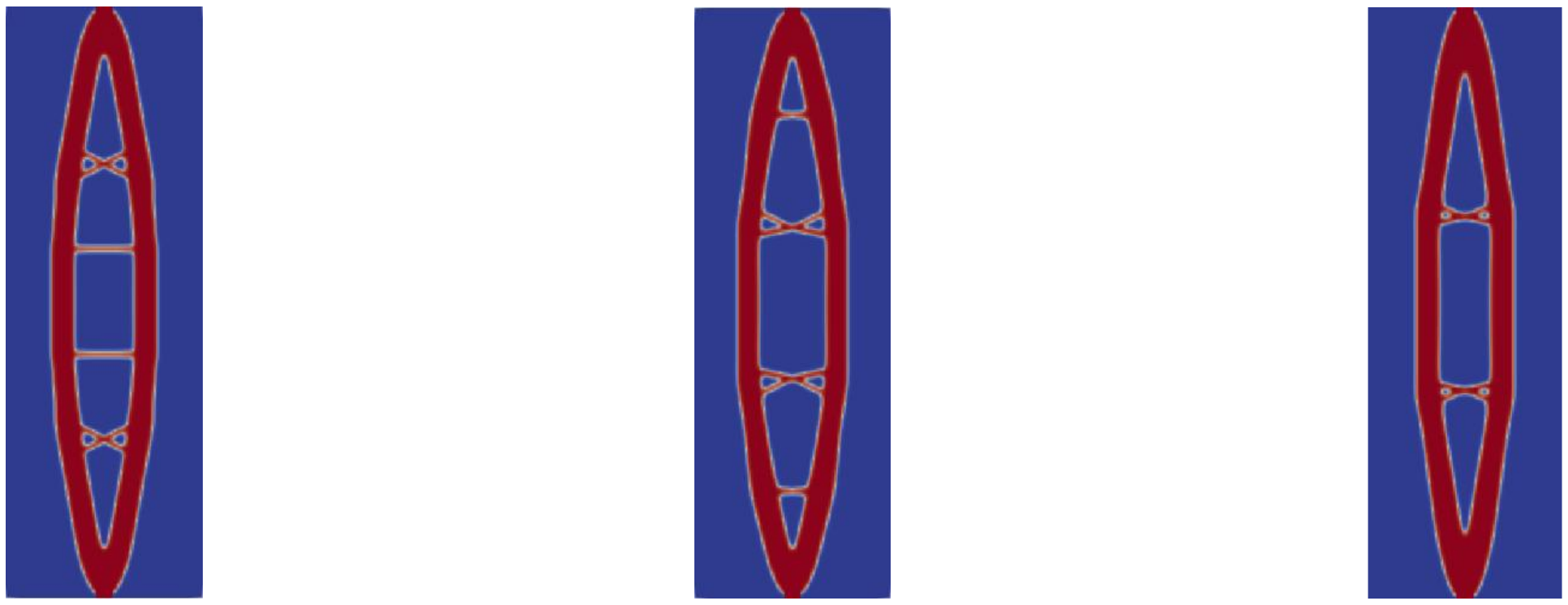

(a) Robust design 1 ($l_{cx}$ = 60 mm, $l_{cy}$ = ∞)　(b) Robust design 2 ($l_{cx}$ = 100 mm, $l_{cy}$ = ∞)　(c) Robust design 3 ($l_{cx}$ = $l_{cy}$ = 400 mm)

Figure 47. Robust designs of the pinned column with different values of $l_{cx}$ and $l_{cy}$ in $E$-field.

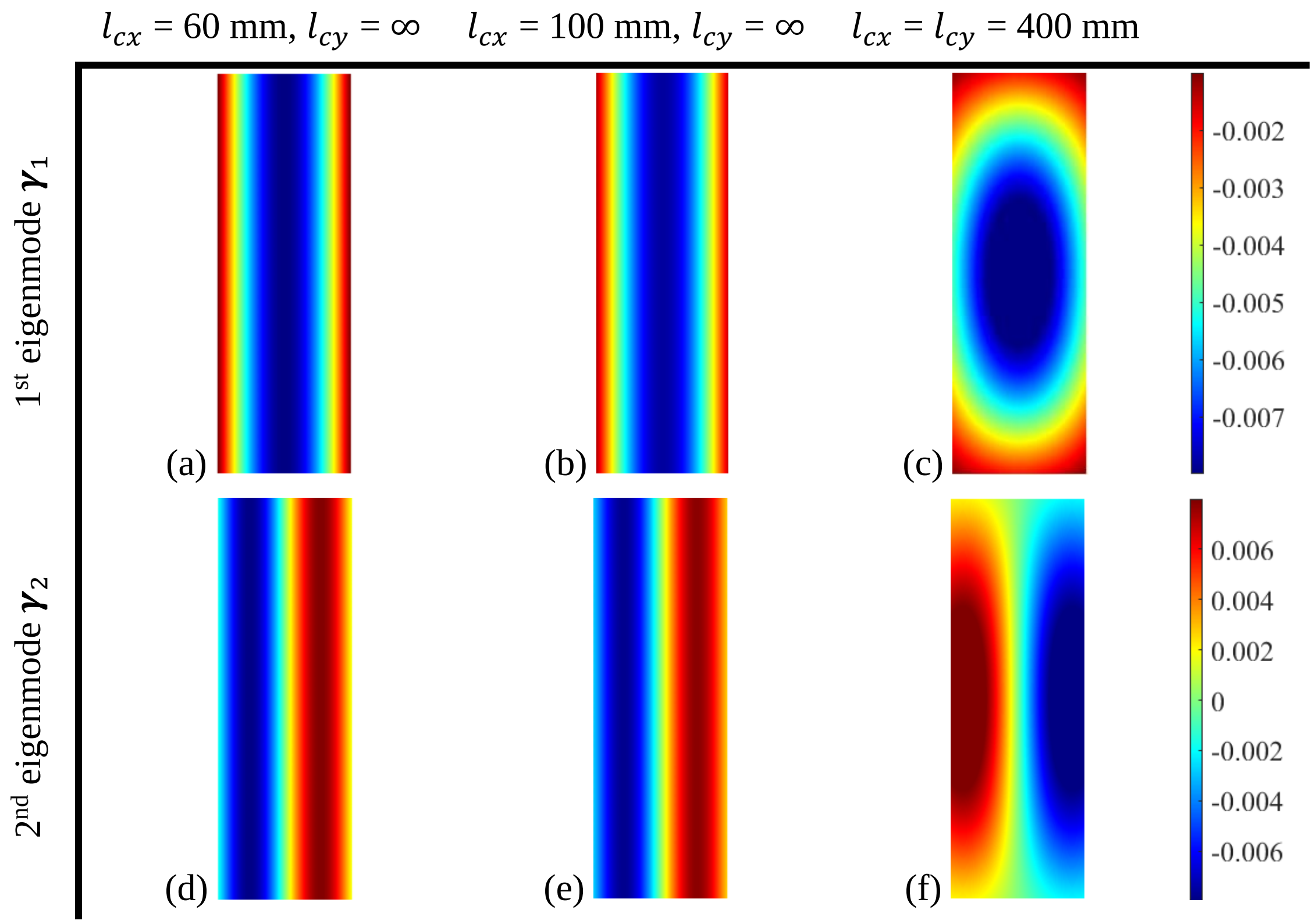

Figure 48. First two eigenmodes in the KL expansion of $E$-field.

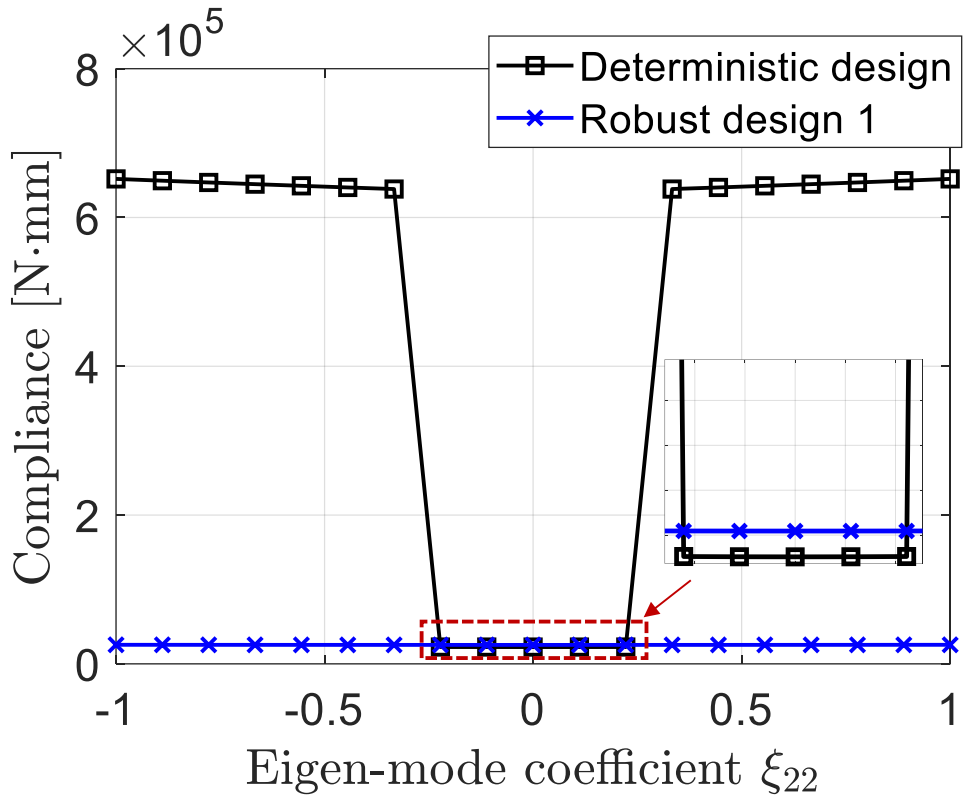

(a) $l_{cx}$ = 60 mm, $l_{cy}$ = ∞

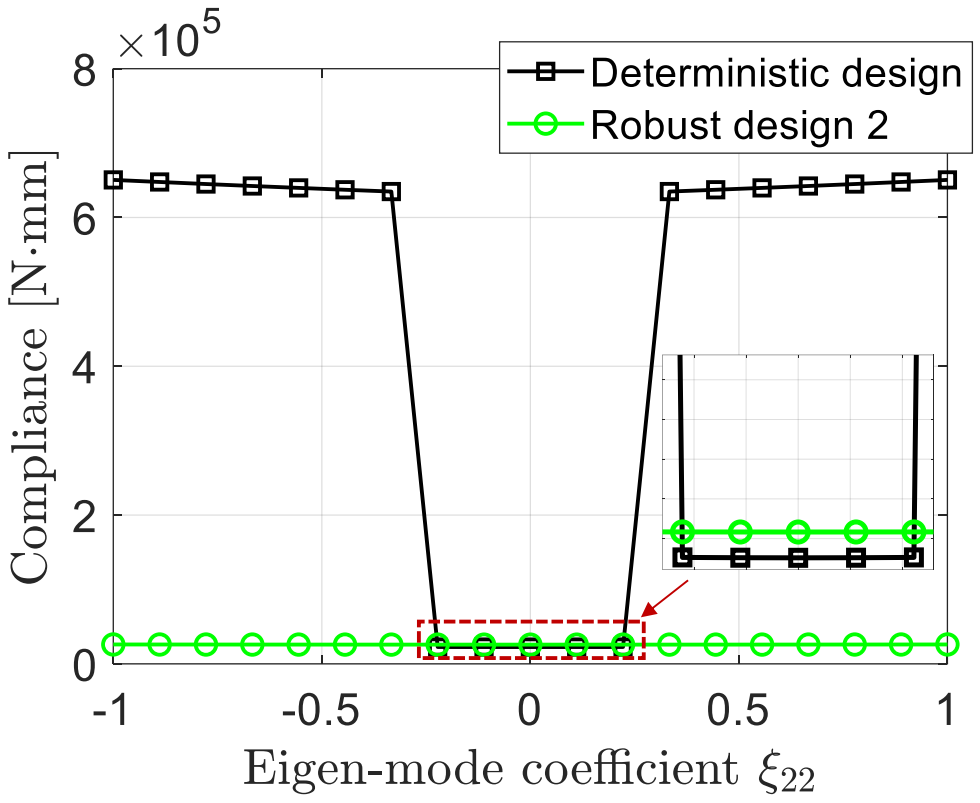

(b) $l_{cx}$ = 100 mm, $l_{cy}$ = ∞

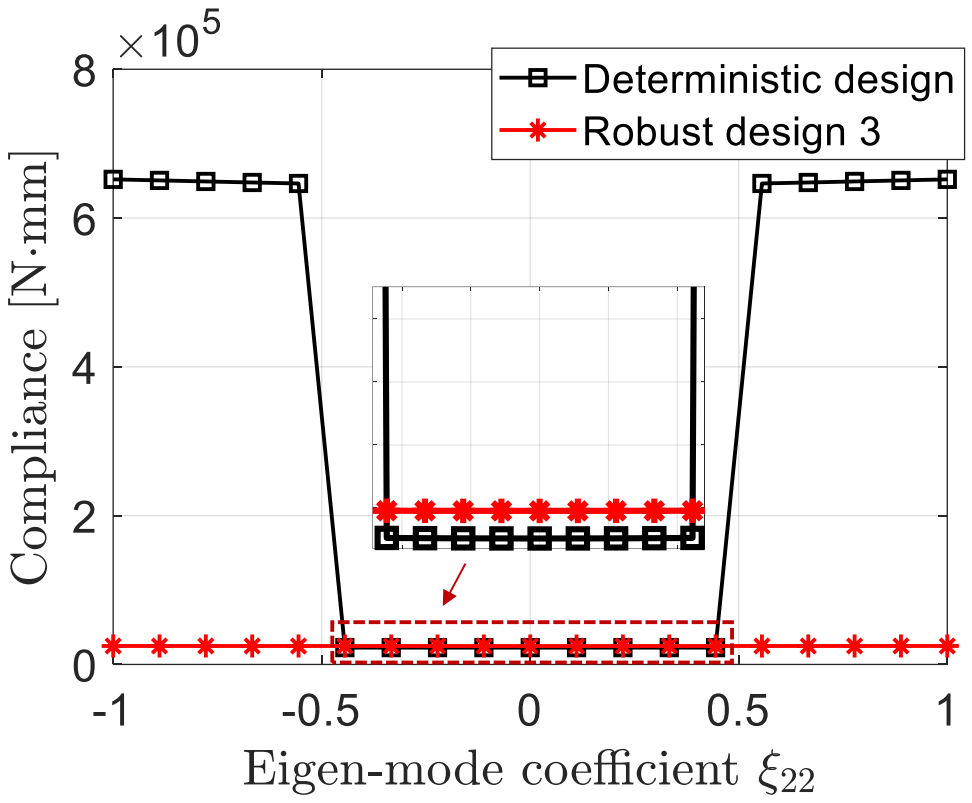

(c) $l_{cx} = l_{cy}$ = 400 mm

Figure 49. Compliance versus eigenmode coefficient curves of deterministic and robust designs for 2nd eigenmode $\boldsymbol{\gamma}_2$ in the KL expansion of $E$-field.

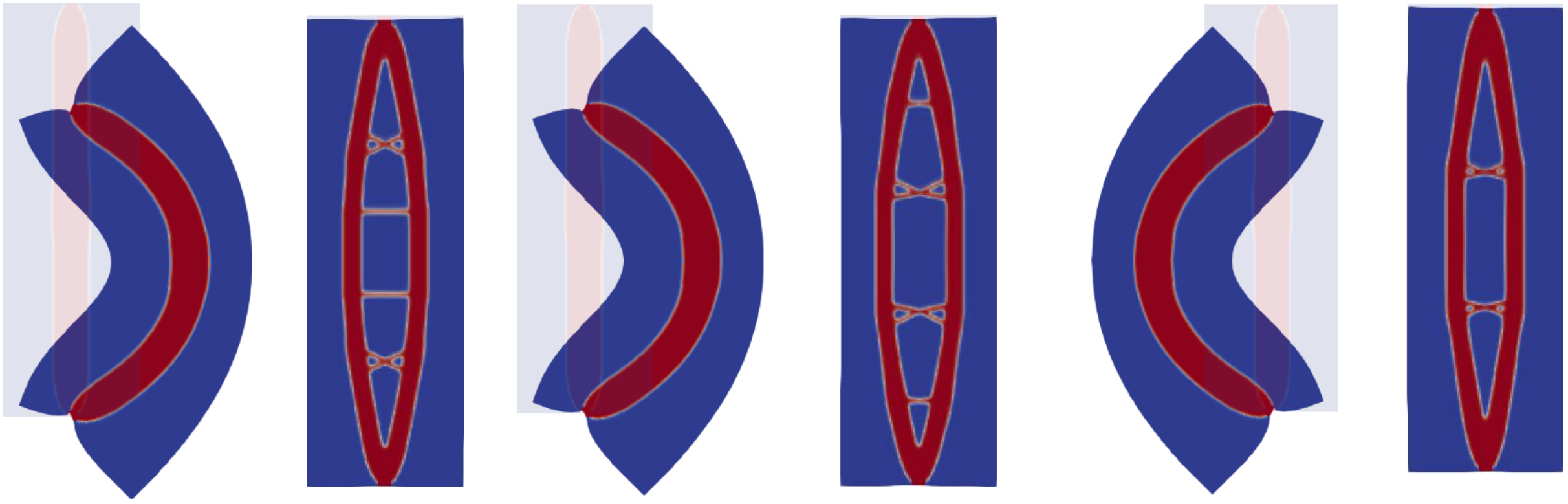

| Deterministic design | Robust design 1 | Deterministic design | Robust design 2 | Deterministic design | Robust design 3 |
|---|---|---|---|---|---|
| (a) $l_{cx}$ = 60 mm, $l_{cy}$ = ∞ | | (b) $l_{cx}$ = 100 mm, $l_{cy}$ = ∞ | | (c) $l_{cx} = l_{cy}$ = 400 mm | |

Figure 50. Deformed shapes of deterministic and robust designs for 2nd eigenmode $\boldsymbol{\gamma}_2$ in the KL expansion of $E$-field with different $l_{cx}$ and $l_{cy}$, and $\xi_{22} = 1.0$.

### 7.3.2 *Effect of geometric uncertainty*

When incorporating only geometric uncertainty, the optimized designs for different values of correlation lengths $l_{cx}$ and $l_{cy}$ of the projection cutoff $\eta$-random field is obtained, as shown in Figure 51. After KL expansion for $\eta$-random fields with different correlation lengths, the first two eigenmodes are shown in Figure 52. Again, Figure 53 plots the curves of compliance versus eigenmode coefficient $\xi_{32}$. In this case, the stiffness of the deterministic design decreases significantly after the $\xi_{32}$ reaches around $\pm 0.33$ for robust designs 4 and 5, and $\pm 0.67$ for robust designs 6 with antisymmetric $\eta$-field. The deformed shapes of those designs examined at the $\xi_{32}$ = 1 exhibits the same phenomenon as that in Figure 50. In addition, the stability analysis results show that all the robust designs are stable up to the design load.

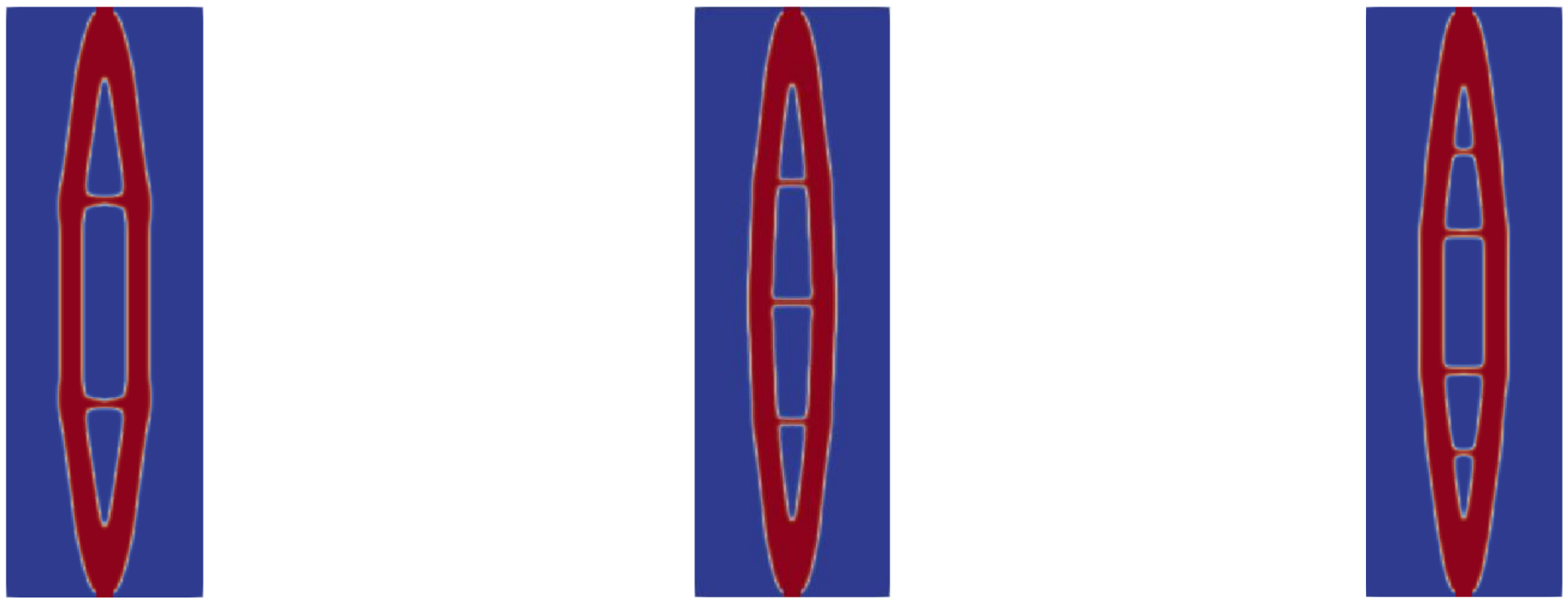

(a) Robust design 4 ($l_{cx}$ = 75 mm, $l_{cy}$ = ∞) (b) Robust design 5 ($l_{cx}$ = 400 mm, $l_{cy}$ = ∞) (c) Robust design 6 ($l_{cx}$ = $l_{cy}$ = 400 mm)

Figure 51. Robust designs with different values of the correlation length $l_{cx}$ in $\eta$-field.

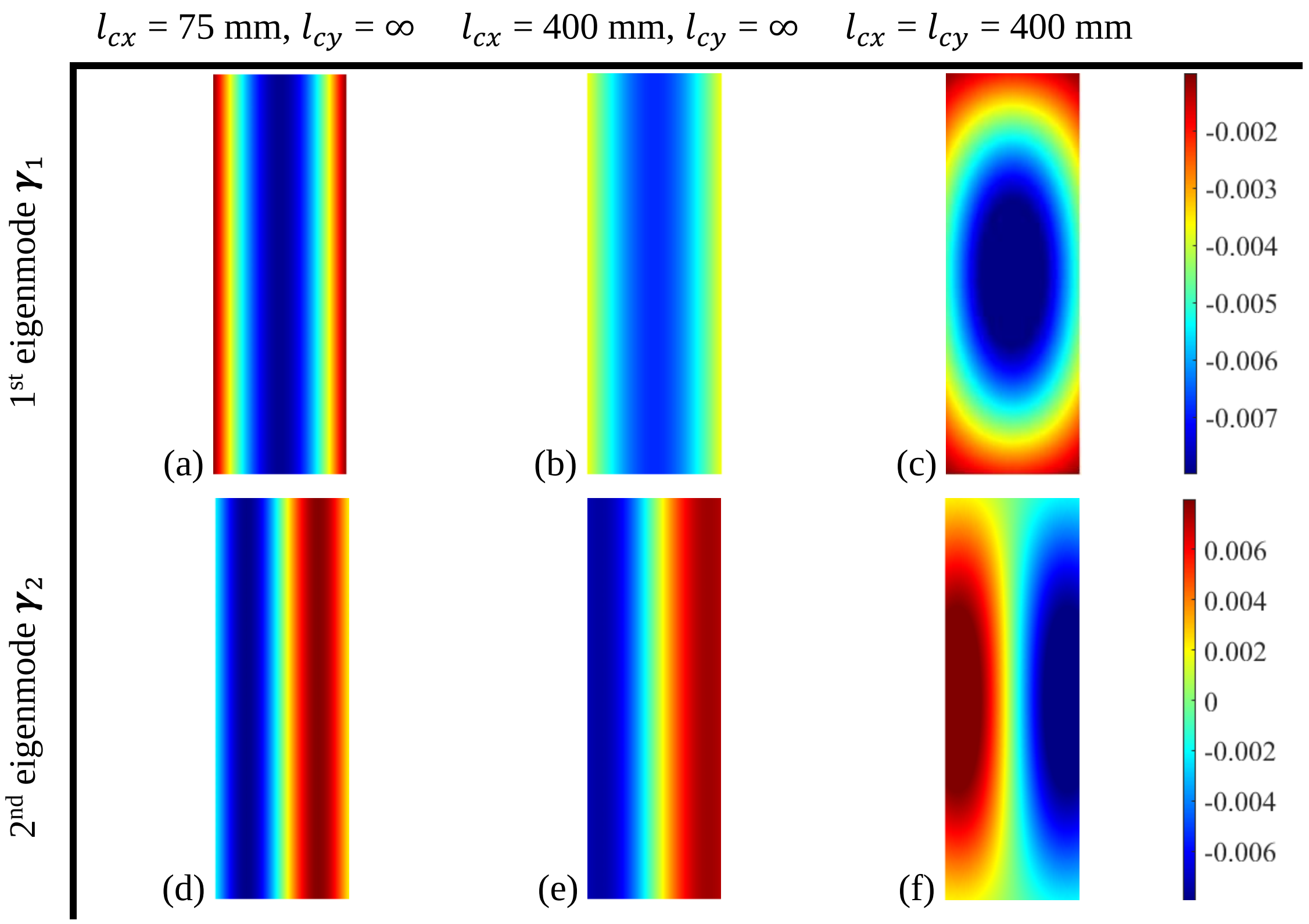

Figure 52. First two eigenmodes in the KL expansion of $\eta$-field.

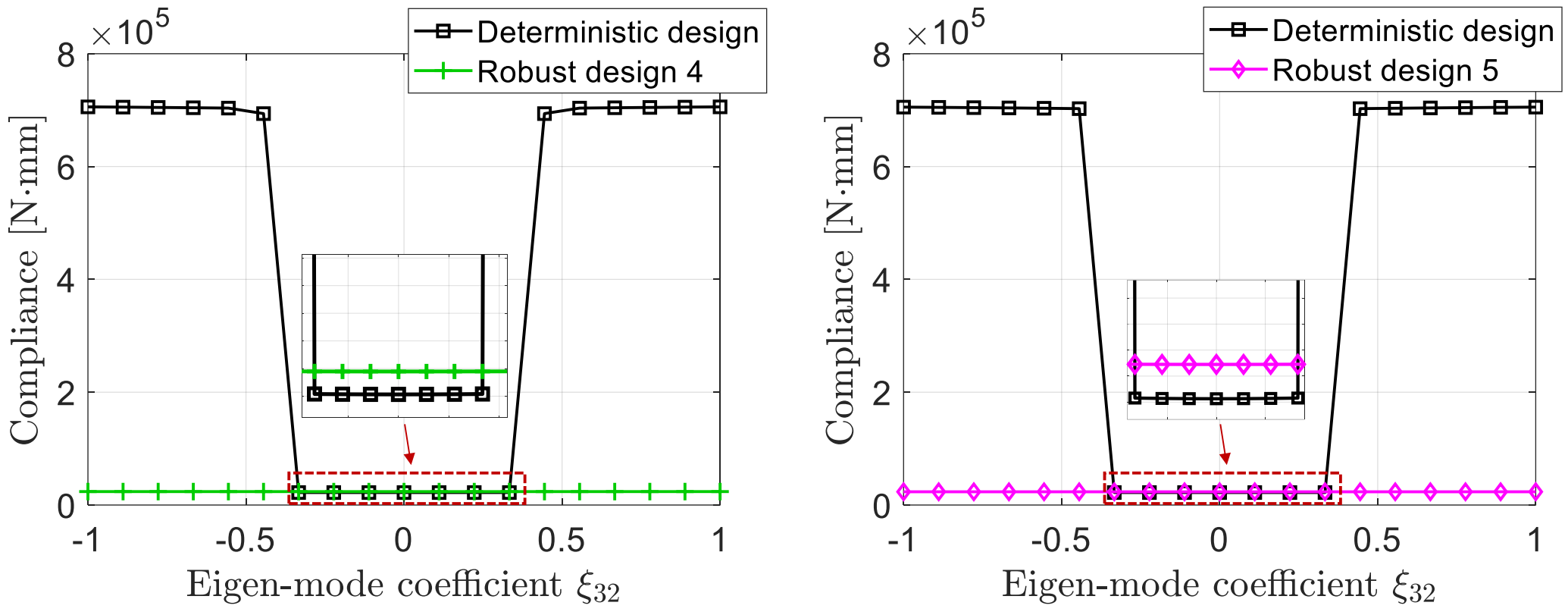

(a) $l_{cx}$ = 75 mm, $l_{cy}$ = ∞

(b) $l_{cx}$ = 400 mm, $l_{cy}$ = ∞

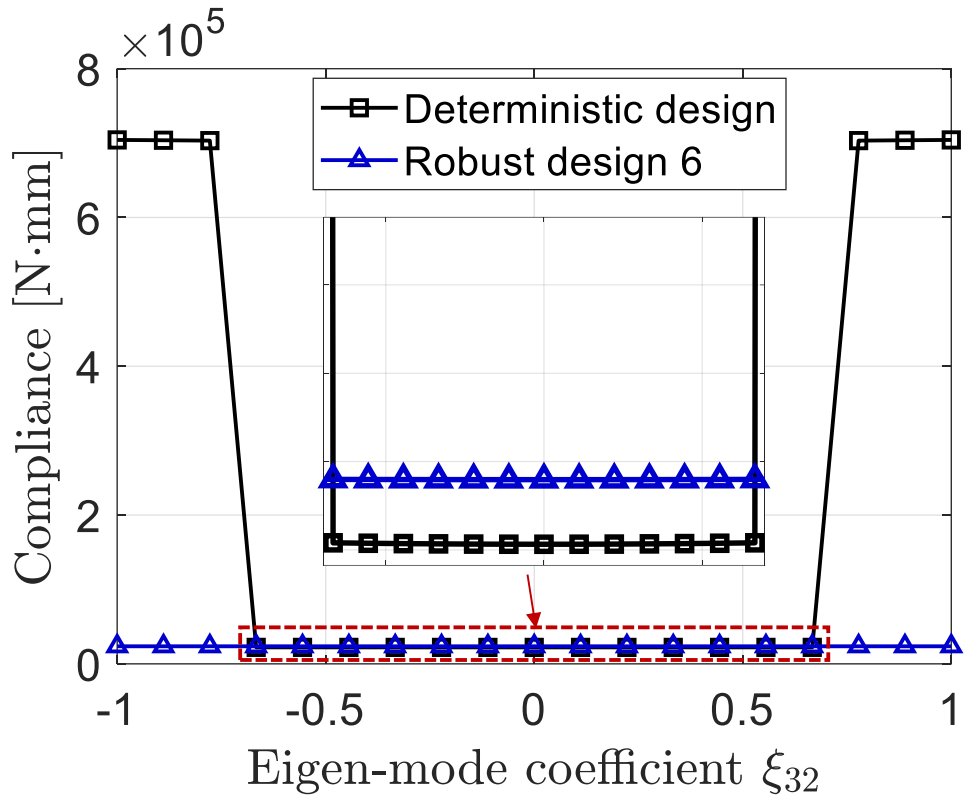

(c) $l_{cx}$ = $l_{cy}$ = 400 mm

Figure 53. Compliance versus eigenmode coefficient curves of deterministic and robust designs for 2nd eigenmode $\boldsymbol{\gamma}_2$ in the KL expansion of $\eta$-field.

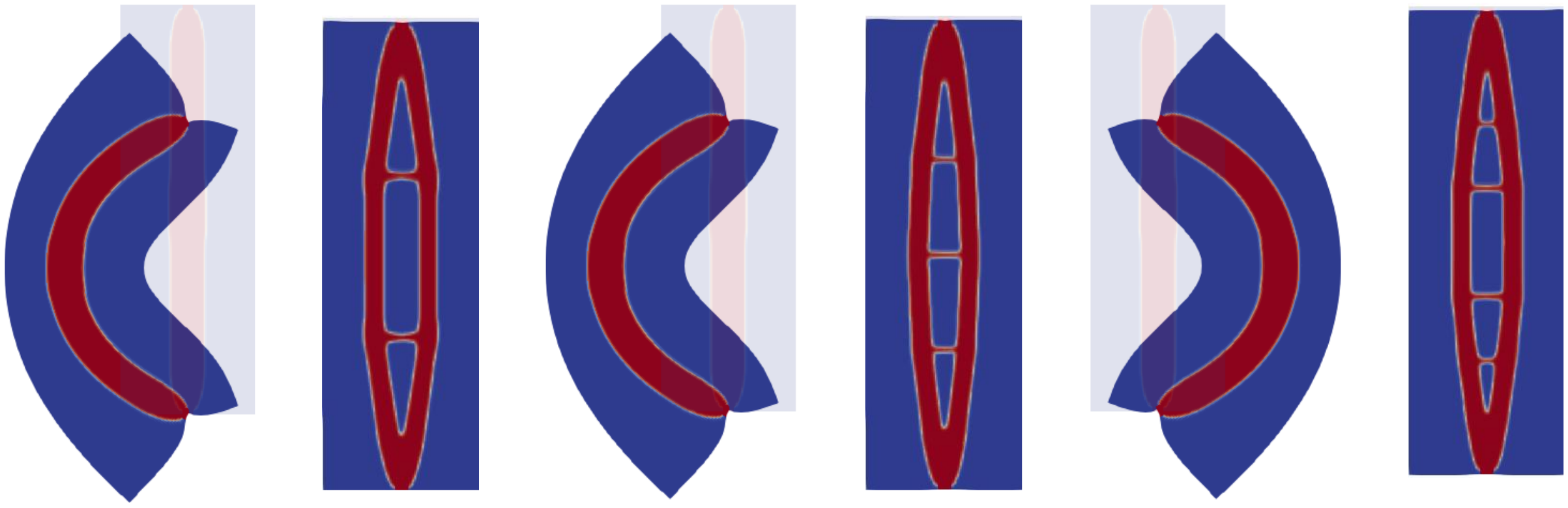

Deterministic design | Robust design 4 | Deterministic design | Robust design 5 | Deterministic design | Robust design 6

(a) $l_{cx}$ = 75 mm, $l_{cy}$ = ∞ (b) $l_{cx}$ = 400 mm, $l_{cy}$ = ∞ (c) $l_{cx}$ = $l_{cy}$ = 400 mm

Figure 54. Deformed shapes of deterministic and robust designs for 2nd eigenmode $\boldsymbol{\gamma}_2$ in the KL expansion of $\eta$-field with different $l_{cx}$ and $l_{cy}$, and $\xi_{32} = 1.0$.

### *7.3.3 Effect of combined material and geometric uncertainties*

Combining the material uncertainty (correlation lengths $l_{cx}$ = 60 mm and $l_{cy}$ =∞) and geometric uncertainty (correlation lengths $l_{cx}$ = 75 mm and $l_{cy}$ = ∞)), the optimized design is obtained, as shown in Figure 55. To compare the robustness of design 1, considering only material uncertainty, and design 7 in terms of geometric uncertainty, the curves of compliance versus coefficient $\xi_{32}$ associated with the antisymmetric $\eta$-field are plotted in Figure 56. As expected, design 7 exhibits higher robustness than design 1, which can be seen from the deformed shapes of the two designs at $\xi_{32}$ = 5.1 in Figure 57. To compare the robustness of design 4, considering only geometric uncertainty, and design 7 w.r.t. the material uncertainty, the curves of compliance versus coefficient $\xi_{22}$ associated with the antisymmetric $E$-field are shown in Figure 58, where it can be seen that design 7 has higher robustness than design 4. The deformations of the two designs in Figure 59 show a higher stiffness of design 7 as compared to design 4 at the coefficient $\xi_{22}$ = 4.4.

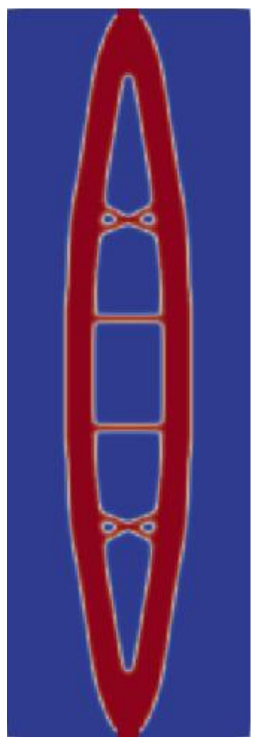

Figure 55. Robust design 7 of the pinned column with combined material and geometric uncertainties.

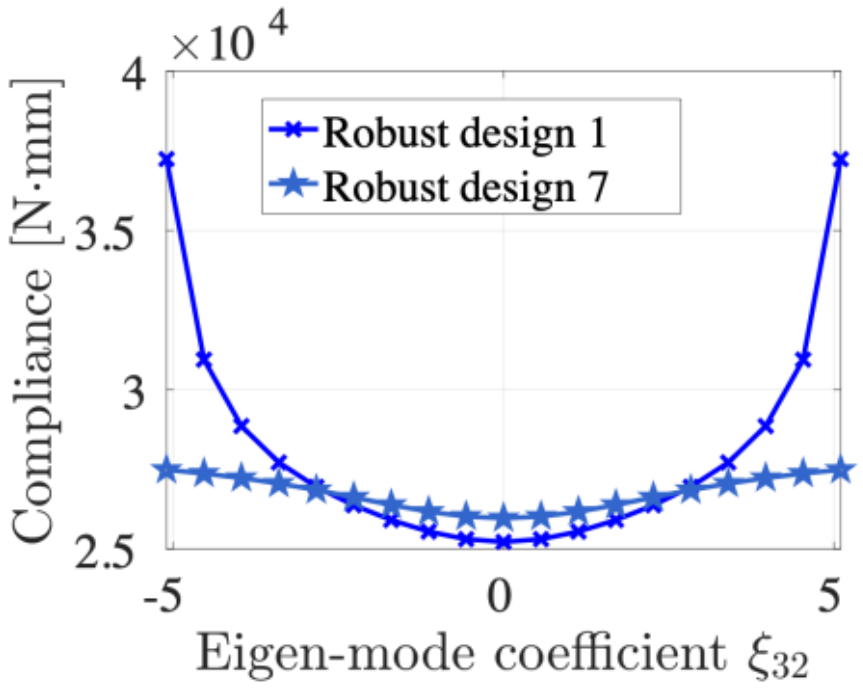

Figure 56. Compliance versus eigenmode coefficient curves of robust design 1 and robust design 7 for 2nd eigenmode $\boldsymbol{\gamma}_2$ in the KL expansion of $\eta$-field with correlation lengths $l_{cx}$ = 75 mm and $l_{cy}$ = ∞.

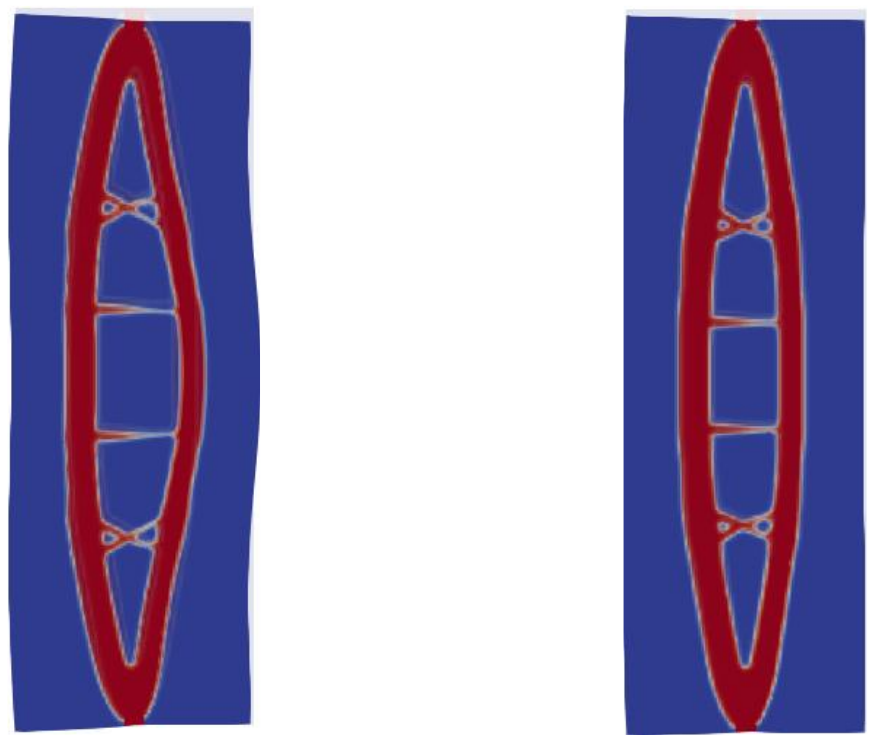

(a) Robust design 1 (b) Robust design 7

Figure 57. Deformed shapes of robust design 1 and robust design 7 for 2nd eigenmode $\boldsymbol{\gamma}_2$ in the KL expansion of $\eta$-field with $l_{cx}$ = 75 mm and $l_{cy}$ = ∞, and $\xi_{32} = 5.1$.

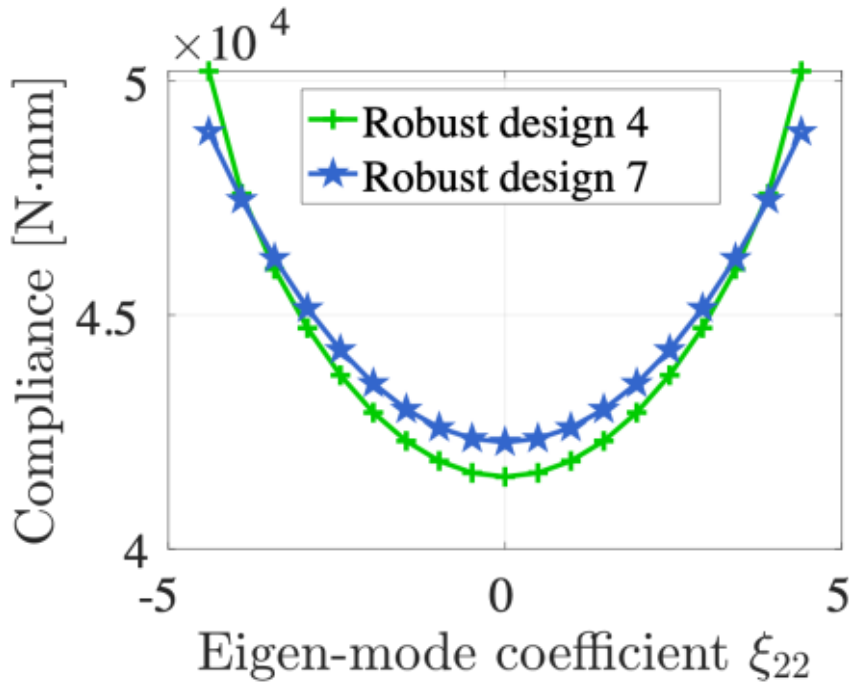

Figure 58. Compliance versus eigenmode coefficient curves of robust design 4 and robust design 7 for 2nd eigenmode $\boldsymbol{\gamma}_2$ in the KL expansion of $E$-field with $l_{cx}$ = 60 mm and $l_{cy}$ = ∞.

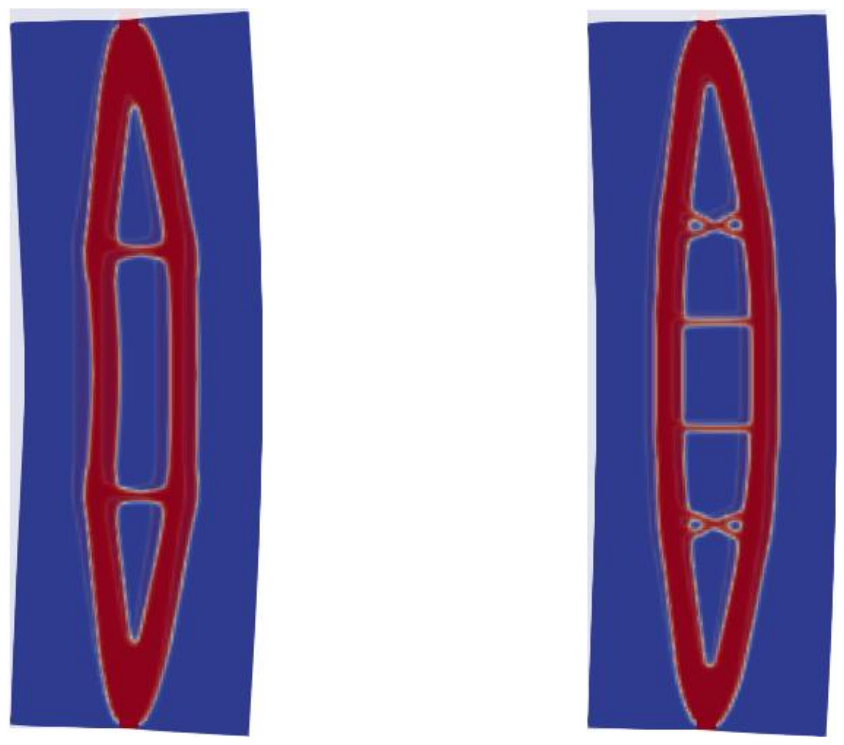

(a) Robust design 4 (b) Robust design 7

Figure 59. Deformed shapes of robust design 4 and robust design 7 for 2nd eigenmode $\boldsymbol{\gamma}_2$ in the KL expansion of $E$-field with $l_{cx}$ = $l_{cy}$ = 400 mm and $\xi_{22} = 4.4$.

## 8 Conclusions

A computational design optimization framework is presented for the topology optimization of robust hyperelastic structures at finite deformations incorporating multiple uncertain sources, i.e., load, material, and geometry. Specifically, a second-order stochastic perturbation method is used to compute the statistics of structural compliance, and an analytical adjoint sensitivity analysis is proposed within a nonlinear stochastic topology optimization framework. The accuracy of the stochastic perturbation method and the adjoint sensitivity calculation is verified numerically via the Monte Carlo and central difference methods, respectively. In addition, the low-density element distortion issue is also handled via an adaptive linear energy interpolation scheme.

Using the proposed framework, the effect of various uncertainties on the optimal topologies is systematically explored. The proposed framework is used to obtain optimized robust designs for compression block, clamped beam, and pinned column examples. For compression block, four different cases of uncertain sources are considered: (a) load uncertainty, (b) material uncertainty, (c) geometric uncertainty, and (d) combined load, material, and geometric uncertainties. With only load, material, or geometric uncertainty, the resulting optimized designs show increased robustness w.r.t the considered uncertainty. In addition, including load, material, and geometric uncertainties in optimization formulation leads to a robust design w.r.t. all these uncertainties. In the clamped beam example, robust designs are explored under load, material, and combined load and material uncertainties. Like the previous case, the design scenarios with single uncertainty are more robust than the deterministic cases. Also, with the increase in the robustness weighting factor in optimization formulation, designs with higher robustness are obtained. Considering load and material uncertainties together also leads to designs that are robust under these combined uncertainties. In the pinned column example, different robust designs are obtained under material or geometric uncertainty and combined material and geometric uncertainties in optimization formulation. The overall trends are like the last two cases, i.e., designs with robust behavior are obtained under the considered uncertainties. Moreover, in this case, considering uncertainties in the optimization process leads to *stable* designs, while an unstable design is obtained via deterministic optimization formulation.

Based on the results of the numerical studies, several conclusions can be drawn: (a) increasing the variance of the uncertain sources in the optimization leads to more robust designs; (b) increasing the weighting factor for the standard deviation in the objective function can also help to improve the robustness of optimized designs; (c) different settings of correlation lengths for the considered uncertainties modeled by random fields may lead to different topologies of the optimized structures; (d) combining multiple sources of uncertainties in the optimization formulation can produce optimal designs that can provide robust performance under these multiple uncertainties; (e) incorporating uncertainties in the optimization may help to obtain stable designs without explicit consideration of stability constraints. In particular, the asymmetric modes in the uncertainty fields promote designs that are stable under these symmetry-breaking modes [60]. However, explicit considerations of stability constraints might be necessary to ensure stable topologies [46, 61], which will be considered in future work. It is emphasized that the scope of this study is limited to the 2$^{nd}$-order perturbation methods, and high-order approaches might be needed depending on the considered applications [14]. However, the presented approach can be canonically extended to such cases, albeit at the cost of additional higher-order tensorial derivatives.

## Acknowledgments

The presented work is supported in part by the US National Science Foundation through grant CMMI-1762277. Any opinions, findings, conclusions, and recommendations expressed in this article are those of the authors and do not necessarily reflect the views of the sponsors.

## Data Availability

Data sharing is not applicable to this article as no datasets were generated or analyzed during the current study.

## References

[1] M.P. Bendsøe, N. Kikuchi, Generating optimal topologies in structural design using a homogenization method, Computer Methods in Applied Mechanics and Engineering, 71 (1988) 197-224.

[2] O. Sigmund, K. Maute, Topology optimization approaches, Structural and Multidisciplinary Optimization, 48 (2013) 1031-1055.

[3] J. Wu, O. Sigmund, J.P. Groen, Topology optimization of multi-scale structures: a review, Structural and Multidisciplinary Optimization, 63 (2021) 1455-1480.

[4] J.D. Deaton, R.V. Grandhi, A survey of structural and multidisciplinary continuum topology optimization: post 2000, Structural and Multidisciplinary Optimization, 49 (2014) 1-38.

[5] B. Blakey-Milner, P. Gradl, G. Snedden, M. Brooks, J. Pitot, E. Lopez, M. Leary, F. Berto, A. du Plessis, Metal additive manufacturing in aerospace: A review, Materials & Design, 209 (2021) 110008.

[6] J. Plocher, A. Panesar, Review on design and structural optimisation in additive manufacturing: Towards next-generation lightweight structures, Materials & Design, 183 (2019) 108164.

[7] R. Ranjan, R. Samant, S. Anand, Integration of Design for Manufacturing Methods With Topology Optimization in Additive Manufacturing, Journal of Manufacturing Science and Engineering, 139 (2017).

[8] H. Zhang, J. Guilleminot, L.J. Gomez, Stochastic modeling of geometrical uncertainties on complex domains, with application to additive manufacturing and brain interface geometries, Computer Methods in Applied Mechanics and Engineering, 385 (2021) 114014.

[9] L. Liu, P. Kamm, F. García-Moreno, J. Banhart, D. Pasini, Elastic and failure response of imperfect three-dimensional metallic lattices: the role of geometric defects induced by Selective Laser Melting, Journal of the Mechanics and Physics of Solids, 107 (2017) 160-184.

[10] T.D. Ngo, A. Kashani, G. Imbalzano, K.T.Q. Nguyen, D. Hui, Additive manufacturing (3D printing): A review of materials, methods, applications and challenges, Composites Part B: Engineering, 143 (2018) 172-196.

[11] S.-K. Choi, R.A. Canfield, R.V. Grandhi, Reliability-Based Structural Optimization, Springer, 2007.

[12] M. Papadrakakis, N.D. Lagaros, Reliability-based structural optimization using neural networks and Monte Carlo simulation, Computer Methods in Applied Mechanics and Engineering, 191 (2002) 3491-3507.

[13] M. Papadrakakis, N.D. Lagaros, Y. Tsompanakis, V. Plevris, Large scale structural optimization: Computational methods and optimization algorithms, Archives of Computational Methods in Engineering, 8 (2001) 239-301.

[14] N. Feng, G. Zhang, K. Khandelwal, On the performance evaluation of stochastic finite elements in linear and nonlinear problems, Computers & Structures, 243 (2021) 106408.

[15] G. Stefanou, The stochastic finite element method: Past, present and future, Computer Methods in Applied Mechanics and Engineering, 198 (2009) 1031-1051.

[16] M. Jansen, G. Lombaert, M. Diehl, B.S. Lazarov, O. Sigmund, M. Schevenels, Robust topology optimization accounting for misplacement of material, Structural and Multidisciplinary Optimization, 47 (2013) 317-333.
[17] J. Zhao, C. Wang, Robust Topology Optimization of Structures Under Loading Uncertainty, AIAA Journal, 52 (2014) 398-407.
[18] D.I. Papadimitriou, C. Papadimitriou, Robust and Reliability-Based Structural Topology Optimization Using a Continuous Adjoint Method, ASCE-ASME Journal of Risk and Uncertainty in Engineering Systems, Part A: Civil Engineering, 2 (2016) B4016002.
[19] X. Peng, J. Li, S. Jiang, Z. Liu, Robust topology optimization of continuum structures with loading uncertainty using a perturbation method, Engineering Optimization, 50 (2018) 584-598.
[20] Z. Meng, Y. Wu, X. Wang, S. Ren, B. Yu, Robust topology optimization methodology for continuum structures under probabilistic and fuzzy uncertainties, International Journal for Numerical Methods in Engineering, 122 (2021) 2095-2111.
[21] S. Bai, Z. Kang, Robust topology optimization for structures under bounded random loads and material uncertainties, Computers & Structures, 252 (2021) 106569.
[22] B. Kriegesmann, Robust design optimization with design-dependent random input variables, Structural and Multidisciplinary Optimization, 61 (2020) 661-674.
[23] M. Schevenels, B.S. Lazarov, O. Sigmund, Robust topology optimization accounting for spatially varying manufacturing errors, Computer Methods in Applied Mechanics and Engineering, 200 (2011) 3613-3627.
[24] B.S. Lazarov, M. Schevenels, O. Sigmund, Topology optimization considering material and geometric uncertainties using stochastic collocation methods, Structural and Multidisciplinary optimization, 46 (2012) 597-612.
[25] M. Tootkaboni, A. Asadpoure, J.K. Guest, Topology optimization of continuum structures under uncertainty – A Polynomial Chaos approach, Computer Methods in Applied Mechanics and Engineering, 201-204 (2012) 263-275.
[26] G.A. da Silva, E.L. Cardoso, Topology optimization of continuum structures subjected to uncertainties in material properties, International Journal for Numerical Methods in Engineering, 106 (2016) 192-212.
[27] V. Keshavarzzadeh, F. Fernandez, D.A. Tortorelli, Topology optimization under uncertainty via non-intrusive polynomial chaos expansion, Computer Methods in Applied Mechanics and Engineering, 318 (2017) 120-147.
[28] G.A. da Silva, A.T. Beck, E.L. Cardoso, Topology optimization of continuum structures with stress constraints and uncertainties in loading, International Journal for Numerical Methods in Engineering, 113 (2018) 153-178.
[29] S.A.L. Rostami, A. Ghoddosian, Topology optimization of continuum structures under hybrid uncertainties, Structural and Multidisciplinary Optimization, 57 (2018) 2399-2409.
[30] A.P. Torres, J.E. Warner, M.A. Aguiló, J.K. Guest, Robust topology optimization under loading uncertainties via stochastic reduced order models, International Journal for Numerical Methods in Engineering, 122 (2021) 5718-5743.

[31] M. Jansen, G. Lombaert, M. Schevenels, Robust topology optimization of structures with imperfect geometry based on geometric nonlinear analysis, Computer Methods in Applied Mechanics and Engineering, 285 (2015) 452-467.
[32] T. Nishino, J. Kato, Robust topology optimization based on finite strain considering uncertain loading conditions, International Journal for Numerical Methods in Engineering, 122 (2021) 1427-1455.
[33] M.P. Bendsøe, O. Sigmund, Material interpolation schemes in topology optimization, Archive of Applied Mechanics, 69 (1999) 635-654.
[34] O. Sigmund, J. Petersson, Numerical instabilities in topology optimization: A survey on procedures dealing with checkerboards, mesh-dependencies and local minima, Structural optimization, 16 (1998) 68-75.
[35] B. Bourdin, Filters in topology optimization, International Journal for Numerical Methods in Engineering, 50 (2001) 2143-2158.
[36] T.E. Bruns, D.A. Tortorelli, Topology optimization of non-linear elastic structures and compliant mechanisms, Computer Methods in Applied Mechanics and Engineering, 190 (2001) 3443-3459.
[37] L. Li, K. Khandelwal, Volume preserving projection filters and continuation methods in topology optimization, Engineering Structures, 85 (2015) 144-161.
[38] F. Wang, B.S. Lazarov, O. Sigmund, On projection methods, convergence and robust formulations in topology optimization, Structural and Multidisciplinary Optimization, 43 (2011) 767-784.
[39] R. De Borst, M.A. Crisfield, J.J. Remmers, C.V. Verhoosel, Nonlinear finite element analysis of solids and structures, John Wiley & Sons, 2012.
[40] F. Wang, B.S. Lazarov, O. Sigmund, J.S. Jensen, Interpolation scheme for fictitious domain techniques and topology optimization of finite strain elastic problems, Computer Methods in Applied Mechanics and Engineering, 276 (2014) 453-472.
[41] G. Zhang, R. Alberdi, K. Khandelwal, Topology optimization with incompressible materials under small and finite deformations using mixed u/p elements, International Journal for Numerical Methods in Engineering, 115 (2018) 1015-1052.
[42] M.E. Gurtin, E. Fried, L. Anand, The Mechanics and Thermodynamics of Continua, Cambridge University Press, Cambridge, 2010.
[43] G. Zhang, K. Khandelwal, Computational design of finite strain auxetic metamaterials via topology optimization and nonlinear homogenization, Computer Methods in Applied Mechanics and Engineering, 356 (2019) 490-527.
[44] G. Zhang, K. Khandelwal, Topology optimization of dissipative metamaterials at finite strains based on nonlinear homogenization, Structural and Multidisciplinary Optimization, 62 (2020) 1419-1455.
[45] G. Zhang, K. Khandelwal, Design of dissipative multimaterial viscoelastic-hyperelastic systems at finite strains via topology optimization, International Journal for Numerical Methods in Engineering, 119 (2019) 1037-1068.

[46] G. Zhang, K. Khandelwal, T. Guo, Finite strain topology optimization with nonlinear stability constraints, Computer Methods in Applied Mechanics and Engineering, 413 (2023) 116119.
[47] G. Zhang, K. Khandelwal, Gurson–Tvergaard–Needleman model guided fracture-resistant structural designs under finite deformations, International Journal for Numerical Methods in Engineering, 123 (2022) 3344-3388.
[48] P.-L. Liu, A. Der Kiureghian, Multivariate distribution models with prescribed marginals and covariances, Probabilistic Engineering Mechanics, 1 (1986) 105-112.
[49] R.G. Ghanem, P.D. Spanos, Stochastic Finite Elements: A Spectral Approach, Springer Science & Business Media, 2012.
[50] B. Sudret, A. Der Kiureghian, Stochastic finite element methods and reliability: a state-of-the-art report, Department of Civil and Environmental Engineering, University of California Berkeley, 2000.
[51] W. Betz, I. Papaioannou, D. Straub, Numerical methods for the discretization of random fields by means of the Karhunen–Loève expansion, Computer Methods in Applied Mechanics and Engineering, 271 (2014) 109-129.
[52] C.A. Schenk, G.I. Schuëller, Uncertainty assessment of large finite element systems, Springer Science & Business Media, 2005.
[53] Y. Sato, K. Izui, T. Yamada, S. Nishiwaki, M. Ito, N. Kogiso, Reliability-based topology optimization under shape uncertainty modeled in Eulerian description, Structural and Multidisciplinary Optimization, 59 (2019) 75-91.
[54] M. Kaminski, The stochastic perturbation method for computational mechanics, John Wiley & Sons, 2013.
[55] B. Pokusiński, M. Kamiński, Numerical convergence and error analysis for the truncated iterative generalized stochastic perturbation-based finite element method, Computer Methods in Applied Mechanics and Engineering, 410 (2023) 115993.
[56] K. Svanberg, The method of moving asymptotes—a new method for structural optimization, International Journal for Numerical Methods in Engineering, 24 (1987) 359-373.
[57] R. Alberdi, G. Zhang, L. Li, K. Khandelwal, A unified framework for nonlinear path-dependent sensitivity analysis in topology optimization, International Journal for Numerical Methods in Engineering, 115 (2018) 1-56.
[58] P. Michaleris, D.A. Tortorelli, C.A. Vidal, Tangent operators and design sensitivity formulations for transient non-linear coupled problems with applications to elastoplasticity, International Journal for Numerical Methods in Engineering, 37 (1994) 2471-2499.
[59] P. Wriggers, Nonlinear finite element methods, Springer Science & Business Media, 2008.
[60] K. Ikeda, K. Murota, Imperfect bifurcation in structures and materials, Springer Cham, 2019.
[61] G. Zhang, K. Khandelwal, T. Guo, Topology optimization of stability-constrained structures with simple/multiple eigenvalues, International Journal of Numerical Methods in Engineering, (2023).

## Appendix A : Derivatives for uncertainty quantification and sensitivity analysis

Table A-1 introduces simplified notations for scalar products and their derivatives, which will be adopted in the following derivations.

Table A-1. List of notations for scalar products and their derivatives.

| | | |
|---|---|---|
| $a_0 = (\gamma E)\vert_{\boldsymbol{\xi}=\mathbf{0}}$ | $b_0 \stackrel{\text{def}}{=} (\gamma^2 E)\vert_{\boldsymbol{\xi}=\mathbf{0}}$ | $c_0 \stackrel{\text{def}}{=} (E_L(1-\gamma^2))\vert_{\boldsymbol{\xi}=\mathbf{0}}$ |
| $a_k \stackrel{\text{def}}{=} \left.\frac{d(\gamma E)}{d\xi_k}\right\vert_{\boldsymbol{\xi}=\mathbf{0}}$ | $b_k \stackrel{\text{def}}{=} \left.\left(\gamma E \frac{d\gamma}{d\xi_k}\right)\right\vert_{\boldsymbol{\xi}=\mathbf{0}}$ | $c_k \stackrel{\text{def}}{=} \left.\frac{d(E_L(1-\gamma^2))}{d\xi_k}\right\vert_{\boldsymbol{\xi}=\mathbf{0}}$ |
| $a_{kl} \stackrel{\text{def}}{=} \left.\frac{d^2(\gamma E)}{d\xi_l d\xi_k}\right\vert_{\boldsymbol{\xi}=\mathbf{0}}$ | $b_{kl} \stackrel{\text{def}}{=} \left.\left(\gamma E \frac{d^2\gamma}{d\xi_l d\xi_k}\right)\right\vert_{\boldsymbol{\xi}=\mathbf{0}}$ | $c_{kl} \stackrel{\text{def}}{=} \left.\frac{d^2(E_L(1-\gamma^2))}{d\xi_l d\xi_k}\right\vert_{\boldsymbol{\xi}=\mathbf{0}}$ |
| $a_\rho \stackrel{\text{def}}{=} \left.\frac{\partial(\gamma E)}{\partial\hat{\rho}}\right\vert_{\boldsymbol{\xi}=\mathbf{0}}$ | $b_\rho \stackrel{\text{def}}{=} \left.\frac{\partial}{\partial\hat{\rho}}(\gamma^2 E)\right\vert_{\boldsymbol{\xi}=\mathbf{0}}$ | $c_\rho \stackrel{\text{def}}{=} \left.\frac{\partial}{\partial\hat{\rho}}\big(E_L(1-\gamma^2)\big)\right\vert_{\boldsymbol{\xi}=\mathbf{0}}$ |
| $a_{k\rho} \stackrel{\text{def}}{=} \left.\frac{\partial}{\partial\hat{\rho}}\left(\frac{\partial(\gamma E)}{\partial\xi_k}\right)\right\vert_{\boldsymbol{\xi}=\mathbf{0}}$ | $b_{k\rho} \stackrel{\text{def}}{=} \left.\frac{\partial}{\partial\hat{\rho}}\left(\gamma E \frac{\partial\gamma}{\partial\xi_k}\right)\right\vert_{\boldsymbol{\xi}=\mathbf{0}}$ | $c_{k\rho} \stackrel{\text{def}}{=} \left.\frac{\partial}{\partial\hat{\rho}}\left(\frac{\partial\big(E_L(1-\gamma^2)\big)}{\partial\xi_k}\right)\right\vert_{\boldsymbol{\xi}=\mathbf{0}}$ |
| $a_{kl\rho} \stackrel{\text{def}}{=} \left.\frac{\partial}{\partial\hat{\rho}}\left(\frac{d^2(\gamma E)}{d\xi_l d\xi_k}\right)\right\vert_{\boldsymbol{\xi}=\mathbf{0}}$ | | $c_{kl\rho} \stackrel{\text{def}}{=} \left.\frac{\partial}{\partial\hat{\rho}}\left(\left.\frac{\partial^2(E_L(1-\gamma^2))}{\partial\xi_l\partial\xi_k}\right\vert\right)\right\vert_{\boldsymbol{\xi}=\mathbf{0}}$ |
| $\gamma_0 \stackrel{\text{def}}{=} \gamma\vert_{\boldsymbol{\xi}=\mathbf{0}}$ | $\gamma_\rho \stackrel{\text{def}}{=} \left.\frac{\partial\gamma}{\partial\hat{\rho}}\right\vert_{\boldsymbol{\xi}=\mathbf{0}}$ | $\gamma_k \stackrel{\text{def}}{=} \left.\frac{d\gamma}{d\xi_k}\right\vert_{\boldsymbol{\xi}=\mathbf{0}}$ |

### *A.1 Derivative $\partial\not{f}_0/\partial\boldsymbol{x}$*

This derivative is given by

$$\frac{\partial\not{f}_0}{\partial\boldsymbol{x}} = \frac{\partial\mathbb{E}[f]}{\partial\boldsymbol{x}} + \alpha\frac{\partial\sqrt{Var[f]}}{\partial\boldsymbol{x}} \quad \text{with}$$

$$\frac{\partial\mathbb{E}[f]}{\partial\boldsymbol{x}} = \frac{\partial f^{[0]}}{\partial\boldsymbol{x}} + \frac{1}{2}\sum_{k=1}^{m}\left(\frac{\partial f_{kk}^{[2]}}{\partial\boldsymbol{x}}\right) \quad \text{and} \tag{A-1}$$

$$\frac{\partial\sqrt{Var[f]}}{\partial\boldsymbol{x}} = \frac{1}{2\sqrt{Var[f]}}\left(2\sum_{k=1}^{m} f_k^{[1]}\frac{\partial f_k^{[1]}}{\partial\boldsymbol{x}} + \sum_{k=1}^{m}\sum_{l=1}^{m} f_{kl}^{[2]}\frac{\partial f_{kl}^{[2]}}{\partial\boldsymbol{x}}\right)$$

Since $\partial f^{[0]}/\partial\boldsymbol{x} = \mathbf{0}$, $\partial f_k^{[1]}/\partial\boldsymbol{x} = \mathbf{0}$ and $\partial f_{kl}^{[2]}/\partial\boldsymbol{x} = \mathbf{0}$, it follows that

$$\frac{\partial \not{f}_0}{\partial \boldsymbol{x}} = \boldsymbol{0} \tag{A-2}$$

### A.2 Derivative $\partial \not{f}_0 / \partial \boldsymbol{u}^{[0]}$

This derivative can be calculated as

$$\frac{\partial \not{f}_0}{\partial \boldsymbol{u}^{[0]}} = \frac{\partial \mathbb{E}[f]}{\partial \boldsymbol{u}^{[0]}} + \alpha \frac{\partial \sqrt{Var[f]}}{\partial \boldsymbol{u}^{[0]}} \quad \text{with}$$

$$\frac{\partial \mathbb{E}[f]}{\partial \boldsymbol{u}^{[0]}} = \frac{\partial f^{[0]}}{\partial \boldsymbol{u}^{[0]}} + \frac{1}{2}\sum_{k=1}^{m}\left(\frac{\partial f_{kk}^{[2]}}{\partial \boldsymbol{u}^{[0]}}\right) = \boldsymbol{F}_{\text{ext}}^{[0]^T} + \frac{1}{2}\sum_{k=1}^{m} \boldsymbol{F}_{\text{ext}(kk)}^{[2]^T} \quad \text{and} \tag{A-3}$$

$$\frac{\partial \sqrt{Var[f]}}{\partial \boldsymbol{u}^{[0]}} = \frac{1}{2\sqrt{Var[f]}}\left(2\sum_{k=1}^{m} f_k^{[1]} \boldsymbol{F}_{\text{ext}(k)}^{[1]^T} + \sum_{k=1}^{m}\sum_{l=1}^{m} f_{kl}^{[2]} \boldsymbol{F}_{\text{ext}(kl)}^{[2]^T}\right)$$

### A.3 Derivative $\partial \not{f}_0 / \partial \boldsymbol{u}_k^{[1]}$

This derivative can be written as

$$\frac{\partial \not{f}_0}{\partial \boldsymbol{u}_k^{[1]}} = \frac{\partial \mathbb{E}[f]}{\partial \boldsymbol{u}_k^{[1]}} + \alpha \frac{1}{2\sqrt{Var[f]}} \frac{\partial Var[f]}{\partial \boldsymbol{u}_k^{[1]}} \quad \text{with}$$

$$\frac{\partial \mathbb{E}[f]}{\partial \boldsymbol{u}_k^{[1]}} = \boldsymbol{F}_{\text{ext}(k)}^{[1]^T} \quad \text{and} \tag{A-4}$$

$$\frac{\partial Var[f]}{\partial \boldsymbol{u}_k^{[1]}} = 2 f_k^{[1]} \boldsymbol{F}_{\text{ext}}^{[0]^T} + \sum_{p=1}^{m}\sum_{q=1}^{m} f_{pq}^{[2]} \left(\delta_{qk} \boldsymbol{F}_{\text{ext}(p)}^{[1]^T} + \delta_{pk} \boldsymbol{F}_{\text{ext}(q)}^{[1]^T}\right)$$

### A.4 Derivative $\partial \not{f}_0 / \partial \boldsymbol{u}_{kl}^{[2]}$

This derivative can be written as

$$\frac{\partial \not{f}_0}{\partial \boldsymbol{u}_{kl}^{[2]}} = \frac{\partial \mathbb{E}[f]}{\partial \boldsymbol{u}_{kl}^{[2]}} + \alpha \frac{1}{2\sqrt{Var[f]}} \frac{\partial Var[f]}{\partial \boldsymbol{u}_{kl}^{[2]}} \quad \text{with}$$

$$\frac{\partial \mathbb{E}[f]}{\partial \boldsymbol{u}_{kl}^{[2]}} = \frac{1}{2} \boldsymbol{F}_{\text{ext}}^{[0]^T} \delta_{kl} \tag{A-5}$$

$$\frac{\partial Var[f]}{\partial \boldsymbol{u}_{kl}^{[2]}} = f_{kl}^{[2]} \boldsymbol{F}_{\text{ext}}^{[0]^T}$$

### A.5 Derivative $\partial R^{[0]}/\partial x$

This derivative is calculated as

$$\frac{\partial \boldsymbol{R}^{[0]}}{\partial \boldsymbol{x}} = \frac{\partial \boldsymbol{F}_{int}^{[0]}}{\partial \hat{\boldsymbol{\rho}}} \boldsymbol{W}$$

$$\text{with} \quad \frac{\partial \boldsymbol{F}_{int}^{[0]}}{\partial \hat{\boldsymbol{\rho}}} = \left[ \dots \quad \frac{\partial \boldsymbol{F}_{int}^{[0]}}{\partial \hat{\rho}_e} \quad \dots \right], \quad e = 1, \dots, n_{ele} \tag{A-6}$$

$$\text{and} \quad \frac{\partial \boldsymbol{F}_{int}^{[0]}}{\partial \hat{\rho}_e} = \underset{e=1}{\overset{n_{ele}}{\mathcal{A}}} \frac{\partial \boldsymbol{F}_{int}^{e[0]}}{\partial \hat{\rho}_e}$$

where the derivatives of the element's internal force are

$$\frac{\partial \boldsymbol{F}_{int}^{e[0]}}{\partial \hat{\rho}_e} = \frac{\partial (E\gamma)}{\partial \hat{\rho}_e} \boldsymbol{p}^e + E\gamma \frac{\partial \gamma}{\partial \hat{\rho}_e} \boldsymbol{k}_1^e . \boldsymbol{u}^e + \frac{\partial \left( E_L (1 - \gamma^2) \right)}{\partial \hat{\rho}_e} \boldsymbol{k}_L^e . \boldsymbol{u}^e \tag{A-7}$$

where the scalar derivatives are given in Appendix A.11.

### A.6 Derivatives $\partial R_k^{[1]}/\partial x$

These derivatives are computed by

$$\frac{\partial \boldsymbol{R}_k^{[1]}}{\partial \boldsymbol{x}} = \frac{\partial \boldsymbol{F}_{int(k)}^{[1]}}{\partial \hat{\boldsymbol{\rho}}} \boldsymbol{W}$$

$$\text{with} \quad \frac{\partial \boldsymbol{F}_{int(k)}^{[1]}}{\partial \hat{\boldsymbol{\rho}}} = \left[ \dots \quad \frac{\partial \boldsymbol{F}_{int(k)}^{[1]}}{\partial \hat{\rho}_e} \quad \dots \right], \quad e = 1 \dots, n_{ele} \tag{A-8}$$

$$\text{and} \quad \frac{\partial \boldsymbol{F}_{int(k)}^{[1]}}{\partial \hat{\rho}_e} = \underset{e=1}{\overset{n_{ele}}{\mathcal{A}}} \frac{\partial \boldsymbol{F}_{int(k)}^{e[1]}}{\partial \hat{\rho}_e}$$

where

$$\begin{aligned} \frac{\partial \boldsymbol{F}_{int(k)}^{e[1]}}{\partial \hat{\rho}_e} = a_{k\rho} \boldsymbol{p}^e + \boldsymbol{k}_1^e . \left( \left[ a_k \gamma_\rho + b_{k\rho} \right] \boldsymbol{u}^{e[0]} + b_\rho \boldsymbol{u}_k^{e[1]} \right) \\ + \boldsymbol{k}_2^e : \left( b_k \gamma_\rho \boldsymbol{u}^{e[0]} \otimes \boldsymbol{u}^{e[0]} + b_0 \gamma_\rho \boldsymbol{u}^{e[0]} \otimes \boldsymbol{u}_k^{e[1]} \right) \\ + \boldsymbol{k}_L^e . \left( c_{k\rho} \boldsymbol{u}^{e[0]} + c_\rho \boldsymbol{u}_k^{e[1]} \right) \end{aligned} \tag{A-9}$$

### A.7 Derivatives $\partial \boldsymbol{R}_k^{[1]} / \partial \boldsymbol{u}^{[0]}$

These derivatives are computed by

$$\frac{\partial \boldsymbol{R}_k^{[1]}}{\partial \boldsymbol{u}^{[0]}} = \frac{\partial \boldsymbol{F}_{int(k)}^{[1]}}{\partial \boldsymbol{u}^{[0]}} = \mathop{\mathcal{A}}_{e=1}^{n_{ele}} \frac{\partial \boldsymbol{F}_{int(k)}^{e[1]}}{\partial \boldsymbol{u}^{e[0]}} \tag{A-10}$$

where the element term is calculated as

$$\frac{\partial \boldsymbol{F}_{int(k)}^{e[1]}}{\partial \boldsymbol{u}^{e[0]}} = (\gamma_0 a_k + b_k)\boldsymbol{k}_1^e + b_0 \boldsymbol{k}_2^e . \boldsymbol{v}_k + c_k \boldsymbol{k}_L^e \tag{A-11}$$

### A.8 Derivatives $\partial \boldsymbol{R}_{kl}^{[2]} / \partial \boldsymbol{x}$

These derivatives are given by

$$\frac{\partial \boldsymbol{R}_{kl}^{[2]}}{\partial \boldsymbol{x}} = \frac{\partial \boldsymbol{F}_{int(kl)}^{[2]}}{\partial \widehat{\boldsymbol{\rho}}} \boldsymbol{W}$$

with $$\frac{\partial \boldsymbol{F}_{int(kl)}^{[2]}}{\partial \widehat{\boldsymbol{\rho}}} = \left[ \dots \quad \frac{\partial \boldsymbol{F}_{int(kl)}^{[2]}}{\partial \hat{\rho}_e} \quad \dots \right], \quad e = 1, \dots, n_{ele} \tag{A-12}$$

and $$\frac{\partial \boldsymbol{F}_{int(kl)}^{[2]}}{\partial \hat{\rho}_e} = \mathop{\mathcal{A}}_{e=1}^{n_{ele}} \frac{\partial \boldsymbol{F}_{int(kl)}^{e[2]}}{\partial \hat{\rho}_e}$$

in which

$$\begin{aligned} \frac{\partial \boldsymbol{F}_{int(kl)}^{e[2]}}{\partial \hat{\rho}_e} = {} & a_{kl\rho} \boldsymbol{p}_0^e + a_0 \gamma_\rho \boldsymbol{k}_{30}^e \\ & \vdots \left(\boldsymbol{u}^{e[0]} \otimes \boldsymbol{v}_k \otimes \boldsymbol{v}_l\right) \\ & + \boldsymbol{k}_1^e . \left(a_{kl}\gamma_\rho \boldsymbol{u}^{e[0]} + a_{k\rho}\boldsymbol{v}_l + a_k \boldsymbol{v}_{l\rho} + a_{l\rho}\boldsymbol{v}_k + a_l \boldsymbol{v}_{k\rho} + a_\rho \boldsymbol{v}_{kl} + a_0 \boldsymbol{v}_{kl\rho}\right) \\ & + \boldsymbol{k}_2^e : \left\{a_k \gamma_\rho \boldsymbol{u}^{e[0]} \otimes \boldsymbol{v}_l + a_l \gamma_\rho \boldsymbol{u}^{e[0]} \otimes \boldsymbol{v}_k + a_\rho (\boldsymbol{v}_k \otimes \boldsymbol{v}_l)\right. \\ & \left. + a_0\left[\boldsymbol{v}_{l\rho} \otimes \boldsymbol{v}_k + \boldsymbol{v}_{k\rho} \otimes \boldsymbol{v}_l + \gamma_\rho\left(\boldsymbol{u}^{e[0]} \otimes \boldsymbol{v}_{kl}\right)\right]\right\} + c_{kl\rho}\boldsymbol{k}_L^e . \boldsymbol{u}^{e[0]} + c_{k\rho}\boldsymbol{k}_L^e . \boldsymbol{u}_l^{e[1]} \\ & + c_{l\rho}\boldsymbol{k}_L^e . \boldsymbol{u}_k^{e[1]} + c_\rho \boldsymbol{k}_L^e . \boldsymbol{u}_{kl}^{e[2]} \end{aligned} \tag{A-13}$$

with

$$\boldsymbol{v}_{k\rho} \stackrel{\text{def}}{=} \left(\frac{\partial^2\gamma}{\partial\hat{\rho}\partial\xi_k}\right)\Bigg|_{\boldsymbol{\xi}=\boldsymbol{0}} \boldsymbol{u}^{e[0]} + \left(\frac{\partial\gamma}{\partial\hat{\rho}}\right)\Bigg|_{\boldsymbol{\xi}=\boldsymbol{0}} \boldsymbol{u}_k^{e[1]}$$

$$\boldsymbol{v}_{kl\rho} \stackrel{\text{def}}{=} \left(\frac{\partial^3\gamma}{\partial\hat{\rho}\partial\xi_l\partial\xi_k}\right)\Bigg|_{\boldsymbol{\xi}=\boldsymbol{0}} \boldsymbol{u}^{e[0]} + \left(\frac{\partial^2\gamma}{\partial\hat{\rho}\partial\xi_k}\right)\Bigg|_{\boldsymbol{\xi}=\boldsymbol{0}} \boldsymbol{u}_l^{e[1]} + \left(\frac{\partial^2\gamma}{\partial\hat{\rho}d\xi_l}\right)\Bigg|_{\boldsymbol{\xi}=\boldsymbol{0}} \boldsymbol{u}_k^{e[1]} + \left(\frac{\partial\gamma}{\partial\hat{\rho}}\right)\Bigg|_{\boldsymbol{\xi}=\boldsymbol{0}} \boldsymbol{u}_{kl}^{e[2]} \tag{A-14}$$

where the scalar derivatives can be found in Appendix A.11.

### A.9 Derivatives $\partial\boldsymbol{R}_{\boldsymbol{kl}}^{[\boldsymbol{2}]}/\partial\boldsymbol{u}^{[\boldsymbol{0}]}$

These derivatives are computed as

$$\frac{\partial\boldsymbol{R}_{kl}^{[2]}}{\partial\boldsymbol{u}^{[0]}} = \frac{\partial\boldsymbol{F}_{int(kl)}^{[2]}}{\partial\boldsymbol{u}^{[0]}} = \mathop{\mathcal{A}}_{e=1}^{n_{ele}} \frac{\partial\boldsymbol{F}_{int(kl)}^{e[2]}}{\partial\boldsymbol{u}^{e[0]}} \tag{A-15}$$

where

$$\frac{\partial\boldsymbol{F}_{int(kl)}^{e[2]}}{\partial\boldsymbol{u}^{e[0]}} = (a_{kl}\gamma_0 + a_k\gamma_l + a_l\gamma_k + b_{kl})\boldsymbol{k}_{10}^e + \boldsymbol{k}_2^e.\big((a_k\gamma_0 + b_k)\boldsymbol{v}_l + (a_l\gamma_0 + b_l)\boldsymbol{v}_k + a_0\boldsymbol{v}_{kl}\big) + a_0\boldsymbol{k}_3^e{:}(\boldsymbol{v}_k \otimes \boldsymbol{v}_l) + c_{kl}\boldsymbol{k}_L^e \tag{A-16}$$

### A.10 Derivatives $\partial\boldsymbol{R}_{\boldsymbol{kl}}^{[\boldsymbol{2}]}/\partial\boldsymbol{u}_{\boldsymbol{q}}^{[\boldsymbol{1}]}$

These derivatives are computed by

$$\frac{\partial\boldsymbol{R}_{kl}^{[2]}}{\partial\boldsymbol{u}_q^{[1]}} = \mathop{\mathcal{A}}_{e=1}^{n_{ele}} \frac{\partial\boldsymbol{F}_{int(kl)}^{e[2]}}{\partial\boldsymbol{u}_q^{e[1]}} \quad \text{with} \quad \frac{\partial\boldsymbol{F}_{int(kl)}^{e[2]}}{\partial\boldsymbol{u}_q^{e[1]}} = \boldsymbol{0} \quad \text{if } q \neq k \text{ or } q \neq l$$

where

$$\frac{\partial\boldsymbol{F}_{int(kl)}^{e[2]}}{\partial\boldsymbol{u}_k^{e[1]}} = (a_l\gamma_0 + b_l)\boldsymbol{k}_1^e + a_0\boldsymbol{k}_2^e.\boldsymbol{v}_l + c_l\boldsymbol{k}_L^e \tag{A-17}$$

$$\frac{\partial \boldsymbol{F}^{e[2]}_{int(kl)}}{\partial \boldsymbol{u}^{e[1]}_{l}} = (a_k\gamma_0 + b_k)\boldsymbol{k}^e_1 + a_0\boldsymbol{k}^e_2.\boldsymbol{v}_k + c_k\boldsymbol{k}^e_L$$

### *A.11 Scalar derivatives*

This subsection gives derivatives of the scalars $E$, $\gamma$, and $\rho$, and the derivatives of scalar products in Table A-1 can be calculated by the chain rule of differentiation.

#### A.11.1 Derivatives of $E$

The derivative $\partial E/\partial \xi_i$ reads

$$\frac{\partial E}{\partial \xi_k} = [\epsilon + (1-\epsilon)\rho^p]\frac{\partial E_0}{\partial \xi_k} + p(1-\epsilon)\rho^{p-1}E_0\frac{\partial \rho}{\partial \xi_k} \quad \text{with}$$

$$\frac{\partial E_0}{\partial \xi_k} = E_0\sigma \exp\left(\frac{(\ln E_0 - \mu)^2}{2\sigma^2} - \frac{Z^2}{2}\right)\frac{\partial Z}{\partial \xi_k} \tag{A-18}$$

where $\partial Z/\partial \xi_k$ will be nonzero only if $\xi_k \in \boldsymbol{\xi}_2$. $\mu$ and $\sigma^2$ are the mean and variance of $\ln E_0$, respectively, and read

$$\mu = \ln\left(\frac{\mu_0^2}{\sqrt{\mu_0^2 + \sigma_0^2}}\right) \quad \text{and} \quad \sigma^2 = \ln\left(1 + \frac{\sigma_0^2}{\mu_0^2}\right) \tag{A-19}$$

The derivative $\partial^2 E/\partial \xi_k \partial \xi_l$ are obtained by

$$\frac{\partial^2 E}{\partial \xi_l \partial \xi_k} = [\epsilon + (1-\epsilon)\rho^p]\frac{\partial^2 E_0}{\partial \xi_l \partial \xi_k} + p(1-\epsilon)\rho^{p-1}\frac{\partial E_0}{\partial \xi_k}\frac{\partial \rho}{\partial \xi_l}$$
$$+ p(1-\epsilon)\rho^{p-1}E_0\frac{\partial^2 \rho}{\partial \xi_l \partial \xi_k} + p(1-\epsilon)\rho^{p-1}\frac{\partial E_0}{\partial \xi_l}\frac{\partial \rho}{\partial \xi_k} \tag{A-20}$$
$$+ p(p-1)(1-\epsilon)\rho^{p-2}E_0\frac{\partial \rho}{\partial \xi_k}\frac{\partial \rho}{\partial \xi_l} \quad \text{with}$$

$$\frac{\partial^2 E_0}{\partial \xi_l \partial \xi_k} = (\sigma^2 + \ln E_0 - \mu) E_0 \exp\left(\frac{(\ln E_0 - \mu)^2}{\sigma^2} - Z^2\right) \frac{\partial Z}{\partial \xi_k} \frac{\partial Z}{\partial \xi_l}$$

$$- \sigma E_0 Z \exp\left(\frac{(\ln E_0 - \mu)^2}{2\sigma^2} - \frac{Z^2}{2}\right) \frac{\partial Z}{\partial \xi_k} \frac{\partial Z}{\partial \xi_l}$$

The derivative $\partial E/\partial \rho$ is computed by

$$\frac{\partial E}{\partial \rho} = p(1-\epsilon)\rho^{p-1} E_0 \tag{A-21}$$

The derivative $\partial^2 E/\partial \xi_k \partial \hat{\rho}$ is computed as

$$\frac{\partial^2 E}{\partial \xi_k \partial \hat{\rho}} = p(1-\epsilon)\rho^{p-1} E_0 \frac{\partial^2 \rho}{\partial \xi_k \partial \hat{\rho}} + p(1-\epsilon)\rho^{p-1} \frac{\partial E_0}{\partial \xi_k} \frac{\partial \rho}{\partial \hat{\rho}}$$

$$+ p(p-1)(1-\epsilon)\rho^{p-2} E_0 \frac{\partial \rho}{\partial \xi_k} \frac{\partial \rho}{\partial \hat{\rho}} \tag{A-22}$$

The 3rd order derivative $\partial^3 E/\partial \xi_k \partial \xi_l \partial \hat{\rho}$ is obtained by

$$\frac{\partial^3 E}{\partial \xi_k \partial \xi_l \partial \hat{\rho}} = p(1-\epsilon)\rho^{p-1} \left( E_0 \frac{\partial^3 \rho}{\partial \xi_k \partial \xi_l \partial \hat{\rho}} + \frac{\partial E_0}{\partial \xi_l} \frac{\partial^2 \rho}{\partial \xi_k \partial \hat{\rho}} + \frac{\partial E_0}{\partial \xi_k} \frac{\partial^2 \rho}{\partial \xi_l \partial \hat{\rho}} + \frac{\partial^2 E_0}{\partial \xi_k \partial \xi_l} \frac{\partial \rho}{\partial \hat{\rho}} \right)$$

$$+ p(p-1)(1-\epsilon)\rho^{p-2} \left( E_0 \frac{\partial \rho}{\partial \xi_l} \frac{\partial^2 \rho}{\partial \xi_k \partial \hat{\rho}} + \frac{\partial E_0}{\partial \xi_k} \frac{\partial \rho}{\partial \xi_l} \frac{\partial \rho}{\partial \hat{\rho}} + E_0 \frac{\partial^2 \rho}{\partial \xi_k \partial \xi_l} \frac{\partial \rho}{\partial \hat{\rho}} \right. \tag{A-23}$$

$$\left. + E_0 \frac{\partial \rho}{\partial \xi_k} \frac{\partial^2 \rho}{\partial \xi_l \partial \hat{\rho}} + \frac{\partial E_0}{\partial \xi_l} \frac{\partial \rho}{\partial \xi_k} \frac{\partial \rho}{\partial \hat{\rho}} \right) + p(p-1)(p-2)(1-\epsilon)\rho^{p-3} E_0 \frac{\partial \rho}{\partial \xi_l} \frac{\partial \rho}{\partial \xi_k} \frac{\partial \rho}{\partial \hat{\rho}}$$

## A.11.2 Derivatives of $\gamma$

The derivative $\partial \gamma/\partial \rho$ reads

$$\frac{\partial \gamma}{\partial \rho} = \frac{\beta_0 \exp\big(\beta_0(\rho + c)\big)}{(\exp(c\beta_0) + \exp(\beta_0 \rho))^2} \tag{A-24}$$

The derivative $\partial^2 \gamma/\partial \rho^2$ is computed as

$$\frac{\partial^2\gamma}{\partial\rho^2} = \frac{\beta_0^2 \exp\big(\beta_0(\rho + c)\big)\,(\exp(c\beta_0) - \exp(\beta_0\rho))}{(\exp(c\beta_0) + \exp(\beta_0\rho))^3} \qquad \text{(A-25)}$$

The derivative $\partial^3\gamma/\partial\rho^3$ is computed as

$$\frac{\partial^3\gamma}{\partial\rho^3} = \frac{\beta_0^3 \exp\big(\beta_0(\rho + c)\big)\left(\exp(2c\beta_0) + \exp(2\beta_0\rho) - 4\exp\big(\beta_0(c + \rho)\big)\right)}{(\exp(c\beta_0) + \exp(\beta_0\rho))^4} \qquad \text{(A-26)}$$

### A.11.3 Derivatives of $\rho$

The derivative $\partial\rho/\partial\xi_k$ is calculated as

$$\frac{\partial\rho}{\partial\xi_k} = \frac{\beta\big[\tanh(\beta\eta) + \tanh\big(\beta(\hat{\rho} - \eta)\big)\big]\big[\tanh\big(\beta(\hat{\rho} - \eta)\big) - \tanh\big(\beta(1 - \eta)\big)\big]}{\tanh(\beta\eta) + \tanh\big(\beta(1 - \eta)\big)}(\eta_{max} - \eta_{min})\frac{1}{\sqrt{2\pi}}e^{-\frac{\bar{Z}^2}{2}}\frac{\partial\bar{Z}}{\partial\xi_k} \qquad \text{(A-27)}$$

where $\partial\bar{Z}/\partial\xi_k$ will be nonzero only if $\xi_k \in \boldsymbol{\xi}_3$.

The derivative $\partial^2\rho/\partial\xi_k\partial\xi_l$ is obtained by

$$\frac{\partial^2\rho}{\partial\xi_k\partial\xi_l} = \frac{2\beta^2\tanh\big(\beta(\hat{\rho} - \eta)\big)\big[\tanh(\beta\eta) + \tanh\big(\beta(\hat{\rho} - \eta)\big)\big]\big[\tanh\big(\beta(\hat{\rho} - \eta)\big) - \tanh\big(\beta(1 - \eta)\big)\big]}{\tanh(\beta\eta) + \tanh\big(\beta(1 - \eta)\big)}(\eta_{max} - \eta_{min})^2\frac{1}{2\pi}e^{-\bar{Z}^2}\frac{\partial\bar{Z}}{\partial\xi_k}\frac{\partial\bar{Z}}{\partial\xi_l} \qquad \text{(A-28)}$$

The derivative $\partial\rho/\partial\hat{\rho}$ is computed as

$$\frac{\partial\rho}{\partial\hat{\rho}} = \frac{\beta\big[1 - \tanh^2\big(\beta(\hat{\rho} - \eta)\big)\big]}{\tanh(\beta\eta) + \tanh\big(\beta(1 - \eta)\big)} \qquad \text{(A-29)}$$

where

$$\frac{\partial^2 \rho}{\partial \xi_k \partial \hat{\rho}} = \frac{\beta^2\left[1-\tanh^2\left(\beta(\hat{\rho}-\eta)\right)\right]\left[2\tanh\left(\beta(\hat{\rho}-\eta)\right)+\tanh(\beta\eta)-\tanh\left(\beta(1-\eta)\right)\right]}{\tanh(\beta\eta)+\tanh\left(\beta(1-\eta)\right)}(\eta_{max} - \eta_{min})\frac{1}{\sqrt{2\pi}}e^{-\frac{\bar{Z}^2}{2}}\frac{\partial \bar{Z}}{\partial \xi_k} \quad \text{(A-30)}$$

where

$$\frac{\partial^3 \rho}{\partial \xi_k \partial \xi_l \partial \hat{\rho}} = \frac{\partial \rho}{\partial \hat{\rho}} 2\beta^2\left\{\left[\tanh(\beta\eta)+2\tanh\left(\beta(\hat{\rho}-\eta)\right)\right]\left[\tanh\left(\beta(\hat{\rho}-\eta)\right) - \tanh\left(\beta(1-\eta)\right)\right] + \tanh\left(\beta(\hat{\rho}-\eta)\right)\left[\tanh(\beta\eta)+\tanh\left(\beta(\hat{\rho}-\eta)\right)\right]\right\}(\eta_{max} - \eta_{min})^2\frac{1}{2\pi}e^{-\bar{Z}^2}\frac{\partial \bar{Z}}{\partial \xi_k}\frac{\partial \bar{Z}}{\partial \xi_l} \quad \text{(A-31)}$$

## Appendix B : Energy function and its derivatives w.r.t. deformation gradient

### *B.1 1st PK Stress*

The 1st PK stress reads

$$\boldsymbol{P} = \frac{\partial \psi(\boldsymbol{C})}{\partial \boldsymbol{F}} = \kappa J(J-1)\boldsymbol{F}^{-T} + \mu J^{-2/3}\left(-\frac{1}{3}\boldsymbol{F}^{-T}\mathrm{tr}(\boldsymbol{C}) + \boldsymbol{F}\right) \tag{B-1}$$

### *B.2 Tangent moduli* $\mathbb{A}_{\mathbf{4}} = \boldsymbol{\partial P}/\boldsymbol{\partial F}$

The material tangent $\mathbb{A}_4$ is computed as

$$\begin{aligned}\mathbb{A}_4 = \frac{\partial^2 \psi}{\partial \boldsymbol{F} \partial \boldsymbol{F}} &= \left[\kappa(J^2 - J) - \frac{1}{3}\mu J^{-\frac{2}{3}}\mathrm{tr}(\boldsymbol{C})\right]\frac{\partial \boldsymbol{F}^{-T}}{\partial \boldsymbol{F}} \\ &+ \left[\kappa(2J^2 - J) + \frac{2}{9}\mu J^{-\frac{2}{3}}\mathrm{tr}(\boldsymbol{C})\right]\boldsymbol{F}^{-T} \otimes \boldsymbol{F}^{-T} - \frac{2}{3}\mu J^{-\frac{2}{3}}\boldsymbol{F}^{-T} \otimes \boldsymbol{F} \\ &- \frac{2}{3}\mu J^{-\frac{2}{3}}\boldsymbol{F} \otimes \boldsymbol{F}^{-T} + \mu J^{-\frac{2}{3}}\mathbb{I}_4 \quad \text{with}\end{aligned} \tag{B-2}$$

$$\frac{\partial \boldsymbol{F}^{-T}}{\partial \boldsymbol{F}} = -\boldsymbol{F}^{-T} \boxdot \boldsymbol{F}^{-1}$$

where the operator $\boxdot$ is defined as $(\boldsymbol{A} \boxdot \boldsymbol{B})_{ijkl} \stackrel{\mathrm{def}}{=} A_{il}B_{jk}$ in which $\boldsymbol{A}$ and $\boldsymbol{B}$ are second-order tensors.

### *B.3 Sixth order tensor* $\mathbb{A}_{\mathbf{6}} = \boldsymbol{\partial^2 P}/\boldsymbol{\partial F \partial F}$

The 2nd order derivative of $\boldsymbol{P}$ with respect to deformation gradient $\boldsymbol{F}$ is computed as follows

$$\mathbb{A}_6 = \frac{\partial^2 \boldsymbol{P}}{\partial \boldsymbol{F} \partial \boldsymbol{F}} = \left[\kappa(J^2 - J) - \frac{1}{3}\mu J^{-2/3}\mathrm{tr}(\boldsymbol{C})\right]\bar{\mathcal{T}} - \frac{2}{3}\mu J^{-\frac{2}{3}}\frac{\partial \boldsymbol{F}^{-T}}{\partial \boldsymbol{F}} \otimes \boldsymbol{F}$$

$$+ \left[\kappa(2J^2 - J) + \frac{2}{9}\mu J^{-\frac{2}{3}}\mathrm{tr}(\boldsymbol{C})\right]\frac{\partial \boldsymbol{F}^{-T}}{\partial \boldsymbol{F}} \otimes \boldsymbol{F}^{-T}$$

$$+ \left[\kappa(2J^2 - J) + \frac{2}{9}\mu J^{-\frac{2}{3}}\mathrm{tr}(\boldsymbol{C})\right]\frac{\partial (\boldsymbol{F}^{-T} \otimes \boldsymbol{F}^{-T})}{\partial \boldsymbol{F}} + \frac{4}{9}\mu J^{-\frac{2}{3}}\boldsymbol{F}^{-T} \otimes \boldsymbol{F}^{-T} \otimes \boldsymbol{F}$$

$$+ \left[\kappa(4J^2 - J) - \frac{4}{27}\mu J^{-\frac{2}{3}}\mathrm{tr}(\boldsymbol{C})\right]\boldsymbol{F}^{-T} \otimes \boldsymbol{F}^{-T} \otimes \boldsymbol{F}^{-T} - \frac{2}{3}\mu J^{-\frac{2}{3}}\frac{\partial (\boldsymbol{F}^{-T} \otimes \boldsymbol{F})}{\partial \boldsymbol{F}}$$

$$+ \frac{4}{9}\mu J^{-\frac{2}{3}}\boldsymbol{F}^{-T} \otimes \boldsymbol{F} \otimes \boldsymbol{F}^{-T} - \frac{2}{3}\mu J^{-\frac{2}{3}}\frac{\partial (\boldsymbol{F} \otimes \boldsymbol{F}^{-T})}{\partial \boldsymbol{F}} + \frac{4}{9}\mu J^{-\frac{2}{3}}\boldsymbol{F} \otimes \boldsymbol{F}^{-T} \otimes \boldsymbol{F}^{-T}$$

$$- \frac{2}{3}\mu J^{-\frac{2}{3}}\mathbb{I}_4 \otimes \boldsymbol{F}^{-T}$$

where (B-3)

$$\left[\frac{\partial (\boldsymbol{F}^{-T} \otimes \boldsymbol{F}^{-T})}{\partial \boldsymbol{F}}\right]_{ijklmn} = -F_{lk}^{-1}F_{jm}^{-1}F_{ni}^{-1} - F_{ji}^{-1}F_{lm}^{-1}F_{nk}^{-1} \qquad \text{and}$$

$$\bar{\mathcal{T}}_{ijklmn} = F_{jk}^{-1}F_{lm}^{-1}F_{ni}^{-1} + F_{li}^{-1}F_{jm}^{-1}F_{nk}^{-1} \qquad \text{and}$$

$$\left[\frac{\partial (\boldsymbol{F} \otimes \boldsymbol{F}^{-T})}{\partial \boldsymbol{F}}\right]_{ijklmn} = \delta_{im}\delta_{jn}F_{lk}^{-1} - F_{ij}F_{lm}^{-1}F_{nk}^{-1}$$

$$\left[\frac{\partial (\boldsymbol{F}^{-T} \otimes \boldsymbol{F})}{\partial \boldsymbol{F}}\right]_{ijklmn} = \delta_{km}\delta_{ln}F_{ji}^{-1} - F_{kl}F_{ni}^{-1}F_{jm}^{-1}$$

### B.4 Eighth order tensor $\mathbb{A}_8 = \partial^3 \boldsymbol{P}/\partial \boldsymbol{F}\partial \boldsymbol{F}\partial \boldsymbol{F}$

By differentiating Eq. (B-3), the 3rd order partial derivative $\partial^3 \boldsymbol{P}/\partial \boldsymbol{F}\partial \boldsymbol{F}\partial \boldsymbol{F}$ is obtained by

$$\mathbb{A}_8 = \frac{\partial^3 \boldsymbol{P}}{\partial \boldsymbol{F} \partial \boldsymbol{F} \partial \boldsymbol{F}} = \left[\kappa(J^2 - J) - \frac{1}{3}\mu J^{-2/3}\mathrm{tr}(\boldsymbol{C})\right]\frac{\partial \bar{\mathcal{T}}}{\partial \boldsymbol{F}} + \left[\kappa(2J^2 - J) + \frac{2}{9}\mu J^{-2/3}\mathrm{tr}(\boldsymbol{C})\right]\bar{\mathcal{T}} \otimes \boldsymbol{F}^{-T}$$
$$- \frac{2}{3}\mu J^{-\frac{2}{3}}\bar{\mathcal{T}} \otimes \boldsymbol{F} - \frac{2}{3}\mu J^{-\frac{2}{3}}\frac{\partial}{\partial \boldsymbol{F}}\left(\frac{\partial \boldsymbol{F}^{-T}}{\partial \boldsymbol{F}} \otimes \boldsymbol{F}\right) + \frac{4}{9}\mu J^{-\frac{2}{3}}\frac{\partial \boldsymbol{F}^{-T}}{\partial \boldsymbol{F}} \otimes \boldsymbol{F} \otimes \boldsymbol{F}^{-T}$$
$$+ \left[\kappa(2J^2 - J) + \frac{2}{9}\mu J^{-\frac{2}{3}}\mathrm{tr}(\boldsymbol{C})\right]\frac{\partial}{\partial \boldsymbol{F}}\left(\frac{\partial \boldsymbol{F}^{-T}}{\partial \boldsymbol{F}} \otimes \boldsymbol{F}^{-T}\right)$$
$$+ \left[\kappa(4J^2 - J) - \frac{4}{27}\mu J^{-\frac{2}{3}}\mathrm{tr}(\boldsymbol{C})\right]\frac{\partial \boldsymbol{F}^{-T}}{\partial \boldsymbol{F}} \otimes \boldsymbol{F}^{-T} \otimes \boldsymbol{F}^{-T} + \frac{4}{9}\mu J^{-\frac{2}{3}}\frac{\partial \boldsymbol{F}^{-T}}{\partial \boldsymbol{F}} \otimes \boldsymbol{F}^{-T} \otimes \boldsymbol{F}$$
$$+ \left[\kappa(2J^2 - J) + \frac{2}{9}\mu J^{-\frac{2}{3}}\mathrm{tr}(\boldsymbol{C})\right]\frac{\partial^2(\boldsymbol{F}^{-T} \otimes \boldsymbol{F}^{-T})}{\partial \boldsymbol{F} \partial \boldsymbol{F}}$$
$$+ \left[\kappa(4J^2 - J) - \frac{4}{27}\mu J^{-\frac{2}{3}}\mathrm{tr}(\boldsymbol{C})\right]\frac{\partial(\boldsymbol{F}^{-T} \otimes \boldsymbol{F}^{-T})}{\partial \boldsymbol{F}} \otimes \boldsymbol{F}^{-T} + \frac{4}{9}\mu J^{-\frac{2}{3}}\frac{\partial(\boldsymbol{F}^{-T} \otimes \boldsymbol{F}^{-T})}{\partial \boldsymbol{F}}$$
$$\otimes \boldsymbol{F} + \frac{4}{9}\mu J^{-\frac{2}{3}}\frac{\partial(\boldsymbol{F}^{-T} \otimes \boldsymbol{F}^{-T} \otimes \boldsymbol{F})}{\partial \boldsymbol{F}} - \frac{8}{27}\mu J^{-\frac{2}{3}}\boldsymbol{F}^{-T} \otimes \boldsymbol{F}^{-T} \otimes \boldsymbol{F} \otimes \boldsymbol{F}^{-T}$$
$$+ \left[\kappa(4J^2 - J) - \frac{4}{27}\mu J^{-\frac{2}{3}}\mathrm{tr}(\boldsymbol{C})\right]\frac{\partial(\boldsymbol{F}^{-T} \otimes \boldsymbol{F}^{-T} \otimes \boldsymbol{F}^{-T})}{\partial \boldsymbol{F}}$$
$$+ \left[\kappa(8J^2 - J) + \frac{8}{81}\mu J^{-\frac{2}{3}}\mathrm{tr}(\boldsymbol{C})\right]\boldsymbol{F}^{-T} \otimes \boldsymbol{F}^{-T} \otimes \boldsymbol{F}^{-T} \otimes \boldsymbol{F}^{-T} - \frac{8}{27}\mu J^{-\frac{2}{3}}\boldsymbol{F}^{-T} \otimes \boldsymbol{F}^{-T} \tag{B-4}$$
$$\otimes \boldsymbol{F}^{-T} \otimes \boldsymbol{F} - \frac{2}{3}\mu J^{-\frac{2}{3}}\frac{\partial^2(\boldsymbol{F}^{-T} \otimes \boldsymbol{F})}{\partial \boldsymbol{F} \partial \boldsymbol{F}} + \frac{4}{9}\mu J^{-\frac{2}{3}}\frac{\partial(\boldsymbol{F}^{-T} \otimes \boldsymbol{F})}{\partial \boldsymbol{F}} \otimes \boldsymbol{F}^{-T}$$
$$+ \frac{4}{9}\mu J^{-\frac{2}{3}}\frac{\partial(\boldsymbol{F}^{-T} \otimes \boldsymbol{F} \otimes \boldsymbol{F}^{-T})}{\partial \boldsymbol{F}} - \frac{8}{27}\mu J^{-\frac{2}{3}}\boldsymbol{F}^{-T} \otimes \boldsymbol{F} \otimes \boldsymbol{F}^{-T} \otimes \boldsymbol{F}^{-T}$$
$$- \frac{2}{3}\mu J^{-\frac{2}{3}}\frac{\partial^2(\boldsymbol{F} \otimes \boldsymbol{F}^{-T})}{\partial \boldsymbol{F} \partial \boldsymbol{F}} + \frac{4}{9}\mu J^{-\frac{2}{3}}\frac{\partial(\boldsymbol{F} \otimes \boldsymbol{F}^{-T})}{\partial \boldsymbol{F}} \otimes \boldsymbol{F}^{-T}$$
$$+ \frac{4}{9}\mu J^{-\frac{2}{3}}\frac{\partial(\boldsymbol{F} \otimes \boldsymbol{F}^{-T} \otimes \boldsymbol{F}^{-T})}{\partial \boldsymbol{F}} - \frac{8}{27}\mu J^{-\frac{2}{3}}\boldsymbol{F} \otimes \boldsymbol{F}^{-T} \otimes \boldsymbol{F}^{-T} \otimes \boldsymbol{F}^{-T} - \frac{2}{3}\mu J^{-\frac{2}{3}}\mathbb{I}_4$$
$$\otimes \frac{\partial \boldsymbol{F}^{-T}}{\partial \boldsymbol{F}} + \frac{4}{9}\mu J^{-\frac{2}{3}}\mathbb{I}_4 \otimes \boldsymbol{F}^{-T} \otimes \boldsymbol{F}^{-T}$$

where

$$\left[\frac{\partial \bar{\mathcal{T}}}{\partial \boldsymbol{F}}\right]_{ijklmnpq} = -F_{jp}^{-1}F_{qk}^{-1}F_{lm}^{-1}F_{ni}^{-1} - F_{jk}^{-1}F_{lp}^{-1}F_{qm}^{-1}F_{ni}^{-1} - F_{jk}^{-1}F_{lm}^{-1}F_{np}^{-1}F_{qi}^{-1} - F_{lp}^{-1}F_{qi}^{-1}F_{jm}^{-1}F_{nk}^{-1}$$
$$- F_{li}^{-1}F_{jp}^{-1}F_{qm}^{-1}F_{nk}^{-1} - F_{li}^{-1}F_{jm}^{-1}F_{np}^{-1}F_{qk}^{-1}$$

$$\left[\frac{\partial}{\partial \boldsymbol{F}}\left(\frac{\partial \boldsymbol{F}^{-T}}{\partial \boldsymbol{F}} \otimes \boldsymbol{F}^{-T}\right)\right]_{ijklmnpq} = F_{lp}^{-1}F_{qi}^{-1}F_{jk}^{-1}F_{nm}^{-1} + F_{li}^{-1}F_{jp}^{-1}F_{qk}^{-1}F_{nm}^{-1} + F_{li}^{-1}F_{jk}^{-1}F_{np}^{-1}F_{qm}^{-1}$$

$$\left[\frac{\partial^2\left(\boldsymbol{F}^{-T} \otimes \boldsymbol{F}^{-T}\right)}{\partial \boldsymbol{F}\partial \boldsymbol{F}}\right]_{ijklmnpq}$$

$$= F_{lp}^{-1}F_{qk}^{-1}F_{jm}^{-1}F_{ni}^{-1} + F_{lk}^{-1}F_{jp}^{-1}F_{qm}^{-1}F_{ni}^{-1} + F_{lk}^{-1}F_{jm}^{-1}F_{np}^{-1}F_{qi}^{-1} + F_{jp}^{-1}F_{qi}^{-1}F_{lm}^{-1}F_{nk}^{-1}$$

$$+ F_{ji}^{-1}F_{lp}^{-1}F_{qm}^{-1}F_{nk}^{-1} + F_{ji}^{-1}F_{lm}^{-1}F_{np}^{-1}F_{qk}^{-1}$$

$$\left[\frac{\partial\left(\boldsymbol{F}^{-T} \otimes \boldsymbol{F}^{-T} \otimes \boldsymbol{F}^{-T}\right)}{\partial \boldsymbol{F}}\right]_{ijklmnpq} = -F_{jp}^{-1}F_{qi}^{-1}F_{lk}^{-1}F_{nm}^{-1} - F_{ji}^{-1}F_{lp}^{-1}F_{qk}^{-1}F_{nm}^{-1} - F_{ji}^{-1}F_{lk}^{-1}F_{np}^{-1}F_{qm}^{-1}$$

$$\left[\frac{\partial^2\left(\boldsymbol{F} \otimes \boldsymbol{F}^{-T}\right)}{\partial \boldsymbol{F}\partial \boldsymbol{F}}\right]_{ijklmnpq} = -\delta_{im}\delta_{jn}F_{lp}^{-1}F_{qk}^{-1} - \delta_{ip}\delta_{jq}F_{lm}^{-1}F_{nk}^{-1} + F_{ij}F_{lp}^{-1}F_{qm}^{-1}F_{nk}^{-1} + F_{ij}F_{lm}^{-1}F_{np}^{-1}F_{qk}^{-1}$$

$$\left[\frac{\partial\left(\boldsymbol{F} \otimes \boldsymbol{F}^{-T} \otimes \boldsymbol{F}^{-T}\right)}{\partial \boldsymbol{F}}\right]_{ijklmnpq} = \delta_{ip}\delta_{jq}F_{lk}^{-1}F_{nm}^{-1} - F_{ij}F_{lp}^{-1}F_{qk}^{-1}F_{nm}^{-1} - F_{ij}F_{lk}^{-1}F_{np}^{-1}F_{qm}^{-1}$$

$$\left[\frac{\partial}{\partial \boldsymbol{F}}\left(\frac{\partial \boldsymbol{F}^{-T}}{\partial \boldsymbol{F}} \otimes \boldsymbol{F}\right)\right]_{ijklmnpq} = F_{qi}^{-1}F_{lp}^{-1}F_{jk}^{-1}F_{mn} + F_{li}^{-1}F_{jp}^{-1}F_{qk}^{-1}F_{mn} - \delta_{mp}\delta_{nq}F_{li}^{-1}F_{jk}^{-1}$$

$$\left[\frac{\partial\left(\boldsymbol{F}^{-T} \otimes \boldsymbol{F}^{-T} \otimes \boldsymbol{F}\right)}{\partial \boldsymbol{F}}\right]_{ijklmnpq} = -F_{jp}^{-1}F_{qi}^{-1}F_{lk}^{-1}F_{mn} - F_{ji}^{-1}F_{lp}^{-1}F_{qk}^{-1}F_{mn} + \delta_{mp}\delta_{nq}F_{ji}^{-1}F_{lk}^{-1}$$

$$\left[\frac{\partial^2\left(\boldsymbol{F}^{-T} \otimes \boldsymbol{F}\right)}{\partial \boldsymbol{F}\partial \boldsymbol{F}}\right]_{ijklmnpq} = -\delta_{km}\delta_{ln}F_{qi}^{-1}F_{jp}^{-1} - \delta_{kp}\delta_{lq}F_{ni}^{-1}F_{jm}^{-1} + F_{kl}F_{qi}^{-1}F_{np}^{-1}F_{jm}^{-1} + F_{kl}F_{ni}^{-1}F_{qm}^{-1}F_{jp}^{-1}$$

$$\left[\frac{\partial\left(\boldsymbol{F}^{-T} \otimes \boldsymbol{F} \otimes \boldsymbol{F}^{-T}\right)}{\partial \boldsymbol{F}}\right]_{ijklmnpq} = -F_{ji}^{-1}F_{qm}^{-1}F_{np}^{-1}F_{kl} - F_{qi}^{-1}F_{jp}^{-1}F_{nm}^{-1}F_{kl} + \delta_{kp}\delta_{lq}F_{ji}^{-1}F_{nm}^{-1}$$

## Appendix C : Verification of the proposed RTO framework

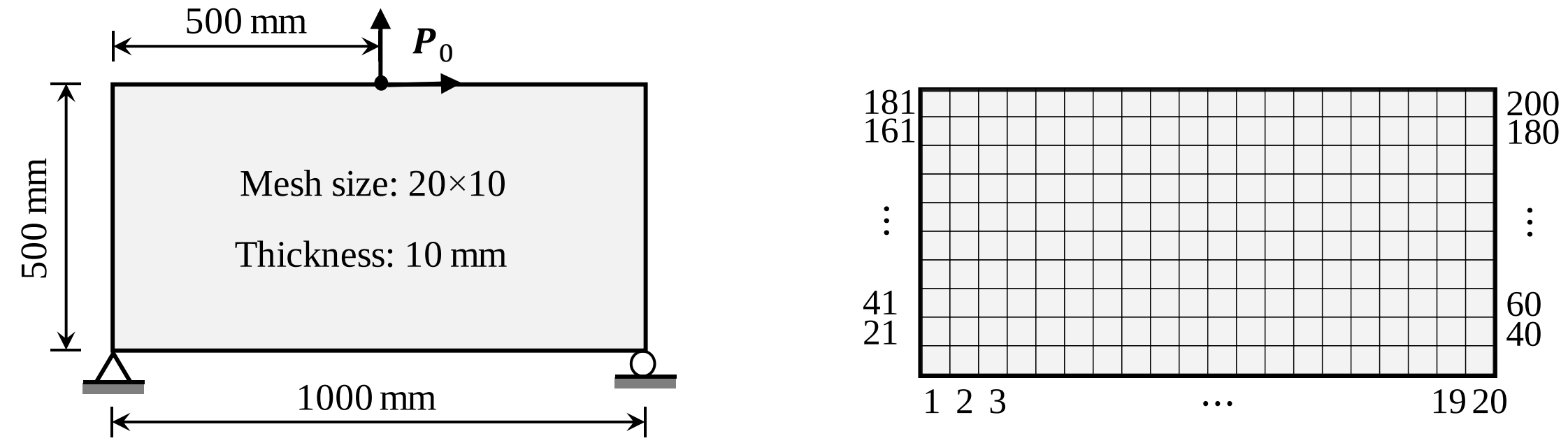

(a) Dimension, load, and boundary conditions (b) Element numbering

Figure C-1. Simply supported beam for the verification of sensitivity analysis and uncertainty quantification.

A simply supported beam in Figure C-1 is used to verify the stochastic perturbation method in Section 5 as well as the analytical sensitivity in Section 6. The beam is discretized by 200 four-node quadrilateral plane strain finite elements. The uncertainties related to loading, material, and geometry are all included. Specifically, the uncertainty in loads is modeled by a random vector $\boldsymbol{P}_0$ following a two-dimensional Gaussian distribution with mean $\boldsymbol{\mu}_P = [2 \quad -10]^T$ N and covariance matrix $\boldsymbol{\Sigma}_P = \sigma_P^2 \begin{bmatrix} 1 & 0 \\ 0 & 1 \end{bmatrix}$, which results in $m_1 = 2$ random variables. Young's modulus of the solid material in SIMP is considered to be uncertain and modeled with a random field with correlation lengths $l_{cx} = l_{cy} = 800$ mm and marginal PDF $E_0 \sim \text{Lognormal}(50, 6.26)$, which corresponds to 53.29% coefficient of variation. The projection filter threshold is taken as a random field with the same correlation function as Young's modulus and marginal PDF $\eta(\boldsymbol{X}, \omega) \sim \mathcal{U}(0,1)$. After KL expansion, there is one random variable arising from the material uncertainty ($m_2 = 2$) and one random variable from geometric uncertainty ($m_3 = 2$), from which the first eigenmode of the correlation matrix takes up 96.37% of the sum of all eigenvalues. The density filter radius is set to be $r = 75$ mm.

### *C.1 Verification of stochastic perturbation method*

To verify the accuracy of the 2nd order perturbation method, the mean and standard deviation of the compliance for the simply supported beam in Figure C-1 are computed using 10,000 Monte Carlo (MC) samples. The design variable for each element is set to be $x_e$ = 0.5, $e = 1, \ldots, n_{ele}$. From Table C-1, it can be shown that the results from the 2nd order perturbation method match closely with those from the MC method with all relative errors smaller than 0.4% when the standard deviation $\sigma_P$ is smaller than 4 N, i.e., coefficient of variation of the vertical load is smaller than 40%. With the above verification study, the correctness of the implementation of the stochastic perturbation method is established.

Table C-1. Comparison of the statistics of the compliance from stochastic perturbation and MC methods.

| $\sigma_P$ | Statistical measures (Unit: N·mm) | Stochastic perturbation method | MC method | Relative error (%) |
|---|---|---|---|---|
| 1 N | Mean | 1.4861 | 1.4868 | 0.0471 |
| | Standard deviation | 0.3119 | 0.3114 | 0.1606 |
| 2 N | Mean | 1.5131 | 1.5122 | 0.0595 |
| | Standard deviation | 0.4380 | 0.4392 | 0.2732 |
| 4 N | Mean | 1.5669 | 1.5682 | 0.0829 |
| | Standard deviation | 0.6212 | 0.6199 | 0.3716 |

### *C.2 Verification of sensitivity analysis*

To verify the adjoint sensitivity analysis framework, the design sensitivity $d\oint_0/d\boldsymbol{x}$ for the simply supported beam is verified using the central difference method (CDM). The design variable for each element is set to be $x_e = 0.5$, $e = 1, \ldots, n_{ele}$. The standard deviation of the random vector $\boldsymbol{P}_0$ is set to be $\sigma_P$ = 4 N. In CDM, the design sensitivity for the *e^th^* finite element is approximated by

$$\frac{d\oint_0}{dx_e} \approx \frac{\oint_0(x_1, \cdots, x_e + h, \cdots, x_{nele}) - \oint_0(x_1, \cdots, x_e - h, \cdots, x_{nele})}{2h} \quad \text{(C-1)}$$

where the mean and variance quantities in $\not{f}_0(x_1, \cdots, x_e + h, \cdots, x_{nele})$ and $\not{f}_0(x_1, \cdots, x_e - h, \cdots, x_{nele})$ are computed by evaluating Eqns. (29) and (30), respectively.

A perturbation value of $h = 10^{-6}$ is used for the CDM. As shown in Figure C- 2, the sensitivities computed using the adjoint method match closely with those from the CDM with relative errors less than around $10^{-4}$. Hence, the adjoint sensitivity analysis can be safely used in gradient-based optimizers.

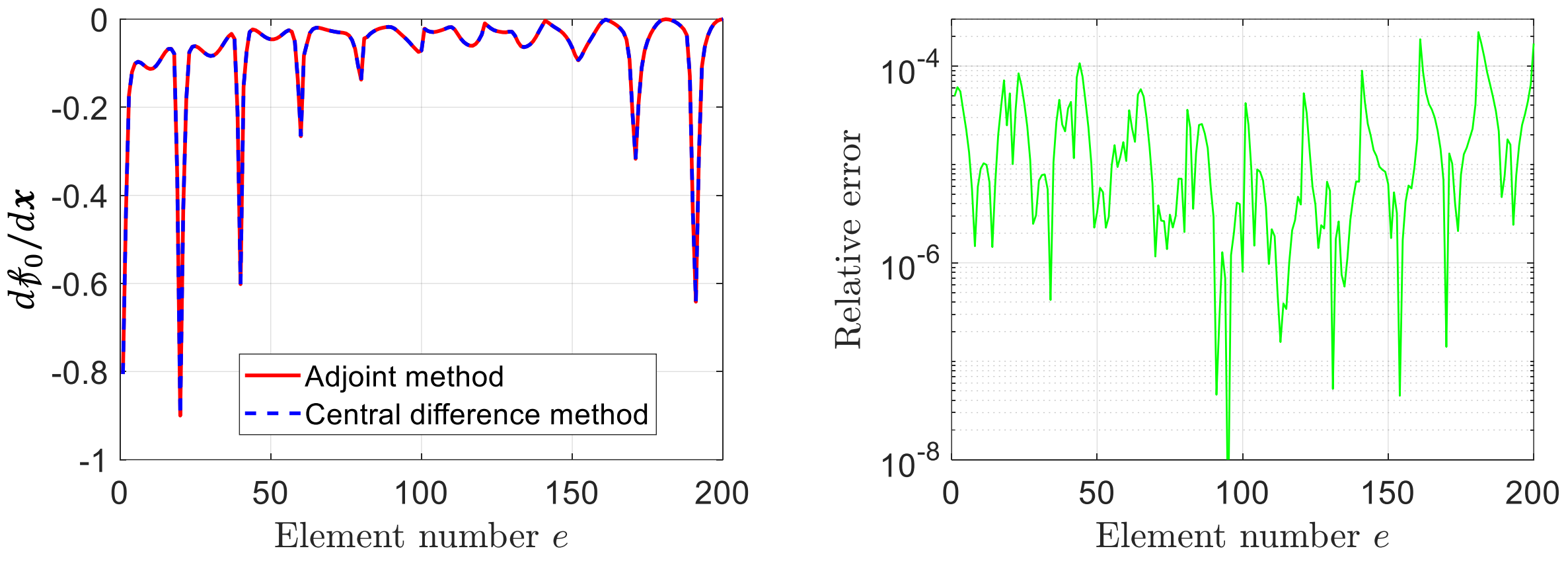

(a) $d\not{f}_0/d\boldsymbol{x}$ (b) Relative error of $d\not{f}_0/d\boldsymbol{x}$

Figure C- 2. Sensitivity comparison between the adjoint method and the central difference method.